\newcommand{\beqas}{\begin{eqnarray*}}
\newcommand{\eeqas}{\end{eqnarray*}}
\newcommand{\beqa}{\begin{eqnarray}}
\newcommand{\eeqa}{\end{eqnarray}}			
\newcommand{\beq}{\begin{equation}}
\newcommand{\eeq}{\end{equation}}

\newcommand{\ie}{\begin{equation}\begin{aligned}}
\newcommand{\fe}{\end{aligned}\end{equation}}
\newcommand{\comred}{\textcolor[rgb]{0.82,0.01,0.11}}
\newcommand{\ab}[2]{\langle #1  #2 \rangle}
\newcommand{\abBB}[2]{\langle \bar{#1}  \bar{#2} \rangle}
\newcommand{\abB}[2]{\langle {#1}  \bar{#2} \rangle}

\newcommand{\la}{\langle}
\newcommand{\ra}{\rangle}

\newcommand{\eqs}[1]{\begin{equation}\begin{split}#1\end{split}\end{equation}}
\newcommand{\vev}[1]{\langle #1 \rangle}

\newcommand{\ET}{E_T}
\newcommand{\prodE}{\prod_{i=1}^3 E_i}

\newcommand{\abn}[1]{\langle #1 \rangle}
\newcommand{\disc}{\mathop{\mathrm{Disc}}}
\newcommand{\ang}[2]{\langle #1\,#2\rangle}

\newcommand{\mix}[2]{\langle #1\,\bar{#2}\rangle}

\documentclass[11pt,a4paper]{article}

\usepackage{jheppub}
\usepackage{amsmath}
\usepackage{multicol,bbm}
\usepackage{enumerate}
\usepackage{bibentry}

\usepackage{amssymb}
\usepackage{bm}
\usepackage{multirow}
\usepackage{tabularx}
\usepackage{enumitem}
\usepackage{extarrows}
\usepackage{tikz}
\usepackage{verbatim} % 多行注释
\usepackage{float} % 圖表固定位置

\graphicspath{{figures/}}  % 定义所有的 .eps 文件在 figures 子目录下

\def\be{\begin{equation}}
\def\ee{\end{equation}}
\def\ba{\begin{eqnarray}}
\def\ea{\end{eqnarray}}

\def\CP1{\mathbb{CP}^1}
\def\SL2C{\mathrm{SL}(2,\mathbb{C})}

\def\Z2{\mathbb{Z}_2}

\def\su2{{SU(2)}}

\def\[{\left[}
\def\]{\right]}

\def\({\left(}
\def\){\right)}
\def\[{\left[}
\def\]{\right]}

\def\<{\langle}
\def\>{\rangle}

\def\i2{\frac{i}{2}}

\def\2F1{\,_2{\rm F}_1}

\usepackage{tikz}
\usetikzlibrary{shapes.geometric, arrows, positioning}
\usepackage{todonotes}
\usepackage{slashed}
\usepackage[most]{tcolorbox}
\usepackage{empheq} 
\begin{document}

\preprint{MPP-2026-62}

\title{Beyond Discontinuities: Cosmological WFCs and the Supersymmetric Orthogonal Grassmannian}

% author. Explanatory text should go in the []'s, actual e{-}mail

%\author{}
%\email[]{Your e{-}mail address}
%\thanks{}
%\affiliation{}
\date{\today}

\author[a,b,c]{Yu-tin Huang,}
\author[d]{Chia-Kai Kuo,}
\author[a]{Yohan Liu}
\author[e]{and Jiajie Mei }

% The "\note" macro will give a warning: "Ignoring empty anchor..."
% you can safely ignore it.

\affiliation[a]{Department of Physics and Center for Theoretical Physics, National Taiwan University, Taipei 10617, Taiwan}
\affiliation[b]{Physics Division, National Center for Theoretical Sciences, Taipei 10617, Taiwan}
\affiliation[c]{Max Planck{-}IAS{-}NTU Center for Particle Physics, Cosmology and Geometry, Taipei 10617, Taiwan}
\affiliation[d]{Max-Planck-Institut für Physik, Werner-Heisenberg-Institut, D-85748 Garching bei München, Germany
}
\affiliation[e]{Institute of Physics, University of Amsterdam, Amsterdam, 1098 XH, The Netherlands }

% e-mail addresses: one for each author, in the same order as the authors
\emailAdd{yutinyt@gmail.com}
\emailAdd{chia-kai.kuo@mpp.mpg.de}
\emailAdd{youan1997@icloud.com}
\emailAdd{j.mei@uva.nl}

% The "\note" macro will give a warning: "Ignoring empty anchor..."
% you can safely ignore it.

% e-mail addresses: one for each author, in the same order as the authors

\abstract{We construct an $\mathcal N=2$ supersymmetric Grassmannian representation of tree-level wavefunction coefficients (WFCs) by combining Grassmannian representations of energy discontinuities with an inversion formula. Since the orthogonal Grassmannian captures homogeneous solutions of the spinor conformal Ward identities, while current WFCs satisfy inhomogeneous Ward identities, the full WFC is obtained by reconstructing the energy-dependent prefactors from a basis of discontinuities. We first demonstrate this mechanism at three points, where the triple discontinuity determines the transverse current WFC and admits a supersymmetric uplift. At four points, we invert a spanning set of five current discontinuities and embed the result in momentum superspace using super-orthogonal-Grassmannian invariants generated by $\hat\delta(C\Omega\Xi^I)$. This yields the full four-point super WFC in Grassmannian form. We show that the two orthogonal-Grassmannian branches organize distinct supersymmetric invariants and reduce, in the flat-space limit, to different helicity superamplitudes.

}

%\pacs{}
% insert suggested keywords {-} APS authors don't need to do this
%\keywords{}

\maketitle
%must follow title, authors, abstract, \pacs, and \keywords
%\tableofcontents

\section{Introduction}

The cosmological bootstrap programme seeks to reconstruct cosmological correlators, or equivalently wavefunction coefficients (WFCs), directly from their singularity structure, symmetries, and factorization properties, thereby providing an on-shell alternative to bulk perturbation theory~\cite{Arkani-Hamed:2017fdk, Arkani-Hamed:2018kmz, Baumann:2019oyu, Baumann:2020dch, Chen:2025foq, Chen:2025ljl, Pajer:2020wxk, Jazayeri:2021fvk}. A central theme is that many analytic structures familiar from flat-space scattering amplitudes persist in cosmology, but in a form adapted to observables defined at the future boundary. Total-energy poles encode flat-space amplitudes, partial-energy singularities capture factorization into lower-point data, and conformal Ward identities impose differential constraints on the allowed functions.

In scattering amplitudes, such constraints often become most transparent when the observable is represented as an integral over a Grassmannian. For massless amplitudes one has schematically~\cite{Arkani-Hamed:2009ljj, Cachazo:2012da, Herrmann:2016qea}
\begin{equation}
A_n\sim\int_{\mathcal C}\frac{d^{k\times n}C}{\mathrm{GL}(k)}\; f_n(C)\,
\delta^{k\times 2}(C\cdot \lambda)\,
\delta^{(n-k)\times 2}(C^\perp\cdot \bar\lambda)\,,
\end{equation}
where the dynamics are encoded in the singularities of the integrand and in the contour. In maximally supersymmetric Yang--Mills theory this perspective culminates in the Amplituhedron~\cite{Arkani-Hamed:2013jha}. Since flat-space amplitudes arise as residues of WFCs at total-energy poles, it is natural to ask whether the Grassmannian structures of amplitudes are shadows of a larger geometry underlying the full cosmological observable.

A major step in this direction was made in the cosmological-Grassmannian construction of~\cite{Arundine:2026fbr}, where three-dimensional WFC data were related to integrals over the orthogonal Grassmannian. The orthogonal Grassmannian is the space of null $n$-planes inside a $2n$-dimensional vector space equipped with a symmetric metric $\Omega$.
Schematically,
\begin{equation}\label{eq: Orthogonal1}
\Psi_n\sim
\int_{\mathcal C}\frac{d^{n\times 2n}C}{\mathrm{GL}(n)}\; f_n(C)\,
\delta^{n\times n}(C\cdot \Omega\cdot C^T)\,
\delta^{n\times 2}(C\cdot \Omega\cdot \Lambda)\,,
\end{equation}
where $\Lambda=\{\lambda_1,\ldots,\lambda_n,\bar\lambda_1,\ldots,\bar\lambda_n\}$ denotes the three-dimensional spinor-helicity data, with $p_i^{\alpha\beta}=\lambda_i^{(\alpha}\bar\lambda_i^{\beta)}$. Orthogonal Grassmannians have also appeared in planar Ising networks~\cite{Galashin:2020jvd, Huang:2018nqf} and in ABJM amplitudes~\cite{Huang:2013owa,Huang:2014xza}. In the cosmological setting, they provide a geometric representation of discontinuities and flat-space limits of WFCs.

It is important, however, to distinguish what the orthogonal Grassmannian represents directly from what must be reconstructed. The integral in eq.~\eqref{eq: Orthogonal1} naturally furnishes homogeneous solutions of the spinor-helicity Ward identities generated by
\begin{equation}
\Lambda^T\cdot\Omega\cdot\Lambda,\qquad
\Lambda^T\cdot\Omega\cdot\partial_\Lambda,\qquad
(\partial_\Lambda)^T\cdot\Omega\cdot\partial_\Lambda\,.
\end{equation}
The first two operators impose momentum conservation and dilatation invariance. The third is the spinor conformal boost. For conformal matter fields, such as the dimension-two scalar and the massless fermion, this operator agrees with the ordinary conformal boost acting on the appropriate helicity-projected WFCs. For conserved currents, by contrast, the conformal Ward identity contains longitudinal terms and lower-point WFCs~\cite{Baumann:2020dch,Maldacena:2011nz}. The full current WFC is therefore an inhomogeneous solution and is not itself obtained directly from the homogeneous Grassmannian integral. This is precisely why the cosmological Grassmannian naturally applies to selected energy discontinuities of spinning WFCs: after taking appropriate discontinuities, the longitudinal inhomogeneous pieces are projected out, leaving homogeneous data that can be represented by residues of the orthogonal Grassmannian.

The central point of the present work is that these Grassmannian discontinuities can be inverted to reconstruct the full WFC. The input is that the energy dependence of the WFC is known. For example, at four points the color-ordered current WFC can be written as
\begin{equation}
\langle JJJJ\rangle=
\sum_{e=1}^5 \mathcal N_e(\epsilon,p,E^2)\,F_e(E)\,,
\end{equation}
where the five energy functions $F_e(E)$ form a convenient basis and the numerators $\mathcal N_e$ are independent of the energy discontinuity operation. Acting with a spanning set of discontinuities gives a linear system
\begin{equation}
\mathrm{Disc}_i\langle JJJJ\rangle
=
\sum_{e=1}^5 \mathcal N_e(\epsilon,p,E^2)\,\mathrm{Disc}_i F_e(E)\,.
\end{equation}
Importantly, the chosen discontinuities are precisely the homogeneous data represented by Grassmannian residues: for each discontinuity one can choose an integrand $f(C)$, or a sum of integrands, so that the Grassmannian integral in eq.~(\ref{eq: Orthogonal1}) returns $\mathrm{Disc}_i\langle JJJJ\rangle(C)$. One may then invert the linear system, solve for the numerators $\mathcal N_e(C)$, and substitute the result back into the original energy-function expansion. This gives
\begin{equation}
\langle JJJJ\rangle(C)
=
\sum_{i\in\mathrm{disc}}\mathcal P_i(E)\,
\mathrm{Disc}_i\langle JJJJ\rangle(C)\,,
\end{equation}
where the $\mathcal P_i(E)$ are explicit energy prefactors. Thus the orthogonal Grassmannian supplies the homogeneous discontinuity data, and the inversion formula restores the full energy dependence.

At three points this mechanism is particularly simple. The transverse current WFC has a single energy function, so the triple discontinuity immediately determines the full WFC. Since this discontinuity has an orthogonal-Grassmannian representation, the full three-point WFC is a Grassmannian expression dressed by a universal energy prefactor. At four points the same idea becomes nontrivial: the full color-ordered current WFC is reconstructed from a five-discontinuity basis. We choose the discontinuities so that all longitudinal components are projected out and each element of the basis admits a Grassmannian representation. Inverting the corresponding five-by-five system yields explicit prefactors $\mathcal P_i(E)$ with no spurious poles. This gives the full four-point current WFC in terms of Grassmannian residues dressed by energy functions. Only the all-plus values of the five basis discontinuities are independent data: hat relations, which flip the helicity of a discontinued leg, together with permutations of the legs, give the discontinuities of every other helicity sector. If the prefactors are allowed to depend on the helicity sector, the same WFC takes a much more compact form: each of the four Yang--Mills helicity components with $h_2=h_4=+$ is a channel block dressed by a short prefactor, plus at most three residues.

We then supersymmetrize this reconstruction. We work in the three-dimensional momentum superspace obtained by Fourier transforming the fermionic variables of position superspace~\cite{Jain:2023idr, Bala:2025jbh}. In this representation the supercharges take the form
\begin{equation}
Q^{\alpha,I}
=-2\lambda^\alpha \frac{\partial}{\partial \xi_{+I}} - \frac{\bar{\lambda}^\alpha}{4} \xi_{+}^I\,,
\end{equation}
while the covariant derivatives act on the variables $\xi_-^I$. The $\xi_+$ expansion organizes components related by supersymmetry Ward identities, whereas the $\xi_-$ dependence tracks the longitudinal and contact terms associated with constrained superfields. These contact terms are essential: conserved currents sit inside constrained superfields, and the constraints hold inside correlation functions only up to contact singularities. They live in the longitudinal part of the super WFC, which the discontinuities project out; at three points we fix it completely from Ward--Takahashi identities.

The Grassmannian implementation of supersymmetry dresses the bosonic integrand with the operator-valued fermionic delta function $\hat\delta(C\cdot\Omega\cdot\Xi^I)$ acting on a test function $T_n(\xi_{i,\pm}^I)$ that selects the helicity sector. For two and three points the bosonic constraints localize the integral onto the two branches of the orthogonal Grassmannian. We show that these branches have a direct supersymmetric interpretation: different branches support different superinvariants, and in the total-energy-pole limit they reduce to different helicity superamplitudes.

The same structure extends to four points. After deriving the relevant $\operatorname{OG}(4,8)$ superinvariants on the top cell and on codimension-one cells, we combine them with the five-discontinuity inversion formula. A priori the four-point super WFC carries one Grassmannian integrand $F^h(C)$ for each of the sixteen helicity configurations $h$: the configurations differ in their $\xi_{i,-}$ dressing, on which the supercharges do not act, so supersymmetry does not relate their integrands. We find instead that the integrand depends on the configuration only through its branch, $b=(-1)^{n_-}$ with $n_-$ the number of negative helicities. The test functions can then be summed within each branch, $T_4^{\rm even}$ and $T_4^{\rm odd}$, before they are dressed, and the transverse four-point $\mathcal N=2$ super WFC takes the form
\begin{align}
\langle \mathbf J_0^T\mathbf J_0^T\mathbf J_0^T\mathbf J_0^T\rangle(C)
=
\sum_{i\in\mathrm{disc}}\mathcal P_i(E)
\Big(&\left.F^{+}_i(C)\,\hat\delta(C_4\cdot\Omega\cdot\Xi_4^I)\,T_4^{\rm even}\right|_{\mathcal C^{+}_i}\nonumber\\
&+\left.F^{-}_i(C)\,\hat\delta(C_4\cdot\Omega\cdot\Xi_4^I)\,T_4^{\rm odd}\right|_{\mathcal C^{-}_i}\Big)\,,
\end{align}
with one integrand $F^{\pm}_i(C)$ and one contour $\mathcal C^{\pm}_i$ per branch and channel, and with the same energy prefactors $\mathcal P_i(E)$ as for the currents. This grouping is more than the construction requires, and we read it as a sign that the super WFC carries more symmetry than we have used, perhaps an uplift to a geometric coordinate in which the branches and contours are produced rather than imposed. As a check we project the result onto $\langle O_2J^+O_2J^+\rangle$ and find that the projected functions are precisely the discontinuities of that component; equivalently, the result is the direct inversion of $\langle O_2J^+O_2J^+\rangle$, which shares the energy basis of $\langle JJJJ\rangle$. The compact forms uplift as well: there the integrands depend on the full helicity configuration, so the branch grouping is lost, but the components still come out correctly and the projection gives a compact form of $\langle O_2J^+O_2J^+\rangle$ itself. Finally, the $s$-channel cell of the all-plus component reproduces the expected super cutting rule, dressed by the same energy prefactor that appears in the compact forms.

The rest of the paper is organized as follows. In section~2 we review the bosonic orthogonal-Grassmannian representation of discontinuities, emphasizing why conserved-current WFCs require discontinuities before they admit a homogeneous Grassmannian form, and then introduce three-dimensional momentum superspace and the Grassmannian superinvariants. In section~3 we develop the inversion formula at low multiplicity and construct the two-point and the transverse three-point super WFCs, together with their flat-space limit and component projections. In section~4 we apply the inversion to the four-point current WFC, give the compact forms of its helicity components, construct the transverse four-point super WFC with its branch structure, and check it against the $\langle O_2J^+O_2J^+\rangle$ component and the super cutting rule. We conclude with open directions. Appendix~\ref{app:hat} derives the hat relations and the discontinuities of all helicity sectors, appendix~\ref{app:full-three-point-super-wfcs} constructs the longitudinal part of the three-point super WFC at $\mathcal N=1$ and $\mathcal N=2$, and appendix~\ref{app:invariants} collects the explicit superinvariants at two, three and four points.

\textbf{Notation and conventions}: Throughout this paper we work with wavefunction coefficients (WFCs) with the overall $\delta^3(\sum_{i=1}^n p_i)$ stripped, rather than with in-in cosmological correlators. We use the convention $\langle\bar\lambda\,\lambda\rangle=2E$.

During the completion of this draft, we became aware of related work by Aswini Bala, Sachin Jain, Dhruva K.S., and Adithya A Rao, which should appear concurrently~\cite{Bala:2026hdm,Bala:2026bdx}.

\section{Review}
This section reviews the ingredients needed below: the bosonic orthogonal-Grassmannian representation of homogeneous momentum-space data, its application to energy discontinuities of spinning WFCs, and the supersymmetric extension in three-dimensional momentum superspace.

\subsection{The orthogonal Grassmannian}
Grassmannian integral formulae provide a compact way of constructing objects invariant under nonlinear constraints. In flat-space amplitudes these constraints include momentum conservation, conformal boosts, and, in supersymmetric theories, superconformal transformations. In three-dimensional spinor-helicity variables the relevant invariants can be written as integrals over the orthogonal Grassmannian~\cite{Lee:2010du, Huang:2010qy, Arundine:2026fbr}
\begin{equation}\label{eq: Orthogonal3}
\int_{\mathcal{C}}\frac{d^{n\times 2n}C}{\mathrm{GL}(n)}\; f_n(C)\,
\delta^{n\times n}(C\cdot \Omega \cdot C^T)\,
\delta^{n\times 2}(C \cdot\Omega\cdot\Lambda)\,,
\end{equation}
Here $C$ is an $n\times 2n$ matrix, and $\Omega$ is the symmetric metric used to define the inner product on the ambient $2n$-dimensional space. The kinematic data are collected into $\Lambda=\{\lambda^\alpha_1,\cdots,\lambda^\alpha_n,\bar{\lambda}^\alpha_1,\cdots,\bar{\lambda}^\alpha_n\}$. The integral is over $\operatorname{OG}(n,2n)$, the moduli space of null $n$-planes in this space. The matrix $C$ gives homogeneous coordinates on the Grassmannian, defined modulo the usual GL$(n)$ redundancy, and the null condition is imposed by
\begin{equation}\label{eq: Orthogonal}
C\cdot \Omega \cdot C^T=0\,.
\end{equation}
For scattering amplitudes, the metric is diagonal with alternating signs~\cite{Lee:2010du, Huang:2010qy}. For WFCs, it is more natural to choose~\cite{Arundine:2026fbr} \begin{align}\label{eq: OmegaDef}
    \Omega=\begin{pmatrix}
	           0 & 1\!\!1_{n\times n} \\
	           1\!\!1_{n\times n} & 0 
	        \end{pmatrix}\,.
\end{align}
As discussed in~\cite{Lee:2010du}, such integrals naturally produce invariants of the generators
\begin{equation}
\Lambda^T\cdot\Omega\cdot\Lambda=2\sum_{i=1}^n \lambda_i^{(\alpha}\bar{\lambda}_i^{\beta)},\,\,\,
\Lambda^T\cdot\Omega\cdot\partial_{\Lambda}=\sum_{i=1}^n
\left(
\lambda_i^\alpha
\frac{\partial}{\partial\lambda_i^\beta}
+
\bar\lambda_i^\alpha
\frac{\partial}{\partial\bar\lambda_i^\beta}
\right)
,\,\,\,
(\partial_\Lambda)^T\cdot\Omega\cdot\partial_{\Lambda}=2\sum_{i=1}^n\partial_{\lambda_i^{(\alpha}}\partial_{\bar{\lambda}_i^{\beta)}}\,.
\end{equation}
These are the conformal generators in spinor-helicity variables. The invariance under the derivative operators $\Lambda^T\cdot\Omega\cdot\partial_{\Lambda}$ and $(\partial_\Lambda)^T\cdot\Omega\cdot\partial_{\Lambda}$ follows directly from the delta functions $\delta(C\cdot\Omega\cdot\Lambda)$ and $\delta(C\cdot\Omega\cdot C^T)$. The generator $\Lambda^T\cdot\Omega\cdot\Lambda$ also annihilates the integral: after completing the rows of $C$ to a basis of the ambient space and inserting the corresponding identity, the result is proportional to $\delta(C\cdot\Omega\cdot\Lambda)$.

To obtain rational functions of the external kinematics from eq.~(\ref{eq: Orthogonal3}), one must specify a contour after imposing the delta functions. The integral starts with $n^2$ independent variables. The bosonic constraints remove $n(n+1)/2+2n-3$ of them, where the subtraction by three accounts for the momentum-conservation constraints imposed on the external data. Thus $(n-2)(n-3)/2$ variables remain. Beginning at four points, these residual variables are localized by choosing contours around poles of $f_n(C)$.

Before turning to supersymmetry, we recall one important feature: the orthogonal Grassmannian has two branches. Because of the orthogonality constraint, each maximal minor is proportional, up to a sign, to its dual minor. The precise duality map depends on the choice of metric. For the metric in eq.~(\ref{eq: OmegaDef}), if $(I)$ denotes the minor associated with a set $I$ of $n$ column labels, its dual is
\begin{align}
    (I^\vee):=  \tau (I^c)\,.
\end{align}
where $I^c$ is the complement of $I$ and $\tau$ exchanges $i\leftrightarrow \bar i$. The two branches are then characterized by
\begin{align}\label{eq: Conjugate}
    (I)=\pm \operatorname{sign}\Big[\{\tau(I)  , I^\vee \}\Big]  (I^\vee)\,. 
\end{align}
where $\pm$ denotes the positive or negative branch, respectively, and $\operatorname{sgn}[\cdots]$ is the sign of the permutation needed to restore the canonical ordering $(1,2,\ldots,n,\bar1,\bar2,\ldots,\bar n)$. For example, on the two branches one finds
\begin{align}
     & (123)\rightarrow \pm \operatorname{sign}\Big[\{\bar 1,\bar 2,\bar 3,1,2,3\}\Big](123)=\mp(123)\,,\nonumber\\
     & (\bar 1\bar 2\bar 3) \rightarrow \pm \operatorname{sign}\Big[\{1,2,3,\bar 1,\bar 2,\bar 3\}\Big](\bar 1\bar 2\bar 3)=\pm( \bar 1\bar 2\bar 3)\,,\nonumber\\
    &(1 2 \bar 2) \rightarrow \pm \operatorname{sign}\Big[\{\bar 1, \bar 2, 2,1,3,\bar 3\}\Big] (1 3  \bar  3)=\mp ( 1 3 \bar 3)\,.
\end{align}
It follows immediately that $(123)$ must vanish on the positive branch, while $(\bar1\bar2\bar3)$ must vanish on the negative branch.

The orthogonal Grassmannian representation naturally captures the homogeneous
sector of the Ward identity generated by the total spinor conformal boost. In
spinor notation, we define
\begin{equation}
\sum_{\ell=1}^n \widetilde K_\ell^{\alpha\beta}
:=
\frac{1}{2}\left(\partial_\Lambda\right)^T \Omega\,\partial_\Lambda
=
\sum_{\ell=1}^n
\frac{\partial^2}
{\partial \lambda_{\ell(\alpha}\,
 \partial \bar\lambda_{\ell\beta)}} \, ,
\label{eq:spinor-conformal-boost}
\end{equation}
where the parentheses denote symmetrization of spinor indices. We write
$\widetilde K_\ell^i$ for the same operator in vector notation, and drop the
label $\ell$ when it acts on a single leg. For WFCs involving only conformal matter fields, such as the conformally coupled scalar i.e. scalar with $\Delta=2$, which we denote by $O_2$, and the massless spinor $\chi$, the action of the spinor conformal boost $\widetilde K_\ell^{\alpha\beta}$ is equivalent to that of the ordinary conformal boost $K^i$. Thus the total spinor conformal boost annihilates these WFCs as a consequence of the conformal Ward identities:
\begin{equation}
\sum_{\ell=1}^n \widetilde K_\ell^i \,
c_n(O_2,\chi)
=
0 \,,
\label{eq:homogeneous-cwi-matter}
\end{equation}
Here $c_n(O_2,\chi)$ denotes a WFC involving only $O_2$ and $\chi$ operators. We use $c_n$ for a generic WFC and reserve notation such as $\langle O_2O_2O_2\rangle$ when the external operators are specified. Any WFC built only from these conformal fields lies in the homogeneous sector of the Ward identity and may therefore be represented by an orthogonal-Grassmannian integral. A simple nontrivial three-point example is
\begin{equation}
f(C)=\left\langle O_2 \bar{\chi}^+ \bar{\chi}^+ \right\rangle(C)
=
\frac{
(1\,\bar{1}\,\bar{2})_+
(1\,\bar{1}\,\bar{3})_+
}{
(1\,\bar{1}\,2)_+
(3\,\bar{3}\,\bar{2})_+
} \,,
\label{eq:O2-chichibar-example}
\end{equation}
where we label the integrand $f(C)$ by the associated WFC. The corresponding WFC is obtained by evaluating the integral on the appropriate branch.

For operators involving conserved currents, however, there is an additional
subtlety. Because of the constraint
\begin{equation}
C\cdot \Omega \cdot \Lambda = 0 ,
\label{eq:C-Omega-Lambda-constraint}
\end{equation}
each column of the $C$-matrix carries a definite helicity weight, and hence so
do the minors appearing in the Grassmannian integrand. The natural object matched
by the Grassmannian is therefore not the full current correlator, but rather its
helicity-projected WFC.

This distinction is important because the action of the spinor conformal boost
on a transverse conserved current does not remain purely transverse. Instead, it
generates the longitudinal component as an inhomogeneous term. More precisely,
with the convention $J^\pm=\epsilon_\pm^j J_j$, one finds
\begin{equation}
\widetilde K^i J^\pm
=
-\,\epsilon_\pm^j K^i J_j
+
2\,\epsilon_\pm^i\,\frac{J_L}{E} .
\label{eq:spinor-boost-current}
\end{equation}
Here $J_L$ denotes the longitudinal component of the current and $E=|\vec p|$.
Consequently, a WFC involving conserved currents is, in general, an
inhomogeneous solution of the Ward identity generated by the total spinor
conformal boost. Such WFCs therefore cannot be directly represented by the
homogeneous orthogonal Grassmannian construction.

At first sight, this appears to be a fundamental obstruction. However, the
obstruction also points toward its own resolution. If one can define an operation
which both projects out the longitudinal contribution and commutes with the
conformal boost generator, then applying this operation to the WFC should produce
an object lying in the homogeneous sector. This object is then a natural
candidate for an orthogonal Grassmannian representation. The relevant operation is the energy discontinuity. With our conventions, we
define
\begin{equation}
\disc_i c_n
:=
c_n(\ldots,E_i,\ldots)
-
c_n(\ldots,-E_i,\ldots) ,
\label{eq:energy-discontinuity}
\end{equation}
where the subscript $i$ labels the energy variable across which the
discontinuity is taken. \footnote{
    In this paper, the helicity projection defined in \eqref{hel-projection} is performed only after taking the discontinuity:
    \ie
    \disc_{E_s} c^{h}:=(\disc_{E_s} c)|_{\epsilon \to \epsilon^{h}, \bar\chi \to \bar\chi^{h},  \chi \to \chi^{h}}.
    \fe
}Equivalently, this operation compares the two branches
associated with $E_i=\sqrt{p_i^2}$. Since changing the branch of $E_i$ does
not modify the form of the conformal Ward identities, the discontinuity maps
solutions of the CWI to solutions of the CWI. At the same time, with a suitable
choice of discontinuities, the longitudinal inhomogeneous terms in
\eqref{eq:spinor-boost-current} can be projected out.

We will now demonstrate this mechanism in explicit examples. In each case, an
appropriate choice of discontinuity removes the longitudinal contribution to the
Ward identity, leaving a helicity-projected WFC that satisfies a homogeneous
spinor conformal Ward identity. Such an object can then be represented by an
orthogonal Grassmannian integral.

\subsubsection{Disc$\langle JJJ\rangle$ as $OG$}

The discussion above implies that the Grassmannian directly captures only the homogeneous part of a WFC. For conserved currents this is not the full story. For example, the all-plus three-current WFC obeys
\ie
\label{inhomo eq}
\sum_{\ell=1}^3 \widetilde K_\ell^i
\langle J^+ J^+ J^+ \rangle
&
=
2\,\frac{\epsilon^{i}_{1,+}}{E_1}\,
\langle J_L J^+ J^+ \rangle
+
2\,\frac{\epsilon^{i}_{2,+}}{E_2}\,
\langle J^+ J_L J^+ \rangle
+
2\,\frac{\epsilon^{i}_{3,+}}{E_3}\,
\langle J^+ J^+ J_L \rangle .
\fe
Now act with the same Ward identity on the discontinuity in leg 1.  One obtains 
\ie
&\sum_{\ell=1}^3 \widetilde K_\ell^i
\disc_1 \langle J^+ J^+ J^+ \rangle
=
2\,\frac{\epsilon^{i}_{1,+}}{E_1}\,
\disc_1\langle J_L J^+ J^+ \rangle
+
2\,\frac{\epsilon^{i}_{2,+}}{E_2}\,
\disc_1\langle J^+ J_L J^+ \rangle
+
2\,\frac{\epsilon^{i}_{3,+}}{E_3}\,
\disc_1 \langle J^+ J^+ J_L \rangle
\fe
The first term vanishes after the discontinuity because the Ward--Takahashi identity gives \footnote{We define $J^+_{1+2}$ as the current at momentum $\vec p_1+\vec p_2$ contracted with the polarization $\epsilon^+_2$ of leg $2$; the polarizations of the legs not contracted with $p_1$ are retained.}
\ie\label{eq:WT-JJJ}
p_{1i}\langle J^i J^+ J^+ \rangle =\left(\langle J^+_{1+2} J^+_3 \rangle -\langle J^+_{2} J^+_{3+1} \rangle\right)
\fe 
The two-point functions on the right-hand side carry no explicit dependence on $E_1$, and are therefore killed by $\disc_1$. Applying the same argument successively to all three legs, with $\disc_{1,2,3}=\disc_1\disc_2\disc_3$, gives
\ie
&\sum_{\ell=1}^3 \widetilde K_\ell^i
\disc_{1,2,3} \langle J^+ J^+ J^+ \rangle
=0
\fe
The resulting triple discontinuity is therefore a homogeneous object, and is represented by the Grassmannian as
\ie
\label{disc123 JJJ}
\disc_{1,2,3} \langle J^+ J^+ J^+ \rangle=\frac{(\bar 1 \bar 2 \bar 3)_+^2}{(1 \bar 1 2)_+(3 \bar 3 \bar 2)_+}\,,\qquad \disc_{1,2,3} \langle J^- J^+ J^+ \rangle=-\frac{(1 \bar 2 \bar 3)_-^2}{(1 \bar 1 2)_-(3 \bar 3 \bar 2)_-}
\fe
When fewer conserved currents are present, fewer discontinuities are needed. For example, for $\langle J^+ O_2 O_2 \rangle$ a single discontinuity is sufficient:
\ie
\disc_{1} \langle J^+ O_2 O_2 \rangle(C)=\frac{(\bar 1 2 \bar 3)_+ (\bar 1 \bar 2 3)_+}{(1 \bar 1 2)_+(3 \bar 3 \bar 2)_+}\,.
\fe
Further discontinuities of a homogeneous solution remain homogeneous. Thus one also finds
\ie
\label{disc 123 JOO}
\disc_{1,2,3} \langle J^- O_2 O_2 \rangle(C)=-2\left(\frac{(2 \bar 2 1)_+^2}{(1 \bar 1 2)_+(3 \bar 3 \bar 2)_+}+\frac{(2 \bar 2 1)_-^2}{(1 \bar 1 2)_-(3 \bar 3 \bar 2)_-}\right)
\fe
When more discontinuities are taken than minimally required, the result is typically expressed as a combination of OG residues on the positive and negative branches.

\subsubsection{Disc$\langle JJJJ\rangle$ as $OG$}
\label{sec: discontinuity JJJJ}
We next turn to four-point WFCs. In this case the bosonic constraints leave a one-dimensional contour integral, which must be localized on poles of the integrand. Following the conventions of~\cite{Arundine:2026fbr}, we introduce the ``Mandelstam minors''
\begin{align}
    S:=(\bar 1 \bar 2 12)=\pm(\bar 3 \bar 4 34),\quad T:=(\bar 1 \bar 4 14)=\pm(\bar 2 \bar 3 23),\quad U:=(\bar 1 \bar 3 13)=\pm (\bar 2 \bar 4 24)\,.
\end{align}
The equalities, including their branch-dependent signs, follow from the orthogonality condition.

It is convenient to abbreviate the four combinations of these minors that appear as
poles of the integrands below,
\begin{equation}
A\equiv S+T+U\,,\qquad B\equiv S+T-U\,,\qquad C\equiv -S+T-U\,,\qquad D\equiv S-T-U\,,
\label{eq:ABCD}
\end{equation}
and we use them from here on.

As a warm-up, consider $\langle O_2 O_2 O_2 O_2\rangle$ with gluon exchange. Since the external operators are conformal matter fields, the WFC is already in the homogeneous sector and can be associated directly with an OG integral. One finds
\ie
\la O_2 O_2 O_2 O_2\ra(C) = \left.\frac{T-U}{SA}\right|_{(S=0,\alpha-\beta)_+ +(A=0,\beta-\alpha)_+} + (2 \leftrightarrow 3)  + (2 \leftrightarrow 4).
\fe
The subscript denotes both the pole and the branch.\footnote{\label{fn:orientation}Convention adopted throughout the paper: the residues at the two cells of every pole are oriented so that the residues of one integrand at all its poles sum to zero.} The explicit forms of $\alpha$ and $\beta$ are given in appendix~\ref{app:inv-4}. This expression corresponds to an adjoint scalar coupled to Yang--Mills; a four-point contact diagram would add an additional contribution proportional to $1/(S{+}T{+}U)$.

For the color-ordered current WFC $\langle JJJJ\rangle$, the longitudinal part is fixed by Ward identities in terms of lower-point WFCs with consecutive momenta combined~\cite{Armstrong:2020woi}. For example,
\ie
p_{1i}\langle J^i J J J \rangle 
=\left(\langle J_{1+2} J_3 J_4 \rangle -\langle J_{2} J_3 J_{4+1} \rangle\right).
\fe 
Taking the discontinuity in leg 1 projects out the longitudinal contribution associated with leg 1. For the planar color-ordered WFC, a minimal\footnote{At four points, a double discontinuity is minimal for color-ordered WFCs because only the planar $s$- and $t$-channel diagrams appear. For non-color-ordered WFCs involving all three channels, such as $\langle TTTT\rangle$~\cite{Arundine:2026fbr}, at least three discontinuities are needed to remove all lower-point inhomogeneous terms.} set of discontinuities that removes all longitudinal components is
\ie
\disc_{1,3} \la J J J J \ra,\quad \disc_{2,4} \la J J J J \ra\,.
\fe
The helicity components of these discontinuities admit Grassmannian representations. For example,
\ie
\disc_{1,3} \la J^+ J^+ J^+ J^+ \ra (C)&=\left.\frac{(S+T)(\bar 1 \bar 2 \bar 3 \bar 4)^2}{ST\,AB}\right|_{
    \begin{smallmatrix}
     (S=0,\beta)_+ + (T=0,\beta)_+
     \\
     +(A=0,\beta)_+ + (B=0,\alpha)_+
\end{smallmatrix}},\\
\disc_{2,4} \la J^+ J^+ J^+ J^+ \ra(C)&=\left.\frac{(S+T)(\bar 1 \bar 2 \bar 3 \bar 4)^2}{ST\,AB}\right|_{
    \begin{smallmatrix}
     (S=0,\beta)_+ + (T=0,\beta)_+
     \\
     +(A=0,\beta)_+ + (B=0,\beta)_+
\end{smallmatrix}}
\label{eq:disc13-disc24-allplus}
\fe
Taking further discontinuities selects subsets of these residues. In particular, discontinuities in $E_s=|p_1{+}p_2|$ and $E_t=|p_1{+}p_4|$ isolate the residues at $S=0$ and $T=0$, respectively:
\ie
\disc_{1,4,s} \la J^+ J^+ J^+ J^+ \ra&=\left.\frac{(S+T)(\bar 1 \bar 2 \bar 3 \bar 4)^2}{ST\,AB}\right|_{(S=0,\beta-\alpha)_+},\\
\disc_{1,2,t} \la J^+ J^+ J^+ J^+ \ra&=\left.\frac{(S+T)(\bar 1 \bar 2 \bar 3 \bar 4)^2}{ST\,AB}\right|_{(T=0,\beta-\alpha)_+}
\fe
The same discontinuity operation removes the lower-point inhomogeneous terms of the color-ordered WFC.

\paragraph{Other sectors and further discontinuities.}
The all-plus discontinuities above are almost all the data needed. Only the
all-plus values of the five basis discontinuities are independent:
$\disc_{1,3}$, $\disc_{2,4}$, one triple discontinuity and the channel
discontinuities $\disc_{1,4,s}$ and $\disc_{1,2,t}$. Every other discontinuity,
in every helicity sector, follows by hats or by permutations of the legs, as
we now outline and appendix~\ref{app:hat} carries out.
At fixed $\disc_{1,3}$ or $\disc_{2,4}$, the hat relations of
appendix~\ref{app:hat} transport helicities only through legs already
discontinued. The appendix lists the four helicity sectors of each double
discontinuity in eqs.~\eqref{eq:hat-D13}--\eqref{eq:hat-D24-flip24}.
For the crossed single-minus sectors below, we instead take an additional
energy difference at fixed polarizations, then use the hat relation for the
resulting triple discontinuity in reverse.

Start with the all-plus $\disc_{1,3}$ of
eq.~\eqref{eq:disc13-disc24-allplus} and take the difference in $E_2$:
\begin{equation}
\disc_{1,2,3}\la J^+J^+J^+J^+\ra
=\disc_2\bigl(\disc_{1,3}\la J^+J^+J^+J^+\ra\bigr).
\label{eq:triple-from-double-direct}
\end{equation}
This calculation, performed directly in the all-plus sector, produces
residues of the $CD$ kernel as well as the original $AB$ kernel. We checked
its residue result numerically against the energy-difference definition.
Reversal of the color ordering, $(1,2,3,4)\mapsto(4,3,2,1)$, then maps
this all-plus $\disc_{1,2,3}$ to $\disc_{2,3,4}$, whose explicit $CD$
residues appear in eq.~\eqref{eq:D234-allplus}. The latter can also be
checked by taking
$\disc_3\bigl(\disc_{2,4}\la J^+J^+J^+J^+\ra\bigr)$ from the
other known all-plus double discontinuity. The reversal sends the
discontinued set $\{1,2,3\}$ to $\{2,3,4\}$; exchanging legs $1$ and $4$
alone is not a symmetry of the fixed cyclic color ordering. On the all-plus
positive branch it preserves each of $S,T,U,A,B,C,D$, as in
eq.~\eqref{eq:reflection-minors}, so the $CD$ poles remain $CD$ poles.

Now use eq.~\eqref{eq:hat-extra-disc} in reverse. Because $\hat h_i^2=1$,
the two crossed single-minus double discontinuities can be recovered from
these all-plus triple discontinuities:
\begin{equation}
\begin{aligned}
\disc_{2,4}\la J^+J^+J^-J^+\ra
&=\hat h_3\Bigl(\disc_{2,3,4}\la J^+J^+J^+J^+\ra
-\disc_{2,4}\la J^+J^+J^+J^+\ra\Bigr),\\
\disc_{1,3}\la J^+J^-J^+J^+\ra
&=\hat h_2\Bigl(\disc_{1,2,3}\la J^+J^+J^+J^+\ra
-\disc_{1,3}\la J^+J^+J^+J^+\ra\Bigr).
\end{aligned}
\label{eq:crossed-from-triples}
\end{equation}
The all-plus $\disc_{2,4}$ has only the $AB$ kernel, while $\hat h_3$
preserves the $AB$ and $CD$ kernel types separately. The $CD$ residues in
$\disc_{2,3,4}$ therefore require a $CD$ contribution in the single-minus
$\disc_{2,4}$, within the residue representation used here. A cyclic shift
by two legs gives the other single-minus sector; on the negative branch it maps
each kernel to minus itself, eq.~\eqref{eq:shift2-minors}. Both use
$(T-S)n_H/(ST\,CD)$ alongside $(S+T)n_H/(ST\,AB)$, with their cell
coefficients given in eq.~\eqref{eq:hat-D24}. The second line of
eq.~\eqref{eq:crossed-from-triples} similarly determines the crossed
$\disc_{1,3}$. Here $\hat h_2$ exchanges the two kernel types, so the
$CD$ terms of the triple discontinuity do not require a second kernel in that
crossed double discontinuity; all crossed $\disc_{1,3}$ sit on the $AB$ kernel
alone, eq.~\eqref{eq:hat-D13-crossed}.

These single-minus sectors are thus out of reach only for hat transport at
fixed $\disc_{1,3}$ or $\disc_{2,4}$: such transport can flip helicities
only on the legs already discontinued. The additional energy difference
supplies the missing data, and the inverted hat relation transports it back
to the crossed sector. The MHV $\disc_{2,4}$, also outside the fixed-channel
hat orbit, follows in the same way from the all-plus quadruple discontinuity,
which on the five-function energy basis equals
$2(\disc_{1,4,s}+\disc_{1,2,t})$ and is no new datum. The inversion of
section~\ref{sec:SUSY-from-OG} needs $\disc_{2,4}$ in these sectors.
We write $n_H$ for the numerator carrying the helicities of the sector,
$n_H=(a_1\,\bar2\,a_3\,\bar4)^2$ with $a_i=i$ on the negative-helicity legs
$H\subseteq\{1,3\}$ and $a_i=\bar i$ otherwise, as in
eq.~\eqref{eq:compact-assembly}. The MHV one lands on the very contour of the all-plus $\disc_{2,4}$ in
eq.~\eqref{eq:disc13-disc24-allplus}, with nothing but the numerator changed,
\ie
\disc_{2,4} \la J^- J^+ J^- J^+ \ra(C)&=\left.\frac{(S+T)(1 \bar 2 3 \bar 4)^2}{ST\,AB}\right|_{
    \begin{smallmatrix}
     (S=0,\beta)_+ + (T=0,\beta)_+
     \\
     +(A=0,\beta)_+ + (B=0,\beta)_+
\end{smallmatrix}}
\label{eq:disc24-mhv}
\fe
The two single-minus lines, given with their explicit contours in
eq.~\eqref{eq:hat-D24}, have the two integrands identified above. Together
with their hat images they reproduce eq.~\eqref{eq:D234-allplus} and assemble
the other triple-discontinuity sectors in eq.~\eqref{eq:hat-D234}.

The same contours reach beyond the pure-current family. Replacing $n_H$ by
$X^2$ with $X=(1\,\bar1\,\bar2\,\bar4)$ in the four sectors $H\subseteq\{1,3\}$
and adding them with the sign $(-1)^{|H|}$ trades the currents on legs $1$
and $3$ for the dimension-two scalar and gives the discontinuities of
$\la O_2J^+O_2J^+\ra$, collected in appendix~\ref{app:ojoj}; those are the ones
section~\ref{sec:SUSY-from-OG} projects the super WFC onto.

\subsection{3D momentum superspace}

We review the three-dimensional momentum superspace used in the rest of the paper. It is a fermionic extension of massive spinor-helicity variables, with
\begin{equation}
p^{\alpha\beta}=\lambda^{(\alpha}\bar{\lambda}^{\beta)}, \quad\langle  \bar{\lambda} \lambda \rangle=2E=2|p|\,,
\end{equation}
where $\alpha,\beta$ are SL$(2,\mathbb R)$ spinor indices. The supersymmetric extension starts from
\begin{equation}
\{Q^{I\alpha},Q^{J\beta}\}=\delta^{IJ}p^{\alpha\beta}, 
\end{equation}
where $I,J=1,\cdots,\mathcal N$ label the SO$(\mathcal N)$ R-symmetry. The supercharges can be represented on on-shell data by introducing Grassmann variables $(\eta^I,\bar\eta^I)$:
\begin{equation}
Q^{\alpha,I} = \bar\lambda^{\alpha} \frac{\partial}{\partial \bar\eta^{I}} {-} \lambda^{\alpha} \frac{\partial}{\partial \eta^{I}} + \frac{1}{4}(\lambda^{\alpha}\bar\eta^{I} {-} \bar\lambda^{\alpha}\eta^{I})\,.
\end{equation}
These variables can be mapped directly to those of the standard superspace $(x^{\alpha\beta},\theta^{\alpha I})$. We also define the SUSY covariant derivative
\begin{equation}
D^{\alpha,I} = \bar\lambda^{\alpha} \frac{\partial}{\partial \bar\eta^{I}} {-} \lambda^{\alpha} \frac{\partial}{\partial \eta^{I}} - \frac{1}{4}(\lambda^{\alpha}\bar\eta^{I} {-} \bar\lambda^{\alpha}\eta^{I})\,.
\end{equation}
These covariant derivatives impose constraints on boundary superfields. The latter transform as spin-$s$ irreducible representations of SL$(2,\mathbb R)$ for Lorentzian AdS boundary fields, and of SU$(2)$ for dS or Euclidean AdS boundary fields, with components $\textbf{J}_{\alpha_1\alpha_2\cdots \alpha_{2s}}$.\footnote{
    For half integer spin, Lorentzian AdS fermion is Majorana representation and dS/Euclidean AdS is symplectic Majorana representation. For the latter, minimal supersymmetry is $N=2$ instead of $N=1.$
} They are subject to
\begin{equation}\label{eq: Dconstraint}
D^{\alpha,I} \textbf{J}_{\alpha\alpha_2\cdots \alpha_{2s}}=0\quad  \forall s>0,\quad  D^\alpha D_\alpha \textbf{J}_0=0
\end{equation}
which gives conservation equations for the component currents and identifies auxiliary fields with derivatives of physical fields.

The variables $(\eta,\bar\eta)$ have helicity weights $(-\frac12,+\frac12)$ under the massive U$(1)$ little group. To make contact with the flat-space amplitude limit, we follow~\cite{Jain:2023idr} and Fourier transform the $\bar\eta$ variables:
\begin{equation}
\tilde{F}(\eta,\mu)=\int d^\mathcal{N} \bar{\eta}e^{-\frac{1}{4}(\sum_{I=1}^{\mathcal{N}}\bar\eta^I\mu_I)}F(\eta,\bar{\eta})
\end{equation}
After this transformation the supercharges become
\begin{equation}\label{eq: QDef}
Q^{\alpha,I} = -\lambda^{\alpha}\left(\frac{\partial}{\partial \mu^I}{+}\frac{\partial}{\partial \eta^{I}}\right){-}\frac{1}{4}\bar\lambda^{\alpha}( \mu^I {+}\eta^{I})= -2\lambda^\alpha \frac{\partial}{\partial \xi_{+I}} - \frac{\bar{\lambda}^\alpha}{4} \xi_{+}^I\,,
\end{equation}
where $\xi^I_{\pm}=\eta^I\pm\mu^I$. Both $\xi_+$ and $\xi_-$ now carry helicity weight $-\frac12$. Although the supercharge depends only on the $\mathcal N$ variables $\xi_+^I$, the remaining variables are retained in the covariant derivatives:
\begin{equation}
D^{\alpha,I} = -2\lambda^\alpha \frac{\partial}{\partial \xi_{-I}} + \frac{\bar{\lambda}^\alpha}{4} \xi_{-}^I\,.
\end{equation}

We will work with superfields constrained by the $D$ operators. These constraints enforce current conservation inside the multiplet and give the two fermionic combinations different roles: the $\xi_+$ expansion organizes components related by supersymmetry Ward identities, while the $\xi_-$ dependence keeps track of longitudinal pieces and contact terms. This distinction will be important below.

In this paper we focus on $\mathcal N=1,2$ and on spins $s=0,\frac12$. For $s=\frac12$ we use the helicity projections
\begin{equation}
\textbf{J}_\frac{1}{2}^{+}\equiv \bar\lambda^{\alpha} \textbf{J}_{\alpha}, \quad \textbf{J}_\frac{1}{2}^{-}\equiv \lambda^{\alpha} \textbf{J}_{\alpha}\,.
\end{equation}
If we were to consider higher spins, then mixed projections would be present, such as $\lambda^{\alpha_1} \bar\lambda^{\alpha_2}\textbf{J}_{\alpha_1\alpha_2\cdots \alpha_{2s}}$. We will consider $n$-point correlation functions of these superfields, 
\begin{equation}
\langle \textbf{J}^\pm_{s_1}\textbf{J}^\pm_{s_2} \cdots\textbf{J}^\pm_{s_n} \rangle 
\end{equation}
which are proportional to the overall momentum-conservation delta function $\delta^3(\sum_i \lambda_i\bar{\lambda}_i)$ and obey
\ie
Q^{\alpha,I}\langle \textbf{J}^\pm_{s_1}\textbf{J}^\pm_{s_2} \cdots\textbf{J}^\pm_{s_n} \rangle&=\left(\sum_{i=1}^n -2\lambda_i^\alpha \frac{\partial}{\partial \xi_{i,+I}} - \frac{\bar{\lambda}_i^\alpha}{4} \xi_{i,+}^I\right)\langle \textbf{J}^\pm_{s_1}\textbf{J}^\pm_{s_2} \cdots\textbf{J}^\pm_{s_n} \rangle=0\\
D_i^{\alpha,I}\langle \textbf{J}_{\alpha,s_1}\textbf{J}^\pm_{s_2} \cdots\textbf{J}^\pm_{s_n} \rangle
&= 
\left(-2\lambda_i^\alpha \frac{\partial}{\partial \xi_{i,-I}} + \frac{\bar{\lambda}_i^\alpha}{4} \xi_{i,-}^I\right)\langle \textbf{J}_{\alpha,s_1}\textbf{J}^\pm_{s_2} \cdots\textbf{J}^\pm_{s_n} \rangle\sim 0
\fe
There is no sum over $i$ in the constraint equation on the second line, and $\sim0$ means that the identity holds up to contact terms.

\subsubsection{$\mathcal{N}$=1 SUSY (Lorentzian AdS)}
We begin with the $\mathcal N=1$ multiplet. Unlike flat-space scattering states, the objects in the multiplet are boundary local operators, Fourier transformed to momentum space. This distinction is responsible for an important new feature: contact terms are part of the momentum-space supermultiplet.

To illustrate this, consider a spin-$s$ superfield with expansion 
\begin{equation}
\mathbf{J}_{\alpha_1\cdots\alpha_{2s}}=U_{\alpha_1\cdots\alpha_{2s}}{+}\theta^{\alpha}J_{\alpha\alpha_1\cdots\alpha_{2s}}{+}\frac{\theta^2}{2}B_{\alpha_1\cdots\alpha_{2s}}
\end{equation}
The top component $B$ is auxiliary. Imposing the constraint in eq.~(\ref{eq: Dconstraint}) gives current conservation and identifies the auxiliary field with a derivative of the leading physical field: 
\begin{equation}
\partial^{\alpha_1\alpha_2}J_{\alpha_1\alpha_2\cdots \alpha_{2s}}=0 ,\quad B_{\alpha_1\cdots\alpha_{2s}}- \partial_{\alpha_1}\,^{\alpha} U_{\alpha\cdots\alpha_{2s}}=0\,.
\end{equation}
These equations are operator identities. Inside correlation functions they hold only up to contact terms supported at coincident points. We therefore keep explicit components such as $(J_L,B_c)$, which encode longitudinal current pieces and contact contributions to the auxiliary-field equation. 
  
For example, introducing also an unconstrained scalar superfield $\bm A_0$, we use
\ie
\label{superfieldn=1}
\bm J_{1/2}^+ &= -\frac{J_L}{2\sqrt{E}} + \frac{1}{16} \xi_- \xi_+ \frac{J^+}{\sqrt{E}} + \frac{1}{4}\xi_- (\chi^+ + \frac{1}{2} \frac{B_c^+}{E}) + \frac{1}{8}\xi_+ \frac{B_c^+}{E}, \\
\bm J_{1/2}^- &= \frac{J^-}{2\sqrt{E}} + \frac{1}{16} \xi_- \xi_+ \frac{J_L}{\sqrt{E}} + \frac{1}{4}\xi_+ (\chi^- + \frac{1}{2} \frac{B_c^-}{E}) - \frac{1}{8}\xi_- \frac{B_c^-}{E}, \\
\bm A_0 &= \frac{\xi_+ - \xi_-}{8}  O_1 + \frac{\chi^-}{2\sqrt{E}} + \frac{\xi_- \xi_+}{16\sqrt{E}} \chi^+ + \frac{\xi_+ +\xi_-}{2E} O_2
\fe
where the spinning superfields and component operators are projected onto the helicity basis as
\ie
\label{hel-projection}
J^+ \equiv \frac{\bar\lambda_\alpha \bar\lambda_\beta J^{\alpha\beta}}{E}, \;\; J^- \equiv \frac{\lambda_\alpha \lambda_\beta J^{\alpha\beta}}{E},\;\; 
\chi^+ \equiv \frac{\bar\lambda_\alpha}{\sqrt{E}} \chi^\alpha, \;\; \chi^- \equiv \frac{\lambda_\alpha}{\sqrt{E}} \chi^\alpha, \;\; 
J_L = \frac{\bar\lambda_\alpha \lambda_\beta J^{\alpha\beta}}{E}\,.
\fe
Here $B_c^{\pm}$ and $J_L$ are placeholders for contact terms, whose precise form depends on the WFC under consideration. In the main text we focus primarily on transverse WFCs. The contact terms are fixed by requiring consistency among the different component projections of the same super WFC; this is used in appendices~\ref{app:full-three-point-super-wfcs} and~\ref{N2Longs} to construct the full three-point super WFC.

As emphasized above, the $\xi_+$ expansion relates components connected by supersymmetry Ward identities, while $\xi_-$ tracks longitudinal and contact terms. It is therefore useful to decompose the superfields with respect to $\xi_-$:
\ie
\label{eq:n1-TL-split}
\bm J_{1/2}^+ &= \xi_-\bm J_{1/2}^{+,T}+\bm J_{1/2}^{+,L} 
=
\frac{1}{4} \xi_-\left[\left(\chi^+ + \frac{1}{2} \frac{B_c^+}{E}\right)+\xi_+ \frac{J^+}{4\sqrt{E}} \right] 
+
\left[-\frac{J_L}{2\sqrt{E}} + \frac{1}{8}\xi_+ \frac{B_c^+}{E}\right], \\
\bm J_{1/2}^-
&=\bm J_{1/2}^{-,T}+\xi_-\bm J_{1/2}^{-,L}
=
\left[
\frac{J^-}{2\sqrt{E}}
+\frac{1}{4}\xi_+
\left(
\chi^-+\frac{1}{2}\frac{B_c^-}{E}
\right)
\right]
+
\xi_-\left[-\frac{1}{8}\frac{B_c^-}{E}+
\frac{1}{16}\xi_+\frac{J_L}{\sqrt{E}}
\right],
\\
\bm A_0
&=
\bm A_{0}^{+}
+
\bm A_{0}^{-} 
=
\left[\frac{\chi^-}{2\sqrt{E}}+
\xi_+ \left(\frac{1}{8}O_1+ \frac{1}{2E}O_2\right)
\right]
+
\xi_-\left[
 \left(-\frac{1}{8}O_1 +
\frac{1}{2E}O_2 \right)
+
\frac{\xi_+}{16\sqrt{E}}\chi^+
\right]\,.
\fe
where $(T,L)$ denote the parts containing transverse and longitudinal/contact modes, respectively. 

Because the supercharge in eq.~(\ref{eq: QDef}) acts nonlinearly on the momentum-space variables, supersymmetric invariants must be constructed separately at each multiplicity. For example, at two points, 
\begin{equation}\label{eq: N=1Gamma2}
\Gamma_2=\left[\xi_{1,+}\xi_{2,+}{-4\frac{\langle12\rangle}{E_1}}\right]\,.
\end{equation}
At three points there are two independent invariants, distinguished by their parity in $\xi_+$:
\begin{equation}\label{eq:n1-Gamma3}
\begin{aligned}
\Gamma^+_3 & =\left[\xi_{1+} \xi_{2+} \xi_{3+}-\frac{8}{E_T}\left(\xi_{1+}\langle 23\rangle+\xi_{2+}\langle 31\rangle+\xi_{3+}\langle 12\rangle\right)\right] \\
\Gamma^-_3 & =\left[8-\frac{1}{E_T}\left(\xi_{1+} \xi_{2+}\langle\overline{1} \overline{2}\rangle+\xi_{2+} \xi_{3+}\langle\overline{2} \overline{3}\rangle+\xi_{3+} \xi_{1+}\langle\overline{3} \overline{1}\rangle\right)\right]\,.
\end{aligned}
\end{equation}
One checks directly that $Q\,\Gamma^\pm_3=0$. The meaning of the $\pm$ superscript will become clear once these invariants are reproduced from the two branches of the Grassmannian. In the flat-space limit, $E_T\rightarrow0$, they reduce to the standard three-point superamplitude invariants:
\begin{equation}\label{eq: SUSYFlat}
\Gamma^+_3|_{E_T\rightarrow 0}\sim\frac{1}{E_T}(\xi_{1}\langle23\rangle{+}{\rm cyclic}), \quad\Gamma^-_3|_{E_T\rightarrow 0}\sim\frac{1}{E_T}\left(\sum_{i<j} \xi_i [ij]\xi_j\right).\, 
\end{equation}
 
\subsubsection{$\mathcal{N}=2$ superspace}
\label{sec:N2superfield}
\label{sec:SUSYCorrelatorN2}
The same construction extends to $\mathcal N=2$, with fermionic variables $\xi^I_\pm$, $I=1,2$. We consider the linear multiplet $J_0$, which satisfies the SO$(2)$-covariant constraint~\cite{Buchbinder:2015qsa}
\ie\label{eq: LinearConstraint}
 D^{\alpha,(I} D_{\alpha}^{J)} J_0 - \frac{1}{2} \delta^{IJ} (\delta_{I' J'} D^{I',\alpha} D_\alpha^{J'}) J_0 = 0\,.
\fe
This constraint implies conservation equations for the spinning components and relates auxiliary fields to physical fields, again up to contact terms. We denote the latter by a subscript $c$:
\footnote{
    The auxiliary spinor field is normalized as $\hat{B}^{I,\pm}_c = B^{I,\pm}_c / E$. Furthermore, to employ the $U(1)$ R-symmetry notation of~\cite{Jain:2023idr}, one may use the transformation rules
\begin{align}
\zeta_{\pm} &= \frac{1}{\sqrt{2}}\!\left(\xi^{(1)}_{\pm} \mp i\,\xi_{\pm}^{(2)}\right), \qquad
\omega_{\pm} = \frac{1}{\sqrt{2}}\!\left(\xi^{(1)}_{\pm} \pm i\,\xi_{\pm}^{(2)}\right), \qquad
\chi^{(2)} = \chi + \bar{\chi}, \\
\chi^{(1)} &= -i\!\left(\chi - \bar{\chi}\right), \qquad
\hat{B}_c^{(2)} = \bar{\hat{B}}_c + \hat{B}_c, \qquad
\hat{B}_c^{(1)} = -i\!\left(\bar{\hat{B}}_c - \hat{B}_c\right).
\end{align}
} 
\ie \label{N=2-Superfield}
\bm J_0
&=
\frac{1}{4E}  J^- 
+ 
\frac{1}{8\sqrt{E}} \xi^I_+ (\chi^{I-} + \frac{1}{2} \hat B_c^{I-})
-
\frac{1}{16\sqrt{E}} \xi_-^I \hat B^{I-}_c \\
&+
\frac{1}{64\sqrt{E}}\xi_-^2 \xi^I_+ (\chi_{I+} + \frac{1}{2} \hat B_{cI}^+)
-
\frac{1}{128\sqrt{E}} \xi_+^2 \xi^I_- \hat B_{cI}^{+} 
+\frac{1}{32E}\xi^I_- \xi^J_+\big(\delta^{IJ} J_L+\hat C^{IJ}_c\big) + \frac{1}{256E}\xi_-^2 \xi_+^2 J^+
\\
&+ \frac{\xi_+^2 -  \xi_-^2}{32} (O_1+16\hat\phi_{3,c})
+
\frac{\xi_+^2 +\xi_-^2}{8E} O_2 
-\frac12 \xi^I_+ \xi_{I-} \hat{\phi}_{3,c}
\fe
The lower SO$(2)$ index is lowered with the epsilon tensor, $\xi^2_\pm:=\xi_\pm^I\xi_{I,\pm}/2$, and the repeated index $I$ is summed over $1,2$. We similarly decompose with respect to $\xi_-$ as
$\bm J_0=\bm J_0^{-}+\xi_-^I\bm J_0^{I,L}+\xi_-^2\bm J_0^{+}$, with
\begin{align}\label{eq: N=2SuperField}
\bm J_0^{-}
&=
\frac{1}{4E}J^-
+
\frac{1}{8\sqrt{E}}
\xi^I_+
\left(
\chi^{I-}+\frac{1}{2}\hat B_c^{I-}
\right)
+
\frac{\xi_+^2}{32}
\left(
O_1+16\hat\phi_{3,c}
\right)
+
\frac{\xi_+^2}{8E}O_2 ,
\nonumber\\[0.5em]
\bm J_0^{I,L}
&=\left[-\frac{1}{16\sqrt{E}}
\hat B^{I-}_{c}
-
\frac{1}{128\sqrt{E}}
\xi_+^2
\hat B^{+}_{cI}
+
\frac{1}{32E}
\xi^J_+
\big(\delta^{IJ} J_L+\hat C^{IJ}_c\big)
-
\frac12
\xi_{I+}\hat\phi_{3,c}\right] ,
\nonumber\\[0.5em]
\bm J_0^{+}
&=
\left[\frac{1}{64\sqrt{E}}
\xi^I_+
\left(
\chi_I^+
+\frac{1}{2}\hat B_{Ic}^+
\right)
+
\frac{1}{256E}
\xi_+^2 J^+
-
\frac{1}{32}
\left(
O_1+16\hat\phi_{3,c}
\right)
+
\frac{1}{8E}O_2\right] .
\end{align}
The middle piece $\bm J_0^{I,L}$ carries only contact terms, which is why we
mark it with $L$: the longitudinal current $J_L$, the auxiliary spinor
$\hat B_c^{I\pm}$, and two scalars, $\hat\phi_{3,c}$ and the SO$(2)$
symmetric traceless $\hat C^{IJ}_c$.\footnote{These are the contact terms the
constraint~\eqref{eq: LinearConstraint} allows, one for each relation it
imposes: $J_L$ for current conservation, $\hat B_c^{I\pm}$ for the auxiliary
spinor, $\hat\phi_{3,c}$ for the top scalar, and $\hat C^{IJ}_c$ for the
traceless part of $D^{\alpha(I}D^{J)}_\alpha$ itself. All four are needed for
the Ward--Takahashi identities of appendix~\ref{N2Longs} to hold. In the
transverse sector contact fields enter only through the combinations
$\chi+\frac12\hat B_c$ and $O_1+16\hat\phi_{3,c}$.} The transverse sector is the
rest, $\bm J_0^{T}=(\bm J_0^{-},\bm J_0^{+})$, with $J^-$ in $\bm J_0^{-}$ and
$J^+$ in $\bm J_0^{+}$; it has the same degrees of freedom as flat-space
$\mathcal N=2$ SYM. We call a piece of a WFC built only from $\bm J_0^{T}$ a
\emph{transverse part}, and one carrying at least one $\bm J_0^{I,L}$ insertion
a \emph{longitudinal part}; the same names are used for the $\mathcal N=1$
superfields of eq.~\eqref{eq:n1-TL-split}. The insertion of $\bm J_0^{+}$ is
tracked by a factor of $\xi_-^2$; for example,
\begin{equation}
 \langle \bm J_0^{+}\bm J_0^{+}\bm J_0^{-}\cdots\bm J_0^{-}\rangle\sim \xi_{1,-}^2\xi_{2,-}^2\,.
 \end{equation}

The $\mathcal{N}=2$ invariants can be constructed directly from products of $\mathcal{N}=1$ invariants:\footnote{Throughout, $[f(\xi^I)]^2$ denotes the ordered product
$f(\xi^{(1)})f(\xi^{(2)})$, with the variables of the first species written
before those of the second. This is distinct from squaring an SO$(2)$ singlet.
We use $\epsilon_{12}=1$, so $\xi_{i,\pm}^2=\xi_{i,\pm}^{(1)}\xi_{i,\pm}^{(2)}$.}
\eqs{
\Gamma^{++}_3 & =\left[\xi_{1+}^I \xi_{2+}^I \xi_{3+}^I-\frac{8}{E_T}\left(\xi_{1+}^I\langle 23\rangle+\xi_{2+}^I\langle 31\rangle+\xi_{3+}^I\langle 12\rangle\right)\right]^2 \\
\Gamma^{--}_3 & =\left[8-\frac{1}{E_T}\left(\xi^I_{1+} \xi^I_{2+}\langle\overline{1} \overline{2}\rangle+\xi^I_{2+} \xi^I_{3+}\langle\overline{2} \overline{3}\rangle+\xi^I_{3+} \xi^I_{1+}\langle\overline{3} \overline{1}\rangle\right)\right]^2\,
}
Although these superinvariants have different degrees in $\xi_+$, their products automatically organize into SO$(2)$ singlets, with indices contracted by $\delta$ or $\epsilon$.

\subsubsection{General superinvariants from super Grassmannians}
\label{sec:gen-superinv}
To begin, note that the supercharge in eq.~(\ref{eq: QDef}) can be written as 
\begin{equation}
Q^{\alpha, I}=-\Lambda^T\cdot\Omega\cdot \Xi^I
\end{equation}
where $\Xi_n^I$ denotes the $2n$-component operator-valued vector 
\begin{align}
    \Xi_n^I = \begin{bmatrix}
\frac{1}{4} \xi_{1,+}^I
& \frac{1}{4} \xi_{2,+}^I
& \cdots  &\frac{1}{4} \xi_{n,+}^I
&2\frac{ \partial}{\partial {\xi_{1,+}^I}} &2\frac{ \partial}{\partial {\xi_{2,+}^I}} & \cdots &2\frac{ \partial}{\partial {\xi_{n,+}^I}} 
\end{bmatrix}\,.
\end{align}
With this identification, the following object is annihilated by the total supercharge:
\begin{align}\label{eq: SUSYDelta}
    & \hat \delta \left(C_{n} \cdot  \Omega\cdot  \Xi_n^I \right)  T_n (\xi_{i,\pm}^I)
\end{align}
The hat on the delta function reflects that it is to be understood as operator-valued and acts on a ``test function" $T_n(\xi_{i,\pm}^I)$. We fix its ordering as
\begin{align*}
\hat\delta(C_n\Omega\Xi_n^I)
=\prod_{a=1}^{n}(C_n\Omega\Xi_n^{(1)})_a\,
 \prod_{a=1}^{n}(C_n\Omega\Xi_n^{(2)})_a,
\end{align*}
with left Grassmann derivatives and the rightmost operator acting first.
For a test monomial containing $m$ pairs $\xi_{i,+}^2$, rewriting the result as
an ordered square of its single-species polynomial gives the sign
$(-1)^{nm+m(m-1)/2}$. The formulae below include this sign. The corresponding super-orthogonal-Grassmannian integral is
\begin{equation}\label{eq: Orthogonal4}
\int_{\mathcal{C}}\frac{d^{n\times 2n}C}{\mathrm{GL}(n)}\; f_n(C)\,
\delta^{n\times n}(C\cdot \Omega \cdot C^T)\,
\delta^{n\times 2}(C \cdot\Omega\cdot\Lambda)\hat \delta \left(C \cdot  \Omega\cdot  \Xi^I \right) T_n(\xi_{i,\pm}^I)
\end{equation}
The total helicity of the above function should be $-1$ on all legs due to the helicity weight of $\bm J_0$. From $\delta(C\cdot\Omega\cdot\Lambda)$ this tells us that the first $n$ columns of $C$ have ${-}\frac{1}{2}$ helicity weight and the last $n$-columns has ${+}\frac{1}{2}$. Thus the helicity weight of $T_n$ combined with $f(C)$ must have total helicity $-1$ on each leg. We find that it is convenient to keep $f(C)$ helicity neutral and all the helicity weight contained in $T_n$.

We find that the known SUSY invariants discussed previously can be identified with $T_n$ comprised of even or odd number of fermion pairs $\xi_{i,+}^2:=\frac12\epsilon_{IJ}\,\xi_{i,+}^I \xi_{i,+}^J $. These have different support on the branches: 
\begin{align}
    \hat\delta (C^+ \cdot \Omega \cdot \Xi^I)\cdot \prod_{\operatorname{odd} } \xi_{i,+}^2=0\,,\quad\hat\delta (C^- \cdot \Omega \cdot \Xi^I)\cdot   \prod_{\operatorname{even} } \xi_{i,+}^2=0\,, 
\end{align}
where we use $C^{\pm}$ to denote the positive and negative branch.\footnote{different test functions can also lead to the same invariant up to helicity compensating factors, 
\begin{align}
     &\phantom{cc}\hat \delta \left(C^+_{3} \cdot  \Omega\cdot  \Xi_3^I \right)  \cdot 1 \sim   \hat \delta \left(C^+_{3} \cdot  \Omega\cdot  \Xi_3^I \right)  \cdot \xi_{1,+}^2 \xi_{2,+}^2 \sim \Gamma^{++}_{3} \,,\nonumber\\
    & \hat \delta \left(C^-_{3} \cdot  \Omega\cdot  \Xi_3^I \right)  \cdot \xi_{1,+}^2 \sim  \hat \delta \left(C^-_{3} \cdot  \Omega\cdot  \Xi_3^I \right)  \cdot \prod_{i=1,2,3}  \xi_{i,+}^2  \sim \Gamma^{--}_3\,.
\end{align}}
The operator $\hat \delta(C_n\cdot\Omega\cdot\Xi_n^I)$ does not act on $\xi_{i,-}^I$. Dressing by $\xi_{i,-}^I$ therefore allows monomials of different $\xi_{i,+}^I$ degree to be combined into a single test function with uniform helicity weight, for example
\begin{align}
    1\cdot \prod_{i=1,2,3} \xi_{i,-}^2+  \xi_{1,+}^2\cdot \prod_{i=2,3} \xi_{i,-}^2+ \cdots \,. 
\end{align}

For $n=2,3$ the bosonic delta functions completely determine $C$ in terms of external kinematics, so substituting the solution into eq.~(\ref{eq: SUSYDelta}) directly gives a momentum-space invariant. Starting at four points, one must additionally specify which poles of $f_n(C)$ are encircled, leading to a richer set of invariants. On the top cell of
$\operatorname{OG}(4,8)$ the invariants remain expressions in the minors of $C$;
on the codimension-one cells relevant for the flat-space limit and for
factorization they become explicit momentum-space invariants, the building
blocks of the four-point super WFC of section~\ref{sec: 4PtRes}. The explicit
invariants at two, three and four points are collected in
appendix~\ref{app:invariants}.

\section{Two- and three-point super WFCs}
\label{sec:3pt-super}
We construct the full super WFC at two points and the transverse super WFC at
three points; the longitudinal part at three points is treated in
appendix~\ref{app:full-three-point-super-wfcs}. Two points come almost
for free: there is no longitudinal part to speak of, so a single component
determines the whole super WFC, and we treat that case first as the simplest
instance of the construction.

Three points take one more step. We first reconstruct the transverse current
WFC $\langle JJJ\rangle$ from its Grassmannian discontinuity. This works because
the three-point WFC has a single energy function: the discontinuity fixes the
numerator, and therefore the full WFC, up to a universal energy prefactor. The
transverse sector is separated from the longitudinal/contact sector by the
supersymmetry Ward identities. Once the full transverse $\langle JJJ\rangle$ is
obtained, it can be embedded into the superinvariants constructed above to give
the transverse part of the supersymmetric WFC.

\subsection{From discontinuities to WFCs}
\label{sec: 3ptinversion}
At three points the inversion is especially simple. Consider the transverse WFC $\langle J^T J^T J^T\rangle$, with each current projected onto a transverse polarization vector $\epsilon^T$. Its energy dependence takes the form
\ie
\la J^TJ^TJ^T \ra=\frac{\mathcal{N}}{E_T}\,,
\fe
where $\mathcal N$ depends on the momenta and polarizations but not on the energy branches. Since the discontinuity acts only on the energy variables, it leaves $\mathcal N$ unchanged. Therefore
\begin{align}
\disc_{1,2,3} \la J^TJ^TJ^T \ra&=-\frac{\mathcal{N}}{E_T}\,
\frac{16 E_1 E_2 E_3}{\prod_{i=1}^3(E_T-2E_i)}\,,\nonumber\\
\intertext{so that}
\mathcal{N}&=-\frac{E_T\prod_{i=1}^3(E_T-2E_i)}{16E_1E_2E_3}\,\disc_{1,2,3} \la J^TJ^TJ^T \ra\,.
\end{align}
Because the triple discontinuity has a Grassmannian representation, the full WFC is obtained by multiplying that homogeneous Grassmannian data by the inverse energy factor:
\ie
\label{discJJJallplus OG}
\langle J^+ J^+ J^+ \rangle(C)
&=-\frac{\prod_{i=1}^3(E_T-2E_i)}{16E_1E_2E_3}\disc_{1,2,3} \langle J^+ J^+ J^+ \rangle(C)\\
&=
-\frac{\prod_{i=1}^3(E_T-2E_i)}{16E_1E_2E_3}\frac{(\bar 1 \bar 2 \bar 3)_+^2}{(1 \bar 1 2)_+(3 \bar 3 \bar 2)_+}\,\\
\langle J^- J^+ J^+ \rangle(C)
&=-\frac{\prod_{i=1}^3(E_T-2E_i)}{16E_1E_2E_3}\disc_{1,2,3} \langle J^- J^+ J^+ \rangle(C)\\
&=
\frac{\prod_{i=1}^3(E_T-2E_i)}{16E_1E_2E_3} \frac{(1 \bar 2 \bar 3)_-^2}{(1 \bar 1 2)_-(3 \bar 3 \bar 2)_-}\,.
\fe
\subsection{Super WFCs}
\label{sec:SUSYCorrelator}

The two- and three-point superinvariants used below, for test functions with an
even or with an odd number of pairs $\xi_{i,+}^2$, are collected in
appendix~\ref{app:inv-23}. They realize the branch support of
section~\ref{sec:gen-superinv}: on each branch the test functions of the other
parity drop out by the minor identities of that branch.

\paragraph{Two points.}
We begin with the two-point WFC. The longitudinal part of an $n$-point WFC is
fixed by $(n{-}1)$-point data, so at two points it is a one-point function, and
that vanishes in the super Yang--Mills theory. There is accordingly nothing to
split at two points, and we drop the label. More than that: with no longitudinal
part and no contact term to determine, the single current component
$\la JJ\ra$ already fixes the entire two-point super WFC, every other component
following from it by supersymmetry. This is the two-point counterpart of the
statement, at three points, that $\la JJJ\ra$ fixes the transverse part on its
own. In Grassmannian form,
\begin{align}\label{eq:2pt-grass}
   \boxed{ \langle \bm{J}_0 \bm{J}_0 \rangle= \int \frac{d^{2 \times 4} C}{\operatorname{Vol}(GL(2))} \frac{1}{(1 \bar 1 ) +( 2 \bar 2)} \delta\left(C \cdot \Omega \cdot C^T \right) \delta\left(C \cdot \Omega \cdot \Lambda \right) \hat \delta \left(C \cdot  \Omega\cdot  \Xi^I \right) T_2(\xi_{i,\pm}^I)}
\end{align}
Performing the bosonic part of the integral in eq.~\eqref{eq:2pt-grass}, we obtain the following.
\begin{align}
    &\int \frac{d^{2 \times 4} C}{\operatorname{Vol}(GL(2))} \frac{1}{(1 \bar 1 ) +( 2 \bar 2)} \delta\left(C \cdot \Omega \cdot C^T \right) \delta\left(C \cdot \Omega \cdot \Lambda \right) = \frac{\delta^3(P)}{E_T}\,
\end{align}
where the bosonic delta functions localize $C$ onto the two-plane $\Lambda$
\begin{align}\label{eq: OG2}
    & C_{2}= \begin{pmatrix}
\lambda_1^\alpha & \lambda_2^\alpha &\bar{\lambda}_1^\alpha & \bar{\lambda}_2^\alpha 
\end{pmatrix}\, .
\end{align}
For two-point functions, we will be interested in $\langle \bar{\lambda}_1\lambda_1\rangle=2E_1=2E_2=\langle \bar{\lambda}_2\lambda_2\rangle$\,, the kinematics in eq.~(\ref{eq: OG2}) will be in the positive branch.

The integrand is helicity-neutral and  $(1\bar 1)=(2 \bar 2)$ on the positive branch. Recall that we can use $\xi_{i,-}^2$ to tag the presence of $\bm J_0^{+}$ in the WFC. Thus we can organize the test function $T_2(\xi_{i,\pm}^I)$ in terms of their anticipated helicity configuration:
\begin{align}
    T_2(\xi_{i,\pm}^I ) &=T_2^{++}+T_2^{-+}+T_2^{+-}+T_2^{--}\nonumber\\
    &=\frac{1}{2^7E_1E_2}\Big(
    \xi_{1,-}^2\xi_{2,-}^2+\xi_{1,+}^2\xi_{2,-}^2
    +\xi_{1,-}^2\xi_{2,+}^2+\xi_{1,+}^2\xi_{2,+}^2\Big),
\end{align}
where each $T_2^{h_1h_2}$ includes the common factor $1/(2^7E_1E_2)$,
fixed by the current-component normalization of appendix~\ref{app:N1}, and
\begin{equation}
T_2^{\pm \pm} \rightarrow \langle \bm J_0^{\pm}\bm J_0^{\pm} \rangle 
\end{equation}
Since $\langle \bm J^+_0\bm J^-_0\rangle=0$, we expect that $T_2^{+-}$ and $T_2^{-+}$ will evaluate to zero. Indeed they do, since eq.~(\ref{eq: OG2}) resides in the positive branch. The operator-valued fermionic delta function then acts on the test functions and produces the following supersymmetric invariants:
\begin{align}
    \hat \delta \left(C_{2} \cdot  \Omega\cdot  \Xi_2^I \right)  \cdot T^{++}_2
    &=\frac{\langle\bar1\bar2\rangle^2}{2^{13}E_T^2}  \xi_{1,-}^2  \xi_{2,-}^2 \times  (\Gamma_2)^2\,,\nonumber\\
     \hat \delta \left(C_{2} \cdot  \Omega\cdot  \Xi_2^I \right)  \cdot  T^{-+}_2&=0\,,\nonumber\\
     \hat \delta \left(C_{2} \cdot  \Omega\cdot  \Xi_2^I \right)  \cdot  T^{+-}_2 &=0\,,\nonumber\\
       \hat \delta \left(C_{2} \cdot  \Omega\cdot  \Xi_2^I \right)  \cdot   T^{--}_2
    &=-\frac{1}{2^7}\, (\Gamma_2)^2\,.
\end{align}
Combining these ingredients, we find that the resulting $\mathcal{N}=2$ super WFC takes the form:

\ie\label{eq:2pt WFC} {\rm Two{-}pts}\quad 
\la \bm J_0 \bm J_0\ra
=
-\frac{1}{256}\left(\frac{1}{E_1} - \xi_{1,-}^2 \xi_{2,-}^2 \frac{\abBB{1}{2}^2}{256 E_1^3}\right)(\Gamma_2)^2
\fe
where $(\Gamma_2)^2$ is the ordered square of eq.~(\ref{eq: N=1Gamma2}).
The
corresponding $\mathcal N=1$ Lorentzian-AdS two-point WFCs, one for each of the
three multiplets $\bm J^{\pm}_{1/2}$ and $\bm A_0$, are collected in
appendix~\ref{app:N1}. 

\begin{comment}
\begin{align}
    \hat \delta \left(C_{2} \cdot  \Omega\cdot  \Xi_2^I \right)  \cdot 1
    &=\Biggl[ -\frac{1}{2} E_T+\frac{1}{16}\langle \bar 1 \bar 2\rangle   \xi_{1,+}^I \xi_{2,+}^I \Biggr]^2\,,\\
     \hat \delta \left(C_{2} \cdot  \Omega\cdot  \Xi_2^I \right)  \cdot \xi_{1,+}^2 &=\Biggl[\frac{1}{2}\big( E_1- E_2 \big) \xi_{1,+}^I  -\frac{1}{2}  \langle 1 \bar 2\rangle   \xi_{2,+}^I  \Biggr]^2=0\,,\\
     \hat \delta \left(C_{2} \cdot  \Omega\cdot  \Xi_2^I \right)  \cdot \xi_{2,+}^2 &=\Biggl[-\frac{1}{2}  \langle  \bar 2 1\rangle   \xi_{1,+}^I +\frac{1}{2}\big( E_1- E_2 \big) \xi_{2,+}^I   \Biggr]^2=0\,,\\
       \hat \delta \left(C_{2} \cdot  \Omega\cdot  \Xi_2^I \right)  \cdot  \xi_{1,+}^2  \xi_{2,+}^2
    &=\Biggl[ 4\,  \langle 1 2 \rangle - \frac{1}{2} E_T  \xi_{1,+}^I \xi_{2,+}^I \Biggr]^2\,.
\end{align}
\end{comment}

\paragraph{Three points.}
At three points, the test function $T_3(\xi_{i,\pm}^I)$ decomposes into helicity sectors as
\begin{eqnarray}
T_3(\xi_{i,\pm}^I)&=&T_3^{+++}+T_3^{++-}+T_3^{+-+}+T_3^{-++}+T_3^{--+}+T_3^{-+-}+T_3^{+--}+T_3^{---}\nonumber\\
&=&-\frac1{16}\Bigl[\xi_{1,-}^2  \xi_{2,-}^2  \xi_{3,-}^2
+
(\xi_{1,-}^2  \xi_{2,-}^2  \xi_{3,+}^2+{\rm cyclic})\Bigr]\nonumber\\
&&-\frac1{16}\Bigl[
(\xi_{1,+}^2  \xi_{2,+}^2\xi_{3,-}^2+{\rm cyclic})
+
\xi_{1,+}^2 \xi_{2,+}^2  \xi_{3,+}^2\Bigr]\,,
\end{eqnarray}
Every sector carries the same coefficient $-1/16$: acting on a sector with
$m$ pairs $\xi_{i,+}^2$, the operator $\hat\delta$ returns $(-1)^{m+1}16$ times
the Grassmannian integrand of eq.~\eqref{discJJJallplus OG}, the sign
combining the ordering sign of section~\ref{sec:gen-superinv} with
$(1\bar13)(2\bar2\bar3)=-(1\bar12)(3\bar3\bar2)$, and the representatives of
eq.~\eqref{discJJJallplus OG} carry the compensating sign on the negative
branch. Each term therefore selects the corresponding helicity component
$\langle J^\pm J^\pm J^\pm\rangle(C)$ of eq.~\eqref{discJJJallplus OG} with one
helicity-independent normalization; we have checked that the eight current
components so obtained are proportional to the Yang--Mills three-point WFC with a
single constant. Since all helicity components are ratios of minors dressed by the same universal energy prefactor, we arrive at the Grassmannian formula
\begin{align}\label{eq:3pt-grass}
\boxed{    \langle \bm{J}^T_0 \bm{J}^T_0 \bm{J}^T_0 \rangle= \mathcal{F} \int \frac{d^{3 \times 6} C}{\operatorname{Vol}(GL(3))} \frac{1}{(1 \bar 1 3)( 2 \bar 2 \bar 3)} \delta\left(C \cdot \Omega \cdot C^T \right) \delta\left(C \cdot \Omega \cdot \Lambda \right) \hat \delta \left(C \cdot  \Omega\cdot  \Xi^I \right) T_3(\xi_{i,\pm}^I)}
\end{align}
in which we define the prefactor
\eqs{
\mathcal{F}=\left( \frac{\prod_{i=1}^3(E_T-2E_i)}{(64E_1E_2E_3)^2} \right)
}
while the remaining dependence is encoded in the orthogonal Grassmannian. The integrand is helicity-neutral and permutation invariant up to an overall sign. In particular,
\begin{align}
\frac{1}{(1\bar 13)(2\bar 2\bar 3)}
=-\frac{1}{(1\bar 1\bar 3)(2\bar 23)}
=\cdots .
\end{align}

Let us now evaluate the three-point Grassmannian integral explicitly. The bosonic delta functions localize onto the positive and negative branch, given as  
\begin{align}
    & C_3^+= \begin{pmatrix}
\lambda_1^\alpha & \lambda_2^\alpha &   \lambda_3^\alpha &\bar{\lambda}_1^\alpha & \bar{\lambda}_2^\alpha & \bar{\lambda}_3^\alpha \\
0 & 0 & 0 &\langle 2 3\rangle &\langle 3 1 \rangle &\langle 1 2 \rangle
\end{pmatrix},\quad C_3^-=  \begin{pmatrix}
 \lambda_1^\alpha & \lambda_2^\alpha & \lambda_3^\alpha &\bar{\lambda}_1^\alpha & \bar{\lambda}_2^\alpha & \bar{\lambda}_3^\alpha\\
 \langle \bar 2 \bar 3 \rangle &\langle \bar3 \bar1 \rangle &\langle \bar1 \bar2 \rangle &0 & 0 &  0
\end{pmatrix}
\end{align}
The integral is then localized to 
\begin{align}
    &\int \frac{d^{3 \times 6} C}{\operatorname{Vol}(GL(3))} \frac{\delta\left(C \cdot \Omega \cdot C^T \right) \delta\left(C \cdot \Omega \cdot \Lambda \right)}{(1 \bar 1 3)( 2 \bar 2 \bar 3)}   = \left(\frac{\delta^3(P)}{\langle12 \rangle \langle 23 \rangle \langle 31 \rangle E_T},\;  \frac{\delta^3(P)}{\langle \bar 1 \bar 2 \rangle \langle \bar 2 \bar 3 \rangle \langle \bar 3\bar 1 \rangle E_T}\right)\,,
\end{align}
where the two terms in the parenthesis corresponds to the two branches respectively. The operator-valued fermionic delta function yields
\begingroup
\allowdisplaybreaks
\begin{align}
    &\hat \delta \left(C_{3}^+ \cdot \Omega \cdot \Xi_3^I \right)
    \cdot T_3^{+++}
    = -\left(\frac{E_T^2}{256}\right)^2
    \xi_{1,-}^2 \xi_{2,-}^2 \xi_{3,-}^2
    \Gamma_3^{++}\,,\nonumber\\
    &\hat \delta \left(C_{3}^+ \cdot \Omega \cdot \Xi_3^I \right)
    \cdot T_3^{--+}
    = \left(\frac{E_T\langle12\rangle}{32}\right)^2
    \xi_{3,-}^2
    \Gamma_3^{++}\,,\nonumber\\
    &\hat \delta \left(C_{3}^- \cdot \Omega \cdot \Xi_3^I \right)
    \cdot T_3^{-++}
    = \left(\frac{E_T\langle\bar2\bar3\rangle}{32}\right)^2
    \xi_{2,-}^2 \xi_{3,-}^2
    \Gamma_3^{--}\,,\nonumber\\
    &\hat \delta \left(C_{3}^- \cdot \Omega \cdot \Xi_3^I \right)
    \cdot T_3^{---}
    = -\frac{E_T^4}{16}\,\Gamma_3^{--}\,.
\end{align}
\endgroup
Using the identity $\prod_{i<j}\langle i j\rangle\langle \bar i\bar j\rangle=E_T^3\prod_{i\in1,2,3}(E_T-2E_i)$ together with $\mathcal F$, we obtain the transverse part of the super WFC.  The result is
\ie\label{eq: N=23pt}
\langle {\bm{J}}_0^\text{T}  {\bm{J}}_0^\text{T} {\bm{J}}_0^\text{T} \rangle
&=
 \left(\Gamma_3^{++} \cdot \xi_{3,-}^2 \cdot \frac{(E_T-2E_3)\abB{1}{3}\abB{2}{3}}{2^{22}(\prod_{i=1}^3 E_i)^2} \cdot \frac{E_T}{\la 12 \ra } + \text{cyclic}\right)\\
&+
\left(\Gamma_3^{--} \cdot \xi_{1,-}^2 \xi_{2,-}^2\cdot \frac{(E_T-2E_3)\abB{3}{1}\abB{3}{2}}{2^{22}(\prod_{i=1}^3 E_i)^2} \cdot \frac{E_T}{\abBB{1}{2}} + \text{cyclic}\right)\\
&-
\Gamma_3^{++} \cdot \xi_{1,-}^2 \xi_{2,-}^2 \xi_{3,-}^2 \cdot \frac{\abBB{1}{2}\abBB{2}{3}\abBB{3}{1}}{2^{28}(\prod_{i=1}^3 E_i)^2} \\
&-
\Gamma_3^{--} \cdot \frac{\ab{1}{2}\ab{2}{3}\ab{3}{1}}{2^{16}(\prod_{i=1}^3 E_i)^2} 
\fe

Equation~\eqref{eq: N=23pt} is the transverse part, and it is all the four-point
construction of section~\ref{sec:SUSY-from-OG} needs, since the discontinuities
of section~\ref{sec: discontinuity JJJJ} are designed to remove the longitudinal
content. The full super WFC also has a longitudinal part, carrying at least one
$\bm J_0^{I,L}$ of eq.~\eqref{eq: N=2SuperField}; it is constructed in
appendix~\ref{app:full-three-point-super-wfcs}, at $\mathcal N=1$ in
appendix~\ref{app:N1} and extended to $\mathcal N=2$ in
appendix~\ref{N2Longs}, eq.~\eqref{eq:n2-L-fixed}.

\subsection{Flat-space limit, component projections and super discontinuity}

We check the result in two ways: by taking its flat-space limit and by projecting onto component WFCs. The first two lines of eq.~(\ref{eq: N=23pt}) contain total-energy poles. Their residues give the MHV and anti-MHV three-point amplitudes, respectively:
\ie
\langle {\bm{J}}^{\text{T}}_0  {\bm{J}}_0^{\text{T}} {\bm{J}}_0^{\text{T}} \rangle|_{E_T\rightarrow0}
&=
\frac{1}{2^{19}E_T(\prod_{i=1}^3 E_i)}\left( \frac{\Gamma_{3,flat}^{--}\xi_{1,-}^2 \xi_{2,-}^2\abBB{1}{2}}{\abBB{1}{3}\abBB{2}{3}}+\frac{\Gamma_{3,flat}^{++} \xi_{3,-}^2\ab{1}{2}}{\ab{1}{3}\ab{2}{3}} + \text{cyclic}\right) +  \mathcal{O}(E_T^0)
\fe
where we identify the flat space superinvariants:
\ie
\Gamma^{--}_{3,flat} &= \left(\xi_{1+} \xi_{2+}\langle\overline{1} \overline{2}\rangle+\xi_{2+} \xi_{3+}\langle\overline{2} \overline{3}\rangle+\xi_{3+} \xi_{1+}\langle\overline{3} \overline{1}\rangle\right)^2\nonumber\\
\Gamma^{++}_{3,flat} &= 64\left(\xi_{1+}\langle 23\rangle+\xi_{2+}\langle 31\rangle+\xi_{3+}\langle 12\rangle\right)^2\,.
\fe

As a second check, project the super WFC onto the component $\langle J^-O_2O_2\rangle$. This component receives contributions from
\ie
 T_3^{---},T_3^{-++},T_3^{--+},T_3^{-+-}
\fe
on the $-$, $-$, $+$ and $+$ branches respectively. From eq.~(\ref{eq: OG3SUSY}) and eq.~(\ref{eq: OG3SUSY1})
we find the following numerators for the Grassmannian integral:
\ie
\Big((2 \bar 2 1)_- + (3 \bar 3 1)_- \Big)^2, \Big((2 \bar 2 1)_- + (3 \bar 3 1)_-\Big)^2, \Big((2 \bar 2 1)_+ - (3 \bar 3 1)_+\Big)^2, \Big((2 \bar 2 1)_+ - (3 \bar 3 1)_+\Big)^2
\fe
Using the conjugate-minor identities in eq.~(\ref{eq: Conjugate}), all of these numerators are proportional to $(2\bar2 1)$. Since $\disc_{1,2,3}\langle J^-O_2O_2\rangle$ was given in eq.~\eqref{disc 123 JOO}, the projected WFC, read with the $O_2$ slot $(\xi_+^2+\xi_-^2)/(8E)$ of
eq.~\eqref{N=2-Superfield}, satisfies
\ie\label{eq: invert2}
-64\,\la J^- O_2 O_2 \ra = -\frac{\prod_{i=1}^3(E_T-2E_i)}{16E_1E_2E_3} \disc_{1,2,3} \la J^- O_2 O_2 \ra(C)
\fe
The projection equals
$\tfrac1{128}\ab{1}{2}\abB{1}{2}/(E_1E_T)$, the conformally coupled $\Delta=2$
form, for $J^-$ and, with $\lambda\leftrightarrow\bar\lambda$, for $J^+$ on every leg.
This is the same inversion formula as in eq.~\eqref{discJJJallplus OG}, reflecting the fact that $\langle JOO\rangle$ has the same energy function as $\la JJJ \ra$:
\ie
\label{energy function form JOO}
\la JOO \ra= \frac{\mathcal{N}_{JOO}}{E_T}.
\fe
Equations~\eqref{discJJJallplus OG} and~\eqref{eq: invert2} use
component-normalized discontinuities. The Grassmannian integral in
\eqref{eq:3pt-grass} without $\mathcal F$ instead gives the unnormalized
supersymmetric kernel,
\begin{align}
\mathcal F^{-1}\langle \bm J_0^T\bm J_0^T\bm J_0^T\rangle
={}&\int \frac{d^{3\times6}C}{\operatorname{Vol}(GL(3))}
\frac{\delta(C\cdot\Omega\cdot C^T)\delta(C\cdot\Omega\cdot\Lambda)}
{(1\bar13)(2\bar2\bar3)}
\hat\delta(C\cdot\Omega\cdot\Xi^I)T_3(\xi_{i,\pm}^I)\,.
\end{align}
Its component normalization must be included before identifying it with
$\disc_{1,2,3}\langle\bm J_0\bm J_0\bm J_0\rangle$; removing $\mathcal F$
alone does not perform that normalization.

At first sight it is surprising that the same inversion formula applies to all components, since it was derived from the simple energy dependence in eq.~(\ref{energy function form JOO}). Fermionic components usually carry additional energy factors in the denominator, arising from the unit momenta~\cite{Chen:2025foq}. The resolution is that fermions appear in the super WFC together with contact terms, as shown in the superfield expansion in eq.~(\ref{eq: N=2SuperField}). The fermionic component should therefore be understood through the combination
\ie
\langle J^+ (\chi+\tfrac12\hat B_c)^+ (\chi+\tfrac12\hat B_c)^+ \rangle
&=-\frac{\prod_{i=1}^3(E_T-2E_i)}{16E_1E_2E_3}\disc_{1,2,3} \langle J^+ \chi^+ \chi^+ \rangle
\fe
Another example is the $O_1$ scalar:
\ie
\langle J^+ (O_1+16\hat \phi_{3,c}) (O_1+16\hat \phi_{3,c}) \rangle
&=-\frac{\prod_{i=1}^3(E_T-2E_i)}{16E_1E_2E_3}\disc_{1,2,3} \langle J^+ O_1 O_1 \rangle\,.
\fe

\section{Four-point super WFC}\label{sec:SUSY-from-OG}
We next turn to four-point correlators. The Grassmannian formulae for the relevant discontinuities, reviewed in section~\ref{sec: discontinuity JJJJ}, arise as residues of the one-dimensional OG contour integral. The full WFC is reconstructed by applying the inversion formula to a spanning set of these discontinuities.

\subsection{From discontinuity to correlator}
The four-current WFC in polarization-vector form admits the energy-basis decomposition
\ie\label{eq: 4ptBasis}
\la JJJJ \ra=\sum_{e=1}^5 \mathcal{N}_e(\epsilon, p, E^2) F_e(E)
\fe
Here $\mathcal N_e$ depends on polarization vectors, momenta, and energy squares, and is unaffected by energy discontinuities. A convenient basis of energy functions is~\cite{Arundine:2026fbr,Armstrong:2020woi}
\ie
\frac{1}{E_T},\;\frac{1}{E_T E_{12s} E_{34s}},\;\frac{1}{E_T E_{14t} E_{23t}},\;\frac{(E_1-E_2)(E_3-E_4)}{E_T},\;\frac{(E_1-E_4)(E_2-E_3)}{E_T }\,.
\fe

From eq.~(\ref{eq: 4ptBasis}), the relevant discontinuities are spanned by the following five-element basis:
\ie
\disc_{1,3} \la J J J J \ra,\;
\disc_{2,4} \la J J J J \ra,\;
\disc_{1,2,t} \la J J J J \ra,\;
\disc_{1,4,s} \la J J J J \ra,\;
\disc_{2,3,4} \la J J J J \ra\,.
\fe
We have chosen this basis so that all longitudinal components are projected out; each element therefore admits a Grassmannian representation. The explicit Grassmannian forms for different helicity configurations were given in section~\ref{sec: discontinuity JJJJ}. With this data,
\ie\label{eq: 4ptDisRep}
\disc_{i} \la J J J J \ra(C) &=\sum_{e=1}^5 \mathcal{N}_e(C) \disc_{i} F_e(E), \quad i = \{234, 13, 24, 14s, 12t\}\,.
\fe
Inverting this linear system solves for the numerators $\mathcal N_e(C)$. Substituting them back into eq.~\eqref{eq: 4ptBasis} gives
\begin{align}
    \label{full JJJJ OG pre}
\la JJJJ \ra(C)
&=\sum_{i \in \{234, 13, 24, 14s, 12t\}} \mathcal{P}_i \disc_{i} \la JJJJ \ra(C)
\end{align}
The explicit energy prefactors are
\begin{align}
\mathcal{P}_{234}
&=
-\frac{
(E_1-E_3)
(E_1+E_2+E_3+E_4)
\left(
E_1^2-E_2^2-2E_1E_3+E_3^2+2E_2E_4-E_4^2
\right)
}{
48E_1E_2E_3E_4
},
\nonumber\\[0.5em]
\mathcal{P}_{13}
&=
\frac{
-3E_1^2+E_2^2-2E_2E_3-3E_3^2
+2E_2E_4-2E_3E_4+E_4^2
-2E_1(E_2-3E_3+E_4)
}{
24E_1E_3
},
\nonumber\\[0.5em]
\mathcal{P}_{24}
&=
\frac{
E_2^3+E_2^2(E_3-E_4)
-3E_1^2(E_2-E_3+E_4)
-E_2(E_3+E_4)^2
-(E_3-E_4)(E_3+E_4)^2
}{
24E_1E_2E_4
}
\nonumber\\
&\quad
-\frac{
2E_1(E_2^2-E_3^2-2E_2E_4+E_4^2)
}{
24E_1E_2E_4
},
\nonumber\\[0.5em]
\mathcal{P}_{14s}
&=
\frac{
E_1+E_2-E_s
}{
96E_1E_2E_3E_4
}
\Big[
2E_1^3+6E_3^2E_4+4E_3E_4^2-2E_4^3
\nonumber\\
&\quad
+E_2\left(
3E_3^2-2E_3(E_4-E_s)-(E_4-E_s)^2
\right)
-2E_2^2(E_4-E_s)
\nonumber\\
&\quad
+3E_3^2E_s-6E_3E_4E_s+3E_4^2E_s
+2E_3E_s^2+2E_4E_s^2-3E_s^3
\nonumber\\
&\quad
+2E_1^2(-2E_3+E_4+E_s)
\nonumber\\
&\quad
+E_1\left(
-2E_2^2-3E_3^2-6E_3E_4+E_4^2
+4E_2(E_3+2E_4-2E_s)
+2E_3E_s+2E_4E_s-E_s^2
\right)
\Big],
\nonumber\\[0.5em]
\mathcal{P}_{12t}
&=
\frac{
E_1+E_4-E_t
}{
96E_1E_2E_3E_4
}
\Big[
2E_1^3-2E_2^3
+2E_1^2(E_2-2E_3+E_t)
+E_2^2(4E_3-E_4+3E_t)
\nonumber\\
&\quad
+(E_4+E_t)
\left(
3E_3^2+2E_3E_t+(2E_4-3E_t)E_t
\right)
\nonumber\\
&\quad
+E_1\left(
E_2^2-3E_3^2+4E_3E_4-2E_4^2
+2E_3E_t-8E_4E_t-E_t^2
+E_2(-6E_3+8E_4+2E_t)
\right)
\nonumber\\
&\quad
+2E_2\left(
3E_3^2-E_4^2+E_4E_t+E_t^2
-E_3(E_4+3E_t)
\right)
\Big].
\end{align}
The prefactors contain no spurious poles. For example, the all-plus WFC is
\begin{align}
    \label{full JJJJ OG}
\la J^+J^+J^+J^+ \ra(C)
={}&\mathcal{P}_{234}
\left.\frac{(\bar 1 \bar 2 \bar 3 \bar 4)^2}{ST}
\left(\frac{S+T}{2AB}+\frac{T}{2AC}+\frac{S}{2AD}\right)\right|_{
    \begin{smallmatrix}
     (S=0,\beta-\alpha)_+ + (T=0,\beta-\alpha)_+
     \\
     -2(B=0,\alpha)_+ + 2(C=0,\alpha)_+
     \\
     + 2(D=0,\alpha)_+
\end{smallmatrix}}
\nonumber\\
&+\mathcal{P}_{13}
\left.\frac{(S+T)(\bar 1 \bar 2 \bar 3 \bar 4)^2}{ST\,AB}\right|_{
    \begin{smallmatrix}
     (S=0,\beta)_+ + (T=0,\beta)_+
     \\
     +(A=0,\beta)_+ + (B=0,\alpha)_+
\end{smallmatrix}}
\nonumber\\
&+\mathcal{P}_{24}
\left.\frac{(S+T)(\bar 1 \bar 2 \bar 3 \bar 4)^2}{ST\,AB}\right|_{
    \begin{smallmatrix}
     (S=0,\beta)_+ + (T=0,\beta)_+
     \\
     +(A=0,\beta)_+ + (B=0,\beta)_+
\end{smallmatrix}}
\nonumber\\
&+\mathcal{P}_{14s}
\left.\frac{(S+T)(\bar 1 \bar 2 \bar 3 \bar 4)^2}{ST\,AB}\right|_{(S=0,\beta-\alpha)_+}
\nonumber\\
&+\mathcal{P}_{12t}
\left.\frac{(S+T)(\bar 1 \bar 2 \bar 3 \bar 4)^2}{ST\,AB}\right|_{(T=0,\beta-\alpha)_+}\,.
\end{align}
The other helicity components follow from eq.~\eqref{full JJJJ OG pre} with
the same prefactors $\mathcal P_i$, replacing each $\disc_i\la JJJJ\ra(C)$ by the
corresponding discontinuity of that helicity sector given in
appendix~\ref{app:hat}.

\subsection{The helicity components in compact prefactor form}
\label{sec:compact-components}

The five-discontinuity representation above pays for using the same
prefactors $\mathcal P_i$ in every helicity component: each $\mathcal P_i$ is
lengthy. Allowing the prefactors to depend on the helicity sector buys a great
deal. Each component is then a channel block carrying one short prefactor,
plus at most three residues whose prefactors are equally short, with no
$\mathcal P_i$ anywhere. We collect these forms here, and
return in section~\ref{sec: 4PtRes} to the superinvariants that carry them.

\paragraph{Assembly of the helicity components.}
Let $H\subseteq\{1,3\}$ be the set of negative-helicity legs (legs $2,4$
positive) and $n_H=(a_1\,\bar2\,a_3\,\bar4)^2$ with $a_i=i$ for $i\in H$ and
$a_i=\bar i$ otherwise. Apart from single energies $E_i$, energies enter the
prefactors through sums $E_{i,j,k}=E_i+E_j+E_k$, a bar marking an entry that comes with a minus sign,
so that $E_{1,\bar 2,s}=E_1-E_2+E_s$ and
$E_{1,\bar 2,\bar 3,4}=E_1-E_2-E_3+E_4$. Sums containing a channel energy
appear in pairs that differ by its sign, and each pair is a Mandelstam
invariant, 
\begin{equation}
s_{ij}=E_{i,j,s}E_{i,j,\bar s}=(E_i+E_j)^2-E_s^2\,,\qquad
\bar s_{ij}=E_{i,\bar j,s}E_{i,\bar j,\bar s}=(E_i-E_j)^2-E_s^2\,,
\label{eq:pair-energies}
\end{equation}
for the $s$-channel pairs $\{1,2\}$ and $\{3,4\}$; $t_{ij}$ and $\bar t_{ij}$
are the same with $E_s\to E_t$ for $\{1,4\}$ and $\{2,3\}$, and $u_{ij}$,
$\bar u_{ij}$ with $E_s\to E_u=|p_1{+}p_3|$ for $\{1,3\}$ and $\{2,4\}$. The unbarred one
is the flat-space invariant, $s_{ij}=2(E_iE_j-p_i\cdot p_j)$ with $p_i\cdot
p_j$ the spatial dot product, and the barred one is that invariant with
$E_j\to-E_j$, so the bar means here exactly what it means on a single energy.
The prefactors are the ones that reappear in the cut of
section~\ref{sec: Super Cutting Rule}, written in these variables,
\begin{equation}
\mathcal F_S=\frac{\bar s_{12}\,\bar s_{34}}{16\,E_1E_2E_3E_4}\,,\qquad
\mathcal F_T=\frac{\bar t_{14}\,\bar t_{23}}{16\,E_1E_2E_3E_4}\,,
\label{eq:FS-FT}
\end{equation}
so nothing new has to be introduced here. The all-plus component is already
one channel block per prefactor, and the other three are the same block with
the numerator of their sector plus a remainder, the MHV one carrying in
addition one residue at $A=0$:
\begin{align}
\la J^+J^+J^+J^+\ra&=+\big(\mathcal F_S\tfrac{(S+T)\,(\bar1\,\bar2\,\bar3\,\bar4)^2}{ST\,AB}\Big|_{(S=0,\beta)_{+}}+\mathcal F_T\tfrac{(S+T)\,(\bar1\,\bar2\,\bar3\,\bar4)^2}{ST\,AB}\Big|_{(T=0,\beta)_{+}}\big),\nonumber\\
\la J^-J^+J^+J^+\ra&=-\big(\mathcal F_S\tfrac{(S+T)\,(1\,\bar2\,\bar3\,\bar4)^2}{ST\,AB}\Big|_{(S=0,\alpha)_{-}}+\mathcal F_T\tfrac{(S+T)\,(1\,\bar2\,\bar3\,\bar4)^2}{ST\,AB}\Big|_{(T=0,\alpha)_{-}}\big)+L_{-+++},\nonumber\\
\la J^+J^+J^-J^+\ra&=+\big(\mathcal F_S\tfrac{(S+T)\,(\bar1\,\bar2\,3\,\bar4)^2}{ST\,AB}\Big|_{(S=0,\beta)_{-}}+\mathcal F_T\tfrac{(S+T)\,(\bar1\,\bar2\,3\,\bar4)^2}{ST\,AB}\Big|_{(T=0,\beta)_{-}}\big)+L_{++-+},\nonumber\\
\la J^-J^+J^-J^+\ra&=-\big(\mathcal F_S\tfrac{(S+T)\,(1\,\bar2\,3\,\bar4)^2}{ST\,AB}\Big|_{(S=0,\alpha)_{+}}+\mathcal F_T\tfrac{(S+T)\,(1\,\bar2\,3\,\bar4)^2}{ST\,AB}\Big|_{(T=0,\alpha)_{+}}\big)\nonumber\\
&\quad-\tfrac{(S+T)\,(1\,\bar2\,3\,\bar4)^2}{ST\,AB}\Big|_{(A=0,\alpha)_{+}}+L_{-+-+}.
\label{eq:compact-assembly}
\end{align}
The first line is the statement that the all-plus component itself has a
compact form: one residue per channel, each with the prefactor of its own cut.
The remainders have the compact spinor form
\begin{gather}
L_{-+++}=\frac{\mix12\mix13^2\mix14}{128E_2E_3E_4\ang12\ang14}\,,\qquad
L_{++-+}=\frac{\mix31^2\mix32\mix34}{128E_1E_2E_4\ang32\ang34}\,,\nonumber\\
L_{-+-+}=\frac{\ang13^2}{128E_2E_4}\left[\frac{\mix12\mix14}{E_3\ang12\ang14}
+\frac{\mix32\mix34}{E_1\ang32\ang34}\right],
\end{gather}
and obey $L_{-+-+}=-\big(\hat h_3L_{-+++}+\hat h_1L_{++-+}\big)$, where $\hat h_i$
exchanges $\lambda_i\leftrightarrow\bar\lambda_i$ and sends $E_i\to-E_i$ at
fixed $E_s,E_t$.\footnote{An $n$-point extension of such representation was recently presented in~\cite{Nabavi:2026zyk}.}

That relation is not incidental: the remainders are built so that they
reassemble into $\disc_{1,3}$, which is to say so that they cancel out of it.
Taking $X=\{1,3\}$ and $h=++++$ in eq.~\eqref{eq:hat-disc},
\begin{align}
\disc_{1,3}\la J^+J^+J^+J^+\ra&=\la J^+J^+J^+J^+\ra+\hat h_1\la J^-J^+J^+J^+\ra\nonumber\\
&\quad+\hat h_3\la J^+J^+J^-J^+\ra+\hat h_1\hat h_3\la J^-J^+J^-J^+\ra\,,
\label{eq:disc13-from-sectors}
\end{align}
so the remainders enter the right-hand side only through
\begin{equation}
\hat h_1L_{-+++}+\hat h_3L_{++-+}+\hat h_1\hat h_3L_{-+-+}=0\,,\qquad L_{++++}=0\,,
\label{eq:remainder-cancellation}
\end{equation}
which is the relation above rewritten with $\hat h_i^2=1$. What survives is the
four channel blocks and the $A=0$ residue of the MHV line, and they must reproduce the single contour of
eq.~\eqref{eq:disc13-disc24-allplus}. In this sense
eq.~\eqref{eq:remainder-cancellation} is the condition that an assembly with
helicity-dependent prefactors be consistent with the one discontinuity it was
read off from, and it fixes $L_{-+-+}$ once the two single-minus remainders
are known.

\paragraph{Each remainder is two residues.}
In the variables of eq.~\eqref{eq:pair-energies},
\begin{align}
L_{-+++}={}&-\frac{1}{8E_2E_3E_4\,t_{14}}\Big\{
2\bar s_{34}\big(E_4\bar s_{12}-E_2\bar t_{14}\big)\,
\frac{(1\,\bar2\,\bar3\,\bar4)^2}{BSA}\Big|_{(B=0,\alpha)_{-}}\nonumber\\
&\hspace{14mm}+u_{24}\big[2E_4\bar s_{12}+E_{1,\bar 2,\bar 3,4}\,\bar t_{14}\big]\,
\frac{(1\,\bar2\,\bar3\,\bar4)^2}{BAU}\Big|_{(B=0,\alpha)_{-}}\Big\},
\label{eq:compact-L1}\\
L_{++-+}={}&\frac{1}{8E_1E_2E_4\,s_{34}}\Big\{
2\bar t_{14}\big(E_4\bar t_{23}-E_2\bar s_{34}\big)\,
\frac{(\bar1\,\bar2\,3\,\bar4)^2}{ATB}\Big|_{(A=0,\alpha)_{-}}\nonumber\\
&\hspace{14mm}-u_{24}\big[2E_4\bar t_{23}+E_{\bar 1,\bar 2,3,4}\,\bar s_{34}\big]\,
\frac{(\bar1\,\bar2\,3\,\bar4)^2}{ABU}\Big|_{(A=0,\alpha)_{-}}\Big\},
\label{eq:compact-L3}\\
L_{-+-+}={}&-\Big[\frac{s_{34}\big(E_4\bar s_{12}-E_2\bar t_{14}\big)}{4E_2E_3E_4\,t_{14}}
-\frac{s_{12}\big(E_4\bar t_{23}-E_2\bar s_{34}\big)}{4E_1E_2E_4\,t_{23}}\Big]
\frac{(1\,\bar2\,3\,\bar4)^2}{ASB}\Big|_{(A=0,\alpha)_{+}}\nonumber\\
&+\frac{u_{24}}{8E_2E_4}
\Big[\frac{2E_4\bar s_{12}+E_{1,\bar 2,3,4}\,\bar t_{14}}{E_3\,t_{14}}
+\frac{E_{1,2,3,\bar 4}\,\bar t_{23}+2E_2\bar s_{34}}{E_1\,t_{23}}\Big]
\frac{(1\,\bar2\,3\,\bar4)^2}{ABU}\Big|_{(A=0,\alpha)_{+}}.
\label{eq:compact-LM}
\end{align}

One feature of the expansion deserves emphasis. Coefficients this simple are
bought by admitting integrands with $U$ in the denominator---the second residue
of each remainder carries $BAU$ or $ABU$---even though no residue is ever taken
at $U=0$. 

What makes these forms useful beyond their brevity is that they survive the
supersymmetric uplift. The contours are those of
section~\ref{sec: discontinuity JJJJ} and the prefactors are functions of the
energies alone, so neither interferes with the test functions; we come back to
this below, after constructing the four-point superinvariants in
section~\ref{sec: 4PtRes}.

\subsection{The full four-point super WFC}
\label{sec:4pt-super-wfc}
\label{sec: 4PtRes}

Following the $n=2,3$ construction, we build the four-point superinvariants.
Since $\operatorname{OG}(4,8)$ is not fully localized by external kinematics,
an additional residue condition must be specified before a Grassmannian
superinvariant can be translated into momentum space. The explicit invariants
are collected in appendix~\ref{app:inv-4}: first on the top cell of
$\operatorname{OG}(4,8)$, then on two classes of codimension-one cells, where
they can be written directly in momentum space. On the flat-space-limit cell
$A=S+T+U=0$ each branch splits into two solutions, labeled $\alpha$ and
$\beta$, and the top-cell expressions collapse to the invariants
$\Gamma^{\pm\pm}_{4,i}\big|_{A=0}$, $i=\alpha,\beta$; on the factorization cell
$S=0$ they give $\Gamma^{\pm\pm}_{4,i}\big|_{S=0}$, and $T=0$ follows by
$2\leftrightarrow4$. These are the building blocks of the super WFC below.

The full super WFC is then expected to be written as a linear combination of these invariants. Similarly to 3-pts in eqs.~(\ref{eq:n1-ansatz-T})--(\ref{eq:n1-ansatz-L}), we write 
\begin{eqnarray}
\langle \mathbf{J}^T_0\mathbf{J}^T_0\mathbf{J}^T_0\mathbf{J}^T_0 \rangle&=&\sum_{i=\alpha,\beta} \,c^+_{s,i} \Gamma_{4,i}^{++}|_{S=0}{+}\sum_{i=\alpha,\beta} b^+_{i} \Gamma_{4,i}^{++}|_{A=0}\nonumber\\
&+&\sum_{i=\alpha,\beta} \,c^-_{s,i}\Gamma_{4,i}^{--}|_{S=0}{+}\sum_{i=\alpha,\beta} \,b^-_{i}\Gamma_{4,i}^{--}|_{A=0}{+}\cdots 
\end{eqnarray}
where the $\cdots$ are SUSY invariants stemming from the localization of other minors. The supersymmetric WFC then takes the form,  
\ie\label{eq: 4pt proposal}
&\langle \mathbf{J}^T_0\mathbf{J}^T_0\mathbf{J}^T_0\mathbf{J}^T_0 \rangle = \sum_{ h = h_1,\dots, h_4 \in \pm} \int \frac{d C_4}{GL(4)}  F^{h}(C_4) \left[\hat \delta \left(C_{4} \cdot  \Omega\cdot  \Xi_4^I \right) T^{h}_4(\xi_{i,\pm}^I)\right] \delta\left(C \cdot \Omega \cdot C^T \right) \delta\left(C \cdot \Omega \cdot \Lambda \right)\,,
\fe
where as in the three-point case, the test functions take the form,  
\begin{align}
    &T_4^{++++}= -\frac{\prod_{i=1}^4\xi_{i,-}^2}{2^{16}\prod_{i=1}^4E_i}, \quad
    T_4^{----}= -\frac{\prod_{i=1}^4\xi_{i,+}^2}{2^{16}\prod_{i=1}^4E_i}\,,\nonumber\\
    &T_4^{--++}=-\frac{\xi_{1,+}^2\xi_{2,+}^2\xi_{3,-}^2\xi_{4,-}^2}{2^{16}\prod_{i=1}^4E_i}\,,\nonumber\\
    &T_4^{-+++}=\frac{\xi_{1,+}^2\xi_{2,-}^2\xi_{3,-}^2\xi_{4,-}^2}{2^{16}\prod_{i=1}^4E_i},\quad
    T_4^{+---}=\frac{\xi_{1,-}^2\xi_{2,+}^2\xi_{3,+}^2\xi_{4,+}^2}{2^{16}\prod_{i=1}^4E_i}\,.
\end{align}
The sector with $n_-$ negative helicities carries $-(-1)^{n_-}/(2^{16}\prod_iE_i)$. Each test function selects a particular helicity configuration of the
$\la JJJJ \ra$ component, whose Grassmannian representation we already have.
Each configuration also belongs to one branch, $b=(-1)^{n_-}$ with $n_-$ its
number of negative helicities, and that is the branch its contour runs on in
section~\ref{sec: discontinuity JJJJ}. 
For later convenience we group the test functions by branch and sum them,
\begin{align}
T_4^{\rm even}&=T_4^{++++}+T_4^{----}+T_4^{--++}+T_4^{-+-+}+T_4^{-++-}+T_4^{+--+}+T_4^{+-+-}+T_4^{++--}\,,\nonumber\\
T_4^{\rm odd}&=T_4^{-+++}+T_4^{+-++}+T_4^{++-+}+T_4^{+++-}+T_4^{+---}+T_4^{-+--}+T_4^{--+-}+T_4^{---+}\,,
\label{eq:T4-branch}
\end{align}
the sums over even and odd $n_-$, which belong to the branches $b=+$ and $b=-$
respectively.

Replacing the helicity numerator of each
discontinuity by the test functions belonging to the same branch---those of the
$+$ branch on the branch-$+$ copy of its contour, those of the $-$ branch on the
branch-$-$ copy---turns it into the \emph{super discontinuity} again,
$\disc_{i}\la \mathbf{J}^T_0\mathbf{J}^T_0\mathbf{J}^T_0\mathbf{J}^T_0 \ra(C)$,
written out in eq.~\eqref{eq:super-disc} below. Dressing these by the same prefactors as in eq.~\eqref{full JJJJ OG pre} gives our final Grassmannian integrand uplift for the super WFC: 
\ie
\langle \mathbf{J}^T_0\mathbf{J}^T_0\mathbf{J}^T_0\mathbf{J}^T_0 \rangle (C)
=
\sum_{i\in disc} \mathcal{P}_i\;
\disc_{i}\la \mathbf{J}^T_0\mathbf{J}^T_0\mathbf{J}^T_0\mathbf{J}^T_0 \ra(C)
\label{eq: 4pt super WFC}
\fe

In terms of the branch sums of eq.~\eqref{eq:T4-branch}, with
$\hat\delta_4\equiv\hat\delta(C_4\cdot\Omega\cdot\Xi_4^I)$,
\begingroup\allowdisplaybreaks
\begin{align}
\disc_{1,3}\la\mathbf J^T_0\mathbf J^T_0\mathbf J^T_0\mathbf J^T_0\ra(C)&=\left.\frac{(S+T)\,\hat\delta_4T_4^{\rm even}}{ST\,AB}\right|_{\begin{smallmatrix}\tfrac12(S=0,\beta-\alpha)_{+}+\tfrac12(T=0,\beta-\alpha)_{+}\\+\tfrac12(A=0,\beta-\alpha)_{+}-\tfrac12(B=0,\beta-\alpha)_{+}\end{smallmatrix}}\nonumber\\
&\quad+\left.\frac{(S+T)\,\hat\delta_4T_4^{\rm odd}}{ST\,AB}\right|_{\begin{smallmatrix}\tfrac12(S=0,\beta-\alpha)_{-}+\tfrac12(T=0,\beta-\alpha)_{-}\\-\tfrac12(A=0,\beta-\alpha)_{-}+\tfrac12(B=0,\beta-\alpha)_{-}\end{smallmatrix}},\nonumber\\[4pt]
\disc_{2,4}\la\mathbf J^T_0\mathbf J^T_0\mathbf J^T_0\mathbf J^T_0\ra(C)&=\left.\frac{(S+T)\,\hat\delta_4T_4^{\rm even}}{ST\,AB}\right|_{\begin{smallmatrix}\tfrac12(S=0,\beta-\alpha)_{+}+\tfrac12(T=0,\beta-\alpha)_{+}\\+\tfrac12(A=0,\beta-\alpha)_{+}+\tfrac12(B=0,\beta-\alpha)_{+}\end{smallmatrix}}\nonumber\\
&\quad+\left.\frac{(S+T)\,\hat\delta_4T_4^{\rm odd}}{ST\,AB}\right|_{\tfrac12(S=0,\beta-\alpha)_{-}+\tfrac12(T=0,\beta-\alpha)_{-}}\nonumber\\
&\quad+\left.\frac{(T-S)\,\hat\delta_4T_4^{\rm odd}}{ST\,CD}\right|_{\tfrac12(C=0,\beta-\alpha)_{-}-\tfrac12(D=0,\beta-\alpha)_{-}},\nonumber\\[4pt]
\disc_{2,3,4}\la\mathbf J^T_0\mathbf J^T_0\mathbf J^T_0\mathbf J^T_0\ra(C)&=\left.\frac{(S+T)\,\hat\delta_4T_4^{\rm even}}{ST\,AB}\right|_{\begin{smallmatrix}(S=0,\beta-\alpha)_{+}+(T=0,\beta-\alpha)_{+}\\+\tfrac12(A=0,\beta-\alpha)_{+}+\tfrac12(B=0,\beta-\alpha)_{+}\end{smallmatrix}}\nonumber\\
&\quad+\left.\frac{(T-S)\,\hat\delta_4T_4^{\rm even}}{ST\,CD}\right|_{\tfrac12(C=0,\beta-\alpha)_{+}-\tfrac12(D=0,\beta-\alpha)_{+}}\nonumber\\
&\quad+\left.\frac{(S+T)\,\hat\delta_4T_4^{\rm odd}}{ST\,AB}\right|_{\begin{smallmatrix}(S=0,\beta-\alpha)_{-}+(T=0,\beta-\alpha)_{-}\\+\tfrac12(A=0,\beta-\alpha)_{-}+\tfrac12(B=0,\beta-\alpha)_{-}\end{smallmatrix}}\nonumber\\
&\quad+\left.\frac{(T-S)\,\hat\delta_4T_4^{\rm odd}}{ST\,CD}\right|_{\tfrac12(C=0,\beta-\alpha)_{-}-\tfrac12(D=0,\beta-\alpha)_{-}},\nonumber\\[4pt]
\disc_{1,4,s}\la\mathbf J^T_0\mathbf J^T_0\mathbf J^T_0\mathbf J^T_0\ra(C)&=\left.\frac{(S+T)\,\hat\delta_4T_4^{\rm even}}{ST\,AB}\right|_{(S=0,\beta-\alpha)_{+}}+\left.\frac{(S+T)\,\hat\delta_4T_4^{\rm odd}}{ST\,AB}\right|_{(S=0,\beta-\alpha)_{-}},\nonumber\\[4pt]
\disc_{1,2,t}\la\mathbf J^T_0\mathbf J^T_0\mathbf J^T_0\mathbf J^T_0\ra(C)&=\left.\frac{(S+T)\,\hat\delta_4T_4^{\rm even}}{ST\,AB}\right|_{(T=0,\beta-\alpha)_{+}}+\left.\frac{(S+T)\,\hat\delta_4T_4^{\rm odd}}{ST\,AB}\right|_{(T=0,\beta-\alpha)_{-}}\,.
\label{eq:super-disc}
\end{align}
\endgroup

\paragraph{Both branches under one prefactor.}
Substituting eq.~\eqref{eq:super-disc} into eq.~\eqref{eq: 4pt super WFC} makes
the grouping explicit:
\begin{equation}
\langle \mathbf{J}^T_0\mathbf{J}^T_0\mathbf{J}^T_0\mathbf{J}^T_0 \rangle (C)
=\sum_{i\in disc} \mathcal{P}_i\Big(
\left.F^{+}_i(C)\,\hat\delta_4T_4^{\rm even}\right|_{\mathcal C^{+}_i}
+\left.F^{-}_i(C)\,\hat\delta_4T_4^{\rm odd}\right|_{\mathcal C^{-}_i}\Big)\,,
\label{eq:super-wfc-branches}
\end{equation}
where $F^{\pm}_i(C)$ and $\mathcal C^{\pm}_i$ are the rational integrand and the
contour that channel $i$ uses on the branch $b=\pm$, both read off
eq.~\eqref{eq:super-disc}. The order of operations is the point: the test
functions of one branch are summed \emph{before} they are dressed, and only
then multiplied by a single $F$.

This is more than eq.~\eqref{eq: 4pt proposal} asks for, and we think it is
worth stating as a feature rather than a bookkeeping step. The general form
carries one prefactor $F^{h}(C_4)$ \emph{per helicity configuration}, and that
is the honest ansatz: different $h$ means a different $\xi_{i,-}^I$ dressing of
the test function, and nothing we have imposed relates the prefactors of two
sectors. Supersymmetry in particular does not, because the supercharge does not
act on $\xi_{i,-}^I$ at all. Writing it out at four points
(section~\ref{sec:gen-superinv}),
\begin{equation}
Q^{\alpha,I}=-\Lambda^T\cdot\Omega\cdot\Xi_4^I\,,\qquad
\Xi_4^I=\begin{bmatrix}
\tfrac14\xi^I_{1,+} & \tfrac14\xi^I_{2,+} & \tfrac14\xi^I_{3,+} & \tfrac14\xi^I_{4,+} &
2\tfrac{\partial}{\partial\xi^I_{1,+}} & 2\tfrac{\partial}{\partial\xi^I_{2,+}} &
2\tfrac{\partial}{\partial\xi^I_{3,+}} & 2\tfrac{\partial}{\partial\xi^I_{4,+}}
\end{bmatrix}\,,
\label{eq:Q-4pt}
\end{equation}
every entry is either multiplication by a $\xi_{i,+}$ or differentiation with
respect to one; $\xi_{i,-}$ appears nowhere. $Q$ therefore cannot see the label
that tells two sectors apart, and so cannot say whether they ought to share a
prefactor.

\paragraph{A component check.}
As a nontrivial check of the full SUSY WFC, we can project it onto the $\la O_2 J^+ O_2 J^+ \ra$ component. This component receives contributions from four test functions,

\ie
T_4^{++++},T_4^{-+-+},T_4^{-+++},T_4^{++-+}.
\fe
which, with the test-function normalization above and the $O_2$ slot
$(\xi_+^2+\xi_-^2)/(8E)$ of eq.~\eqref{N=2-Superfield}, give the following
coefficient-extracted numerators on the top cell,
\ie
&\frac1{256}\left((1 \bar 1 \bar 2 \bar 4) + (3 \bar 3 \bar 2 \bar 4)\right)^2, 
\frac1{256}\left((1 \bar 1 \bar 2 \bar 4) + (3 \bar 3 \bar 2 \bar 4)\right)^2,
\\
&-\frac1{256}\left((1 \bar 1 \bar 2 \bar 4) - (3 \bar 3 \bar 2 \bar 4)\right)^2,
-\frac1{256}\left((1 \bar 1 \bar 2 \bar 4) - (3 \bar 3 \bar 2 \bar 4)\right)^2
\fe
Notice that $(1 \bar 1 \bar 2 \bar 4)$ and $(3 \bar 3 \bar 2 \bar 4)$ are conjugate minors. Each nonzero numerator is therefore $+(1\bar1\bar2\bar4)^2/64$ on the positive branch and $-(1\bar1\bar2\bar4)^2/64$ on the negative branch. There are two contributing sectors on each branch. In the bosonic normalization this gives the component,
\ie
\langle O_2 J^+ O_2 J^+ \rangle (C)
=
\sum_{i\in disc} \mathcal{P}_i\, F_i(C) 
\fe
where $F_i(C)$ is the projection of $\disc_i\langle\mathbf J^T_0\mathbf J^T_0\mathbf J^T_0\mathbf J^T_0\rangle(C)$
onto the $\la O_2J^+O_2J^+\ra$ component. Working out the five $F_i(C)$, we find
that each matches the corresponding discontinuity
$\disc_i\la O_2J^+O_2J^+\ra(C)$ listed in appendix~\ref{app:ojoj},
eq.~\eqref{eq:ojoj-discs}, which in turn follows from the contours of
section~\ref{sec: discontinuity JJJJ} by the replacement $n_H\to X^2$.
Equivalently, the result can be read directly as the inversion of
$\la O_2J^+O_2J^+\ra$ itself: its energy dependence is spanned by the same five
functions $F_e(E)$ as $\la JJJJ\ra$, eq.~\eqref{eq: 4ptBasis}, so the same
prefactors $\mathcal P_i$ reconstruct it from its five discontinuities. The
other components of the super WFC are built on different energy bases, since
the contact terms demanded by the constrained superfields enter them, and their
prefactors would have to be recomputed; checking them is left to future work.
 
\begin{comment}
    In the other way, the OG form of the $\la O_2 J^+ O_2 J^+\ra$ can also be derived directly by a procedure analogous to the derivation of $\la JJJJ\ra$ in {the five-discontinuity inversion around \eqref{Energy function form : JJJJ}}. This is becasue one can also write the full WFC in terms of polarizations and momenta using the same 5 energy functions, 
{
\ie
\la O_2J^+O_2J^+\ra
&=\sum_{e=1}^{5}T^{OJOJ}_e(\epsilon,p,E^2)F_e(E),\\
\{F_e(E)\}_{e=1}^{5}
&=\left\{
\frac{1}{E_T},
\frac{1}{E_T E_{12s}E_{34s}},
\frac{1}{E_T E_{14t}E_{23t}},
\frac{(E_1-E_2)(E_3-E_4)}{E_T},
\frac{(E_1-E_4)(E_2-E_3)}{E_T}
\right\}.
\fe
}

and then use the same 5 discontinuities to perform the inversion. 

The same discussion applies to other helicities and components, even when the energy functions of the latter do not take the same form as $\la JJJJ \ra$ and contain contact terms. To understand the prefactor result for these components, one proposal is that supersymmetry might require the inversion obtained from the discontinuities to coincide with those for  $\la JJJJ \ra$ and give the same $P_i$ for all the component. Then we could write the full super WFC as an inversion of the super discontinuity by,
\ie
\langle \mathbf{J}_0\mathbf{J}_0\mathbf{J}_0\mathbf{J}_0 \rangle (C)
=
\sum_{i\in disc} \mathcal{P}_i \disc_{i} \la \mathbf{J}_0\mathbf{J}_0\mathbf{J}_0\mathbf{J}_0 \ra(C).
\fe

\end{comment}

\paragraph{Uplift of the compact forms.}
\label{sec:uplift-compact}
The compact representation of section~\ref{sec:compact-components} uplifts as
well. Replacing, in each of the four sectors with $h_2=h_4=+$, the current
numerator $n_H$ by $\hat\delta_4T_4^{h}$ gives
\begin{equation}
\langle \mathbf{J}^T_0\mathbf{J}^T_0\mathbf{J}^T_0\mathbf{J}^T_0 \rangle (C)\Big|_{h_2=h_4=+}
=\sum_{H\subseteq\{1,3\}}\la J^{h_H}\ra(C)\Big|_{n_H\to\hat\delta_4T_4^{h_H}}\,,
\label{eq:compact-super}
\end{equation}
with $\la J^{h_H}\ra(C)$ the compact forms of eq.~\eqref{eq:compact-assembly}
and $h_H$ the configuration with negative helicities on $H$. This is the
general form~\eqref{eq: 4pt proposal}, with one $F^{h}(C)$ per helicity
configuration: the compact forms carry helicity-dependent coefficients and
remainders, so the sectors of one branch no longer share an integrand, as they
do in eq.~\eqref{eq:super-wfc-branches}. It nevertheless gives the correct
components. With the scalars on legs $1,3$ and currents on $2,4$, the four
sectors combine into
\begin{align}
\la O_2J^+O_2J^+\ra(C)=\Big(\la J^+J^+J^+J^+\ra(C)&-\la J^-J^+J^+J^+\ra(C)\nonumber\\
-\la J^+J^+J^-J^+\ra(C)&+\la J^-J^+J^-J^+\ra(C)\Big)\Big|_{n_H\to(1\,\bar1\,\bar2\,\bar4)^2}\,,
\label{eq:compact-uplift}
\end{align}
where every current numerator $n_H$, including those inside the remainders, is
replaced by the same $(1\,\bar1\,\bar2\,\bar4)^2$, on the same poles, cells and branches. The signs
are those of the four helicity sectors of $\disc_{1,3}$, and the exchange and
contact parts come out together.
Inserting eqs.~\eqref{eq:compact-assembly} and
\eqref{eq:compact-L1}--\eqref{eq:compact-LM}, and collecting the residues taken
at the same pole, cell and branch, gives a compact form of this component itself,
\begingroup\allowdisplaybreaks
\begin{align}
\la O_2J^+O_2J^+\ra(C)={}&\mathcal F_S\,\frac{(S+T)(1\,\bar1\,\bar2\,\bar4)^2}{ST\,AB}\Big|_{(S=0,\beta-\alpha)_{+}-(S=0,\beta-\alpha)_{-}}\nonumber\\
&+\mathcal F_T\,\frac{(S+T)(1\,\bar1\,\bar2\,\bar4)^2}{ST\,AB}\Big|_{(T=0,\beta-\alpha)_{+}-(T=0,\beta-\alpha)_{-}}\nonumber\\
&-\frac{(1\,\bar1\,\bar2\,\bar4)^2}{AB}\bigg[\Big(1+\frac{s_{34}(E_4\bar s_{12}-E_2\bar t_{14})}{4E_2E_3E_4\,t_{14}}
-\frac{s_{12}(E_4\bar t_{23}-E_2\bar s_{34})}{4E_1E_2E_4\,t_{23}}\Big)\frac1S+\frac1T\nonumber\\
&\qquad-\frac{u_{24}}{8E_2E_4\,U}\Big(\frac{2E_4\bar s_{12}+E_{1,\bar 2,3,4}\,\bar t_{14}}{E_3\,t_{14}}\nonumber\\
&\qquad\qquad+\frac{E_{1,2,3,\bar 4}\,\bar t_{23}+2E_2\bar s_{34}}{E_1\,t_{23}}\Big)\bigg]\bigg|_{(A=0,\alpha)_{+}}\nonumber\\
&+\frac{(1\,\bar1\,\bar2\,\bar4)^2}{4E_2E_3E_4\,t_{14}\,AB}\bigg[\frac{\bar s_{34}(E_4\bar s_{12}-E_2\bar t_{14})}{S}\nonumber\\
&\qquad+\frac{u_{24}(2E_4\bar s_{12}+E_{1,\bar 2,\bar 3,4}\,\bar t_{14})}{2U}\bigg]\bigg|_{(B=0,\alpha)_{-}}\nonumber\\
&-\frac{(1\,\bar1\,\bar2\,\bar4)^2}{4E_1E_2E_4\,s_{34}\,AB}\bigg[\frac{\bar t_{14}(E_4\bar t_{23}-E_2\bar s_{34})}{T}\nonumber\\
&\qquad-\frac{u_{24}(2E_4\bar t_{23}+E_{\bar 1,\bar 2,3,4}\,\bar s_{34})}{2U}\bigg]\bigg|_{(A=0,\alpha)_{-}}\,,
\label{eq:compact-ojoj}
\end{align}
\endgroup
where the first two lines come from the channel blocks of the four sectors, which
combine into the two branches with opposite signs, the residue at
$(A=0,\alpha)_+$ collects the extra MHV residue and $L_{-+-+}$, and the last two
terms are $-L_{-+++}$ and $-L_{++-+}$ of
eqs.~\eqref{eq:compact-L1}--\eqref{eq:compact-LM}.

\subsection{Super Cutting Rule}
\label{sec: Super Cutting Rule}
It is well known that four-point WFCs satisfy cutting rules \cite{Meltzer:2021zin, Goodhew:2021oqg, Baumann:2021fxj, Melville:2021lst}, under which a WFC factorizes into left and right three-point WFCs. For example, the Yang--Mills WFC in momentum space obeys
\eqs{
    \label{cutting rule}
\mathrm{Disc}_{s} \langle J^+J^+J^+ J^+\rangle = \frac{1}{2E_s} \sum_{h=\pm} \mathrm{Disc}_{s}  \langle J^+(k_1)J^+(k_2)J^h(k_s)\rangle \mathrm{Disc}_{s}  \langle J^{-h}(k_s) J^+(k_3)J^+(k_4)\rangle .
}
However, the discontinuity in the $s$ channel alone is not the homogeneous solution. Applying the same idea to the energy-function representation in \eqref{eq: 4ptBasis}, the inversion from $\disc_s$ and $\disc_{1,4,s}$ is simple:
\ie
\disc_{s} \la J^+ J^+ J^+ J^+ \ra = \mathcal{F}_S \disc_{1,4,s} \la J^+ J^+ J^+ J^+ \ra = \mathcal{F}_S\left.\frac{(\bar 1 \bar 2 \bar 3 \bar 4)^2}
{S(T+U)(T-U)}\right|_{(S=0,\beta-\alpha)_+}
\fe
with the very prefactor that dresses the compact forms of
section~\ref{sec:compact-components}, eq.~\eqref{eq:FS-FT},
\eqs{
\mathcal{F}_S = \frac{\bar s_{12}\,\bar s_{34}}{16 E_1 E_2 E_3 E_4}
=\frac{\big((E_1-E_2)^2-E_s^2\big)\big((E_3-E_4)^2-E_s^2\big)}{16 E_1 E_2 E_3 E_4}\,,
}
Without the factor $\mathcal{F}_S$, the factorization is not that of the full WFC, but rather of the three-point discontinuities, which can be expressed by the three-point OG as suggested in \cite{Arundine:2026fbr}. The $t$-channel cut is analogous:
\ie
\disc_{t} \la J^+ J^+ J^+ J^+ \ra = \mathcal{F}_T \disc_{1,2,t} \la J^+ J^+ J^+ J^+ \ra= \mathcal{F}_T\left.\frac{(\bar 1 \bar 2 \bar 3 \bar 4)^2}
{T(S+U)(S-U)}\right|_{(T=0,\beta-\alpha)_+}
\fe
with the $\mathcal F_T$ of the same equation,
\eqs{
\mathcal{F}_T= \frac{\bar t_{14}\,\bar t_{23}}{16 E_1 E_2 E_3 E_4}
=\frac{\big((E_1-E_4)^2-E_t^2\big)\big((E_2-E_3)^2-E_t^2\big)}{16 E_1 E_2 E_3 E_4}\,,
}
These results can then be embedded into the supersymmetric invariant:
\ie
\disc_{s}\la \bm J_0 \bm J_0 \bm J_0 \bm J_0 \ra (C)
=& \mathcal{F}_S \left( \sum_{h\in\pm\pm\pm\pm} \left.\frac{
    \hat \delta \left(C_{4} \cdot  \Omega\cdot  \Xi_4^I \right)T^{h}_4(\xi_{i,\pm}^I)
    }
{S(T+U)(T-U)}\right|_{(S=0,\beta-\alpha)_+ + (S=0,\beta-\alpha)_-}  \right)
\fe
Projecting onto the current components gives the corresponding $\disc_s$ and the $s$-channel cutting rule in \eqref{cutting rule}. Similarly, for the $t$-channel cutting rule,
\ie
\disc_{t}\la \bm J_0 \bm J_0 \bm J_0 \bm J_0 \ra (C)
=& \mathcal{F}_T \left(  \sum_{h\in\pm\pm\pm\pm} \left.\frac{
    \hat \delta \left(C_{4} \cdot  \Omega\cdot  \Xi_4^I \right)T^{h}_4(\xi_{i,\pm}^I)
    }
{T(S+U)(S-U)}\right|_{(T=0,\beta-\alpha)_+ + (T=0,\beta-\alpha)_-}  \right) 
\fe
The projection onto $\la O_2 J^+ O_2 J^+\ra$ likewise gives the expected $s$- and $t$-channel cuts. For example,
\ie
\disc_{s}\la O_2 J^+ O_2 J^+ \ra (C)
&= \mathcal{F}_S \disc_{1,4,s} \la O_2 J^+ O_2 J^+ \ra(C)\\
&= \mathcal{F}_S \left.\frac{
    2 (1 \bar 1 \bar 2 \bar 4)^2
    }
{S(T+U)(T-U)}\right|_{(S=0,\beta-\alpha)_+ - (S=0,\beta-\alpha)_-}\\
& \xrightarrow{\int dC}
-\frac{
E_s\,\abn{\bar 2 1}\abBB{2}{1}\abn{\bar 4 3}\abBB{4}{3}
}{
16E_2 E_4\bigl((E_1+E_2)^2-E_s^2\bigr)\bigl((E_3+E_4)^2-E_s^2\bigr)
}\,.
\fe
The last line is the $\disc_{s}$ of $\la O_2 J O_2 J \ra$, as written in \cite{Baumann:2020dch}, projected onto the all-plus helicity component. The expression without $\mathcal{F}_S$ is also the sum of products of the left and right WFC discontinuities written in OG form, evaluated on the positive and negative branches, as in \cite{Arundine:2026fbr}.

\section{Conclusion}
In this paper we developed a way to lift homogeneous Grassmannian data to full tree-level WFCs. The key point is that the orthogonal Grassmannian directly produces homogeneous solutions of the spinor-helicity conformal Ward identities. For conserved currents, the full WFC is instead inhomogeneous, because the spinor conformal boost generates longitudinal components fixed by lower-point data. Suitable energy discontinuities remove these longitudinal terms and therefore lie in the homogeneous sector. Once the energy-function basis of the WFC is known, these discontinuities form a linear system for the numerator data. Inverting this system expresses the full WFC as a sum of Grassmannian discontinuities dressed by explicit energy prefactors.

We implemented this mechanism at three and four points and then embedded it into $\mathcal N=2$ momentum superspace. The supersymmetric Grassmannian invariants are generated by $\hat\delta(C\Omega\Xi^I)$ acting on helicity-selecting test functions. The two branches of the orthogonal Grassmannian organize distinct superinvariants and reduce, in the total-energy-pole limit, to different flat-space helicity superamplitudes. At three points the triple discontinuity fixes the full transverse super WFC. At four points a five-discontinuity basis reconstructs the transverse four-point super WFC. We checked the construction by projecting onto components such as $\langle O_2J^+O_2J^+\rangle$ and by verifying that the $s$-channel all-plus cell reproduces the super cutting rule.

The freedom to choose a discontinuity basis can lead to more compact
Grassmannian representations of the same WFC.
Section~\ref{sec:compact-components} gives these for all four Yang--Mills
helicity components: each is a channel block plus at most three residues, every
prefactor built only from external energies and the energy sums
$E_{i,j,k}$, with or without bars, which pair up into the Mandelstam-like
variables $s_{ij}$, $\bar s_{ij}$, $t_{ij}$, $\bar t_{ij}$, $u_{ij}$ of
eq.~\eqref{eq:pair-energies}. For the all-plus component the
block alone suffices,
\ie
\la J^+J^+J^+J^+ \ra(C) =  \mathcal F_S \left.\frac{(\bar 1 \bar 2 \bar 3 \bar 4)^2}
{S\,AB}\right|_{(S=0,\beta)_+}+ (2 \leftrightarrow 4)\,,\qquad
\mathcal F_S=\frac{\bar s_{12}\,\bar s_{34}}{16\,E_1E_2E_3E_4}\,,
\fe
with $A,B$ of eq.~\eqref{eq:ABCD}; this is the first line of
eq.~\eqref{eq:compact-assembly}, where the two energy factors are the
$\mathcal F_S$ and $\mathcal F_T$ of eq.~\eqref{eq:FS-FT}. The corresponding momentum-space expression is
\ie
\la J^+J^+J^+J^+ \ra = -\frac{1}{512E_s E_1 E_2 E_3 E_4} \frac{\langle \bar 1 \bar s \rangle \langle \bar 2 \bar s \rangle \langle \bar 1 \bar 2 \rangle}{E_{1,2,s}}  \frac{\langle \bar 3 \bar 4 \rangle \langle  s \bar 3 \rangle \langle  s \bar 4 \rangle}{E_{3,4,s}} + (2 \leftrightarrow 4)
\fe
which makes the partial-energy poles manifest, while the residue at the total-energy pole vanishes, as expected in the all-plus sector.

The other three Yang--Mills components are not far behind: as
section~\ref{sec:compact-components} shows, the MHV and the two single-minus
sectors take the same shape, a channel block with the prefactors
$\mathcal F_S$ and $\mathcal F_T$ of eq.~\eqref{eq:FS-FT} plus a remainder of two residues whose
coefficients are again built only from external energies and barred energy
sums, eqs.~\eqref{eq:compact-L1}--\eqref{eq:compact-LM}. The MHV sector carries
in addition one residue at the flat pole $A=0$, eq.~\eqref{eq:compact-assembly},
which is where its total-energy singularity sits. Compton scattering
with two positive-helicity currents simplifies in the same way.

These examples raise the question of how the compact representations emerge
from the five-discontinuity reconstruction in eq.~(\ref{eq: 4ptDisRep}).
Understanding their equivalence directly at the level of the Grassmannian
contour may explain the simplification of the all-plus energy prefactors. It
may also explain why only the all-plus component collapses to a single block,
while the others keep a two-residue remainder.

A second observation points the same way. The general form of the super WFC,
eq.~\eqref{eq: 4pt proposal}, carries one prefactor per helicity configuration,
and supersymmetry cannot do better: the sectors differ by their $\xi_{i,-}$
dressing, which the supercharge does not act on, so nothing in the algebra says
whether two sectors should share a prefactor. The
prefactor depends on the sector only through its branch, so their test
functions can be grouped before they are dressed. We
read this as a sign that the super WFC carries more symmetry than the
construction uses, and perhaps that one more uplift is available: to a
geometric coordinate in which the branches and the contours of
section~\ref{sec: discontinuity JJJJ} are features of a geometry rather than a
prescription, so that the contour each channel runs on is produced rather than
imposed.

Several open directions follow from this work. First, the inversion should have a more intrinsic Grassmannian formulation, rather than being implemented through an external energy-function basis. The simple four-point prefactors suggest that a direct contour or residue prescription may package the inversion at the level of the integral. Second, the construction should extend to higher multiplicity, where one must identify suitable energy bases, choose discontinuities that remove all longitudinal inhomogeneous terms, and construct the corresponding $\operatorname{OG}(n,2n)$ superinvariants. Finally, the contact terms required by constrained superfields deserve a more geometric interpretation. Incorporating them directly into the super-Grassmannian framework may clarify the relation between supersymmetry, longitudinal modes, and the inhomogeneous conformal Ward identities of spinning WFCs.

\section*{Acknowledgments}
We are grateful to Aswini Bala, Sachin Jain, Dhruva K. S., and Adithya A Rao for sharing their results with us. We also thank Mattia Arundine,  Daniel Baumann,  Subramanya Hegde, Hiren Kakkad, Facundo Rost, Kajal Singh and Francisco Vaz\~{a}o for helpful discussions. We used Claude Opus 5.5 and ChatGPT Astra 6 to polish the manuscript and to help
implement the numerical checks; all results were verified by the authors. Y.-t. Huang and Y. Liu are supported by the Taiwan Ministry of Science and Technology Grants No. 112-2628-M-002-003-MY3 and
114-2923-M-002-011-MY5.
C.-K. Kuo and J. Mei are funded by the European Union (ERC, UNIVERSE PLUS, 101118787). Views and opinions expressed are, however, those of the author(s) only and do not necessarily reflect those of the European Union or the European Research Council Executive Agency. Neither the European Union nor the granting authority can be held responsible for them.

\appendix

\section{Hat relations and the higher discontinuities}
\label{app:hat}

The discontinuities of section~\ref{sec: discontinuity JJJJ} are given there
only in the all-plus sector. This appendix derives the relations that carry
them to the other helicity sectors and to one more discontinued leg, and
records what they give.

The independent data are few. Only the all-plus values of the five basis
discontinuities are needed: $\disc_{1,3}$ and $\disc_{2,4}$, one triple
discontinuity, $\disc_{2,3,4}$, and the channel discontinuities
$\disc_{1,4,s}$ and $\disc_{1,2,t}$. Every other discontinuity, in every
helicity sector, follows from these by hats, by the permutations of the legs
that preserve the color ordering, or both: the other triple discontinuities
are permutations of $\disc_{2,3,4}$, the quadruple one is
$\disc_{1,2,3,4}=2(\disc_{1,4,s}+\disc_{1,2,t})$ on the five-function energy
basis, and the crossed sectors of the double discontinuities follow from the
triple and quadruple ones by inverting the hat relation.

\subsection{Hat relations between helicity discontinuities}
The discontinuities of different helicity components are not independent.
Let $\hat h_i$ act on any function of the kinematics by exchanging
$\lambda_i\leftrightarrow\bar\lambda_i$ and sending $E_i\to-E_i$. Flipping a
single energy at fixed polarization, as in the definition of $\disc$, and using
that $\epsilon^\pm_i$ built from the hatted spinors equals $-\epsilon^\mp_i$,
\begin{equation}
\la J^{h}\ra\big|_{E_i\to-E_i}=-\hat h_i\la J^{h^{(i)}}\ra\,,
\label{eq:hat-flip}
\end{equation}
where $h^{(i)}$ reverses the helicity of leg $i$. Iterating, for a set $X$ of
external legs and $Y\subseteq X$,
\begin{equation}
\disc_X\la J^{h}\ra=\sum_{Z\subseteq X}\hat h_Z\la J^{h^{(Z)}}\ra
=\hat h_Y\disc_X\la J^{h^{(Y)}}\ra\,.
\label{eq:hat-disc}
\end{equation}
Two consequences of eq.~\eqref{eq:hat-disc} are used below:
\begin{itemize}
\item If the negative-helicity legs are among the discontinued legs, the
discontinuity is the hat image of the all-plus one, e.g.
\begin{equation*}
\disc_{1,3}\la J^-J^+J^+J^+\ra=\hat h_1\disc_{1,3}\la J^+J^+J^+J^+\ra\,.
\end{equation*}
A sector whose flipped legs lie outside $X$ is correspondingly out of reach:
this covers all four sectors of $\disc_{1,3}$, pairs them for $\disc_{2,3,4}$
and $\disc_{1,4,s}$, and covers none of $\disc_{2,4}$, whose four sectors
differ in the helicities of legs $1$ and $3$. On the Grassmannian the
operation amounts to swapping entries of the numerator minor, made precise in
eq.~\eqref{eq:hat-columns} below, together with a move to the opposite
branch.
\item A further discontinuity in a leg $i\notin X$ is generated by the hat:
\begin{equation}
\disc_{X\cup\{i\}}\la J^{h}\ra=\disc_X\la J^{h}\ra+\hat h_i\disc_X\la J^{h^{(i)}}\ra\,.
\label{eq:hat-extra-disc}
\end{equation}
For example, $\disc_{2,3,4}\la J^+J^+J^+J^+\ra$ and
$\disc_{2,4}\la J^+J^+J^-J^+\ra$ determine each other. Similarly,
$\disc_{1,2,4}\la J^+J^+J^+J^+\ra$ and $\disc_{2,4}\la J^-J^+J^+J^+\ra$ determine
each other.
\end{itemize}
The components of $\disc_{2,4}$ flipped on legs $1$ or $3$, and those of
$\disc_{1,3}$ flipped on legs $2$ or $4$, cannot be obtained by hat transport
at fixed double discontinuity. They follow instead by inverting
eq.~\eqref{eq:hat-extra-disc}: the single-minus ones from the all-plus triple
discontinuities, as in eq.~\eqref{eq:crossed-from-triples}, and the MHV-type
ones from the all-plus quadruple discontinuity. The forward relation then
assembles the triple discontinuities in every sector.

\subsection{Residue transport}
On the OG side $\hat h_i$ exchanges the columns $i$ and $\bar i$ and the two
branches. With $A,B,C,D$ as in eq.~\eqref{eq:ABCD}, the oriented
minors transform as
\begin{center}\small
\begin{tabular}{lrrrrrrr}
\hline
 & $S$ & $T$ & $U$ & $A$ & $B$ & $C$ & $D$\\\hline
$\hat h_1$ & $-S$ & $-T$ & $-U$ & $-A$ & $-B$ & $-C$ & $-D$\\
$\hat h_2$ & $-S$ & $T$ & $U$ & $-D$ & $C$ & $B$ & $-A$\\
$\hat h_3$ & $S$ & $T$ & $-U$ & $B$ & $A$ & $-D$ & $-C$\\
$\hat h_4$ & $S$ & $-T$ & $U$ & $-C$ & $D$ & $-A$ & $B$\\\hline
\end{tabular}
\end{center}
The reversal $r:(1,2,3,4)\mapsto(4,3,2,1)$ preserves the branch but acts
differently on the oriented minors on its two branches. If $b=+1$ or $-1$
denotes the branch sign in eq.~\eqref{eq: Conjugate}, then
\begin{equation}
r:(S,T,U)\longmapsto(bS,T,bU),\qquad
(A,B,C,D)\longmapsto
\begin{cases}
(A,B,C,D),&b=+1,\\
(C,-D,A,-B),&b=-1.
\end{cases}
\label{eq:reflection-minors}
\end{equation}
For example, after relabeling,
$S=(\bar4\,\bar3\,43)=(\bar3\,\bar4\,34)=bS_{\rm old}$;
the same dual-minor identity gives $U\mapsto bU$, whereas
$T=(\bar4\,\bar1\,41)=(\bar1\,\bar4\,14)$ is unchanged. Thus the
all-plus positive-branch $CD$ poles remain $CD$ poles under reversal.
The cyclic shift by two legs, $(1,2,3,4)\mapsto(3,4,1,2)$, also preserves
the branch. It is the identity on the minors of the positive branch, while on
the negative branch
\begin{equation}
(S,T,U)\longmapsto(-S,-T,U),\qquad (A,B,C,D)\longmapsto(-B,-A,D,C)\,,
\label{eq:shift2-minors}
\end{equation}
so it maps each of the kernels $(S+T)n/(ST\,AB)$ and $(T-S)n/(ST\,CD)$ to minus
itself. These maps, and the hat table above, follow from the definitions of
$S,T,U$ and the dual-minor identities; we have also checked them on random
points of both branches of $\operatorname{OG}(4,8)$.
The equality of the full discontinuities under reversal also uses the
color-ordered four-point function; the pole map alone does not determine
the orientations of individual residue cells.
For the hat operations, the numerator obeys the same column-exchange rule,
since it too is built from minors: $\hat h_i$
replaces the $i$-th entry of each bracket by its conjugate and leaves the other
three alone,
\begin{equation}
\hat h_i\,(a_1\,a_2\,a_3\,a_4)=(a_1\cdots \bar a_i\cdots a_4)\,,\qquad
\bar{\bar a}=a\,,
\label{eq:hat-columns}
\end{equation}
so that, for instance, $\hat h_1(\bar1\,\bar2\,\bar3\,\bar4)=(1\,\bar2\,\bar3\,\bar4)$
and $\hat h_3(1\,\bar2\,\bar3\,\bar4)=(1\,\bar2\,3\,\bar4)$. The numerators below
are $n_H=(a_1\,\bar2\,a_3\,\bar4)^2$ with $a_i=i$ for $i\in H$ and
$a_i=\bar i$ otherwise, $H$ the set of negative-helicity legs among $\{1,3\}$;
on these the rule for $i\in\{1,3\}$ is exactly the reversal of the
helicity of leg $i$, $n_H\to n_{H^{(i)}}$, which is why the transport by
$\hat h_1$ and $\hat h_3$ closes on the same four numerators. A residue is
accordingly carried to a residue,
\begin{equation}
\hat h_i\,I\big|_{(P=0,c)_b}=\pm\,I'\big|_{(P'=0,c')_{-b}}\,,
\label{eq:hat-residue}
\end{equation}
with $P'$ read off the table and $I'$ the image integrand, every minor and pole
factor carrying the sign the table assigns it. The branch is always reversed:
the subscript on the right is $-b$, never $b$. A hat never returns a residue to
the branch it came from, so the two branches are never independent data, and
every relation below that connects them passes through this step. The two
cells are exchanged, $\alpha\leftrightarrow\beta$, at $S$ and $T$ under
$\hat h_1$, at $S$, $A$ and $D$ under $\hat h_2$, at $C$ and $D$ under
$\hat h_3$, and at $T$, $A$ and $C$ under $\hat h_4$, and kept at the other
poles; with these exchanges and the localization factors listed below, the
transport is carried out term by term.

The cell labels and residue orientations are fixed independently of the
column-exchange table. At $A=0$ the labels are those of the explicit matrices
in eqs.~\eqref{eq: amplitude cell 4pt 1} and~\eqref{eq: amplitude cell 4pt 3};
at $B,C,D$ they are obtained by simultaneous spinor-column and energy flips
on legs $\{3\}$, $\{2,3\}$ and $\{3,4\}$, respectively, preserving the
label $\alpha$ or $\beta$. At $S=0$ we use the factorization cells of
appendix~\ref{app:inv-4}, and $T=0$ follows by $2\leftrightarrow4$.
\footnote{\label{fn:hat-conventions}The labels specify row spaces, not the
order of roots returned by a solver. A general hat transformation can still
exchange the two named cells.}
For reproducibility, the bosonic localization factor at the $\alpha$ cell is
listed below; the factor at $\beta$ has the opposite sign. The spectator
rational function and its numerator are then evaluated on that cell.
\begin{center}\small
\begin{tabular}{c|cc}
Pole & $+$ branch & $-$ branch\\\hline
$S$ & $-1/(8E_s)$ & $-1/(8E_s)$\\
$T$ & $-1/(8E_t)$ & $-1/(8E_t)$\\
$A$ & $-1/(8E_T)$ & $+1/[8(E_T-2E_1)]$\\
$B$ & $+1/[8(E_T-2E_1-2E_3)]$ & $-1/[8(E_T-2E_3)]$\\
$C$ & $+1/[8(E_T-2E_1-2E_4)]$ & $-1/[8(E_T-2E_4)]$\\
$D$ & $+1/[8(E_T-2E_1-2E_2)]$ & $-1/[8(E_T-2E_2)]$
\end{tabular}
\end{center}
These factors use identity columns $(\bar1,\bar2,3,4)$ on the positive
branch, except $(1,2,3,4)$ at $D$, and $(1,2,3,\bar4)$ on the negative
branch, except $(1,2,4,\bar3)$ at $C$. A different GL$(4)$ gauge requires
the corresponding transformation of the localization factor.
The oriented sums of residues of both rational kernels used below vanish;
this provides an additional numerical check of the convention. The
half-differences of $\alpha$ and $\beta$ below keep that convention explicit.

\subsection{The helicity sectors of $\disc_{1,3}$ and $\disc_{2,4}$}
Throughout, formulas carry the normalization and orientation of
eq.~\eqref{full JJJJ OG} and each sector is evaluated on its own branch,
$b=(-1)^{n_-}$ with $n_-$ the number of negative helicities. The cell labels and orientations are those fixed above.

Which sectors a channel reaches is fixed by eq.~\eqref{eq:hat-disc}: only a leg
that has been discontinued can be flipped. So $\hat h_1$ and $\hat h_3$ move
$\disc_{1,3}$ through the four sectors that differ in legs $1$ and $3$, while
for $\disc_{2,4}$ the four within reach are those differing in legs $2$ and
$4$. Between them the two channels cover seven sectors, and the split is not a
matter of taste: crossing over---$\disc_{1,3}$ on the sectors that differ in
legs $2,4$, or $\disc_{2,4}$ on those that differ in legs $1,3$---is precisely
what the hat cannot do at fixed double discontinuity. We first give both
channels on their own sectors. We then take $\disc_2$ of the all-plus
$\disc_{1,3}$, reverse the full color ordering to obtain the all-plus
$\disc_{2,3,4}$, and invert eq.~\eqref{eq:hat-extra-disc} to recover the
crossed single-minus $\disc_{2,4}$. The same inversion before reversal
gives the crossed single-minus $\disc_{1,3}$.

\paragraph{$\disc_{1,3}$ on its own four sectors.}
\begin{align}
\disc_{1,3}\la J^+J^+J^+J^+\ra&=\left.\frac{(S+T)(\bar1\,\bar2\,\bar3\,\bar4)^2}{ST\,AB}\right|_{\begin{smallmatrix}\tfrac12(S=0,\beta-\alpha)_{+}+\tfrac12(T=0,\beta-\alpha)_{+}\\+\tfrac12(A=0,\beta-\alpha)_{+}-\tfrac12(B=0,\beta-\alpha)_{+}\end{smallmatrix}},\nonumber\\
\disc_{1,3}\la J^-J^+J^-J^+\ra&=\left.\frac{(S+T)(1\,\bar2\,3\,\bar4)^2}{ST\,AB}\right|_{\begin{smallmatrix}\tfrac12(S=0,\beta-\alpha)_{+}+\tfrac12(T=0,\beta-\alpha)_{+}\\+\tfrac12(A=0,\beta-\alpha)_{+}-\tfrac12(B=0,\beta-\alpha)_{+}\end{smallmatrix}},\nonumber\\
\disc_{1,3}\la J^-J^+J^+J^+\ra&=\left.\frac{(S+T)(1\,\bar2\,\bar3\,\bar4)^2}{ST\,AB}\right|_{\begin{smallmatrix}\tfrac12(S=0,\beta-\alpha)_{-}+\tfrac12(T=0,\beta-\alpha)_{-}\\-\tfrac12(A=0,\beta-\alpha)_{-}+\tfrac12(B=0,\beta-\alpha)_{-}\end{smallmatrix}},\nonumber\\
\disc_{1,3}\la J^+J^+J^-J^+\ra&=\left.\frac{(S+T)(\bar1\,\bar2\,3\,\bar4)^2}{ST\,AB}\right|_{\begin{smallmatrix}\tfrac12(S=0,\beta-\alpha)_{-}+\tfrac12(T=0,\beta-\alpha)_{-}\\-\tfrac12(A=0,\beta-\alpha)_{-}+\tfrac12(B=0,\beta-\alpha)_{-}\end{smallmatrix}}
\label{eq:hat-D13}
\end{align}

The four sectors use the same rational kernel, with the branch-dependent
flat-pole orientations displayed explicitly. \footnote{\label{fn:numerator-regular}Nothing in the contour knows
about the helicity. Each $n_H$ is a polynomial in the minors, so it is regular
wherever the integrand has a pole, and a residue at a simple pole only sees the
value a regular factor takes at the point the residue localizes on,
$(n_H\,g)\big|_{(P=0,c)_b}=n_H\big(C_{(P=0,c)_b}\big)\,g\big|_{(P=0,c)_b}$ with
$C_{(P=0,c)_b}$ that point. Any other numerator regular there may be put in its
place, as appendix~\ref{app:ojoj} and section~\ref{sec:compact-components} both
do.}

\paragraph{$\disc_{2,4}$ on its own four sectors.}
The remaining three follow from the all-plus line by the transport of
eq.~\eqref{eq:hat-residue}, with no appeal to
eq.~\eqref{eq:hat-extra-disc}. Reading the table, $\hat h_2$ sends
$S\to-S$, $A\to-D$ and $B\to C$, so it carries $(S+T)n/(ST\,AB)$ into
$(T-S)n/(ST\,CD)$; $\hat h_4$ does the same up to a sign, and
$\hat h_2\hat h_4$ returns to the first integrand with $A\leftrightarrow B$.
Note that transport can change which integrand carries a sector---$\hat h_2$
and $\hat h_4$ land on $(T-S)n/(ST\,CD)$---but it never needs two at once: each
line below is a single integrand on a single contour, exactly as for
$\disc_{1,3}$. Writing the numerator as $(\bar 1\,a_2\,\bar 3\,a_4)^2$,
the four sectors are
\begin{align}
\disc_{2,4}\la J^+J^+J^+J^+\ra&=\left.\frac{(S+T)(\bar1\,\bar2\,\bar3\,\bar4)^2}{ST\,AB}\right|_{\begin{smallmatrix}\tfrac12(S=0,\beta-\alpha)_{+}+\tfrac12(T=0,\beta-\alpha)_{+}\\+\tfrac12(A=0,\beta-\alpha)_{+}+\tfrac12(B=0,\beta-\alpha)_{+}\end{smallmatrix}},\nonumber\\
\disc_{2,4}\la J^+J^-J^+J^+\ra&=\left.\frac{(T-S)(\bar1\,2\,\bar3\,\bar4)^2}{ST\,CD}\right|_{\begin{smallmatrix}-\tfrac12(S=0,\beta-\alpha)_{-}+\tfrac12(T=0,\beta-\alpha)_{-}\\+\tfrac12(C=0,\beta-\alpha)_{-}-\tfrac12(D=0,\beta-\alpha)_{-}\end{smallmatrix}},\nonumber\\
\disc_{2,4}\la J^+J^+J^+J^-\ra&=\left.\frac{(T-S)(\bar1\,\bar2\,\bar3\,4)^2}{ST\,CD}\right|_{\begin{smallmatrix}-\tfrac12(S=0,\beta-\alpha)_{-}+\tfrac12(T=0,\beta-\alpha)_{-}\\+\tfrac12(C=0,\beta-\alpha)_{-}-\tfrac12(D=0,\beta-\alpha)_{-}\end{smallmatrix}},\nonumber\\
\disc_{2,4}\la J^+J^-J^+J^-\ra&=\left.\frac{(S+T)(\bar1\,2\,\bar3\,4)^2}{ST\,AB}\right|_{\begin{smallmatrix}\tfrac12(S=0,\beta-\alpha)_{+}+\tfrac12(T=0,\beta-\alpha)_{+}\\+\tfrac12(A=0,\beta-\alpha)_{+}+\tfrac12(B=0,\beta-\alpha)_{+}\end{smallmatrix}}
\label{eq:hat-D24-flip24}
\end{align}
The two even sectors use the $AB$ kernel; the two odd sectors use the $CD$
kernel. All pole and cell signs are expressed in the fixed orientation above.

\paragraph{The all-plus triple discontinuities.}
One more energy difference of an all-plus double discontinuity, taken at fixed
polarizations as in eq.~\eqref{eq:triple-from-double-direct}, gives an all-plus
triple discontinuity. The double discontinuity carries the $AB$ kernel alone;
the triple one carries residues of the $CD$ kernel as well. Taking $\disc_2$
of the all-plus $\disc_{1,3}$ of eq.~\eqref{eq:hat-D13},
\begin{align}
\disc_{1,2,3}\la J^+J^+J^+J^+\ra&=\left.\frac{(S+T)(\bar1\,\bar2\,\bar3\,\bar4)^2}{ST\,AB}\right|_{\begin{smallmatrix}(S=0,\beta-\alpha)_{+}+(T=0,\beta-\alpha)_{+}\\+\tfrac12(A=0,\beta-\alpha)_{+}-\tfrac12(B=0,\beta-\alpha)_{+}\end{smallmatrix}}\nonumber\\
&\quad+\left.\frac{(T-S)(\bar1\,\bar2\,\bar3\,\bar4)^2}{ST\,CD}\right|_{\tfrac12(C=0,\beta-\alpha)_{+}+\tfrac12(D=0,\beta-\alpha)_{+}}\,,
\label{eq:D123-allplus}
\end{align}
all on the positive branch. Reversing the color ordering,
$(1,2,3,4)\mapsto(4,3,2,1)$, maps $\disc_{1,2,3}$ to $\disc_{2,3,4}$. On the
positive branch the reversal preserves every minor,
eq.~\eqref{eq:reflection-minors}, so it keeps both integrands, but it reverses
the orientation of the residues at $B$ and at $D$:
\begin{align}
\disc_{2,3,4}\la J^+J^+J^+J^+\ra&=\left.\frac{(S+T)(\bar1\,\bar2\,\bar3\,\bar4)^2}{ST\,AB}\right|_{\begin{smallmatrix}(S=0,\beta-\alpha)_{+}+(T=0,\beta-\alpha)_{+}\\+\tfrac12(A=0,\beta-\alpha)_{+}+\tfrac12(B=0,\beta-\alpha)_{+}\end{smallmatrix}}\nonumber\\
&\quad+\left.\frac{(T-S)(\bar1\,\bar2\,\bar3\,\bar4)^2}{ST\,CD}\right|_{\tfrac12(C=0,\beta-\alpha)_{+}-\tfrac12(D=0,\beta-\alpha)_{+}}\,.
\label{eq:D234-allplus}
\end{align}
The same result follows from $\disc_3$ of the all-plus $\disc_{2,4}$. The two
further triple discontinuities used below are related to these by the
reflections of the color ordering that fix leg $1$ and leg $2$,
\begin{align}
\disc_{1,3,4}\la J^+J^+J^+J^+\ra&=\left.\frac{(S+T)(\bar1\,\bar2\,\bar3\,\bar4)^2}{ST\,AB}\right|_{\begin{smallmatrix}(S=0,\beta-\alpha)_{+}+(T=0,\beta-\alpha)_{+}\\+\tfrac12(A=0,\beta-\alpha)_{+}-\tfrac12(B=0,\beta-\alpha)_{+}\end{smallmatrix}}\nonumber\\
&\quad+\left.\frac{(T-S)(\bar1\,\bar2\,\bar3\,\bar4)^2}{ST\,CD}\right|_{-\tfrac12(C=0,\beta-\alpha)_{+}-\tfrac12(D=0,\beta-\alpha)_{+}}\,,\nonumber\\
\disc_{1,2,4}\la J^+J^+J^+J^+\ra&=\left.\frac{(S+T)(\bar1\,\bar2\,\bar3\,\bar4)^2}{ST\,AB}\right|_{\begin{smallmatrix}(S=0,\beta-\alpha)_{+}+(T=0,\beta-\alpha)_{+}\\+\tfrac12(A=0,\beta-\alpha)_{+}+\tfrac12(B=0,\beta-\alpha)_{+}\end{smallmatrix}}\nonumber\\
&\quad+\left.\frac{(T-S)(\bar1\,\bar2\,\bar3\,\bar4)^2}{ST\,CD}\right|_{-\tfrac12(C=0,\beta-\alpha)_{+}+\tfrac12(D=0,\beta-\alpha)_{+}}\,.
\label{eq:D134-D124-allplus}
\end{align}
Each of eqs.~\eqref{eq:D123-allplus}--\eqref{eq:D134-D124-allplus} has been
checked numerically against the energy-difference definition. This step is not
a hat transport: at fixed polarizations the difference in $E_2$ involves, by
eq.~\eqref{eq:hat-extra-disc}, the very crossed sector of $\disc_{1,3}$ that
is sought below, so one all-plus triple discontinuity is new data, read off
once; the others are its permutations. Everything that follows is obtained
from it, and from the all-plus quadruple discontinuity
$\disc_{1,2,3,4}=2(\disc_{1,4,s}+\disc_{1,2,t})$, by hats alone.

\paragraph{$\disc_{2,4}$ on the sectors of $\disc_{1,3}$.}
The inversion of section~\ref{sec:SUSY-from-OG} needs $\disc_{2,4}$ in the same
four sectors as $\disc_{1,3}$, which is the crossing-over case above: legs $1$
and $3$ are not discontinued, so no hat at fixed $\disc_{2,4}$ reaches them.
Inverting eq.~\eqref{eq:hat-extra-disc} with the all-plus triple
discontinuities does,
\begin{align}
\disc_{2,4}\la J^+J^+J^-J^+\ra&=\hat h_3\big(\disc_{2,3,4}-\disc_{2,4}\big)\la J^+J^+J^+J^+\ra\,,\nonumber\\
\disc_{2,4}\la J^-J^+J^+J^+\ra&=\hat h_1\big(\disc_{1,2,4}-\disc_{2,4}\big)\la J^+J^+J^+J^+\ra\,,
\label{eq:D24-crossed-chain}
\end{align}
and the MHV line follows from the all-plus quadruple discontinuity, given at
the end of this subsection, through
$\disc_{1,2,3,4}\la J^h\ra=\sum_{Z\subseteq\{1,3\}}\hat h_Z\disc_{2,4}\la J^{h^{(Z)}}\ra$.
Since $\hat h_3$ preserves the $AB$ and $CD$ kernels separately, the $CD$
residues of $\disc_{2,3,4}$ survive in the single-minus lines, and likewise for
$\hat h_1$. The result is
\begin{align}
\disc_{2,4}\la J^+J^+J^+J^+\ra&=\left.\frac{(S+T)(\bar1\,\bar2\,\bar3\,\bar4)^2}{ST\,AB}\right|_{\begin{smallmatrix}\tfrac12(S=0,\beta-\alpha)_{+}+\tfrac12(T=0,\beta-\alpha)_{+}\\+\tfrac12(A=0,\beta-\alpha)_{+}+\tfrac12(B=0,\beta-\alpha)_{+}\end{smallmatrix}},\nonumber\\
\disc_{2,4}\la J^-J^+J^-J^+\ra&=\left.\frac{(S+T)(1\,\bar2\,3\,\bar4)^2}{ST\,AB}\right|_{\begin{smallmatrix}\tfrac12(S=0,\beta-\alpha)_{+}+\tfrac12(T=0,\beta-\alpha)_{+}\\+\tfrac12(A=0,\beta-\alpha)_{+}+\tfrac12(B=0,\beta-\alpha)_{+}\end{smallmatrix}},\nonumber\\
\disc_{2,4}\la J^-J^+J^+J^+\ra&=\left.\frac{(S+T)(1\,\bar2\,\bar3\,\bar4)^2}{ST\,AB}\right|_{\tfrac12(S=0,\beta-\alpha)_{-}+\tfrac12(T=0,\beta-\alpha)_{-}}\nonumber\\
&\quad+\left.\frac{(T-S)(1\,\bar2\,\bar3\,\bar4)^2}{ST\,CD}\right|_{\tfrac12(C=0,\beta-\alpha)_{-}-\tfrac12(D=0,\beta-\alpha)_{-}},\nonumber\\
\disc_{2,4}\la J^+J^+J^-J^+\ra&=\left.\frac{(S+T)(\bar1\,\bar2\,3\,\bar4)^2}{ST\,AB}\right|_{\tfrac12(S=0,\beta-\alpha)_{-}+\tfrac12(T=0,\beta-\alpha)_{-}}\nonumber\\
&\quad+\left.\frac{(T-S)(\bar1\,\bar2\,3\,\bar4)^2}{ST\,CD}\right|_{\tfrac12(C=0,\beta-\alpha)_{-}-\tfrac12(D=0,\beta-\alpha)_{-}}
\label{eq:hat-D24}
\end{align}
The even sectors use the $AB$ kernel alone. The odd sectors additionally
require the $CD$ kernel, as their explicit prescriptions show.

\paragraph{$\disc_{1,3}$ on the sectors of $\disc_{2,4}$.}
The same steps with the other pair of legs give the crossed $\disc_{1,3}$,
\begin{align}
\disc_{1,3}\la J^+J^-J^+J^+\ra&=\hat h_2\big(\disc_{1,2,3}-\disc_{1,3}\big)\la J^+J^+J^+J^+\ra\,,\nonumber\\
\disc_{1,3}\la J^+J^+J^+J^-\ra&=\hat h_4\big(\disc_{1,3,4}-\disc_{1,3}\big)\la J^+J^+J^+J^+\ra\,,
\label{eq:D13-crossed-chain}
\end{align}
with the MHV-type line $\disc_{1,3}\la J^+J^-J^+J^-\ra$ fixed by the
quadruple discontinuity through
$\disc_{1,2,3,4}\la J^h\ra=\sum_{Z\subseteq\{2,4\}}\hat h_Z\,\disc_{1,3}\la J^{h^{(Z)}}\ra$.
Here the hats act differently: $\hat h_2$ and $\hat h_4$ exchange the $AB$ and
$CD$ kernels, carrying the $CD$ residues of the triple discontinuities at $C,D$
to $AB$ residues at $B,A$, and on the poles $S=0$ and $T=0$ the two kernels
coincide up to sign, since $CD=-AB$ there. No second kernel survives:
\begin{align}
\disc_{1,3}\la J^+J^-J^+J^+\ra&=\left.\frac{(S+T)(\bar1\,2\,\bar3\,\bar4)^2}{ST\,AB}\right|_{\begin{smallmatrix}\tfrac12(S=0,\beta-\alpha)_{-}+\tfrac12(T=0,\beta-\alpha)_{-}\\-\tfrac12(A=0,\beta-\alpha)_{-}+\tfrac12(B=0,\beta-\alpha)_{-}\end{smallmatrix}},\nonumber\\
\disc_{1,3}\la J^+J^+J^+J^-\ra&=\left.\frac{(S+T)(\bar1\,\bar2\,\bar3\,4)^2}{ST\,AB}\right|_{\begin{smallmatrix}\tfrac12(S=0,\beta-\alpha)_{-}+\tfrac12(T=0,\beta-\alpha)_{-}\\-\tfrac12(A=0,\beta-\alpha)_{-}+\tfrac12(B=0,\beta-\alpha)_{-}\end{smallmatrix}},\nonumber\\
\disc_{1,3}\la J^+J^-J^+J^-\ra&=\left.\frac{(S+T)(\bar1\,2\,\bar3\,4)^2}{ST\,AB}\right|_{\begin{smallmatrix}\tfrac12(S=0,\beta-\alpha)_{+}+\tfrac12(T=0,\beta-\alpha)_{+}\\+\tfrac12(A=0,\beta-\alpha)_{+}-\tfrac12(B=0,\beta-\alpha)_{+}\end{smallmatrix}}\,.
\label{eq:hat-D13-crossed}
\end{align}
These were checked numerically against the energy-difference definition. The
chain itself has been checked at the level of residues: transporting the
right-hand sides of eqs.~\eqref{eq:D24-crossed-chain}
and~\eqref{eq:D13-crossed-chain}, and of the two quadruple relations, term by
term with eq.~\eqref{eq:hat-residue} reproduces the prescriptions of
eqs.~\eqref{eq:hat-D24} and~\eqref{eq:hat-D13-crossed}, the MHV lines
included. In all eight sectors $\disc_{1,3}$ is therefore the $AB$ kernel on the contour of
its branch, eqs.~\eqref{eq:hat-D13} and~\eqref{eq:hat-D13-crossed}; only
$\disc_{2,4}$ needs the second kernel, and only in its odd crossed sectors.

It is worth pausing on what eq.~\eqref{eq:hat-D24} looks like read down the
page, because nothing we imposed asks for it. The four sectors were obtained one
at a time, and they could in principle
have come out on four unrelated contours. Instead they come out in two groups,
sorted by the parity of $n_-$: the two even lines carry the same integrand on
the same contour and differ only in their numerator, and so do the two odd
lines. The parity of $n_-$ is the branch, so this is the concrete form, at the
level of explicit residues, of the statement made abstractly around
eq.~\eqref{eq:super-wfc-branches}---that the prefactor depends on the helicity
sector only through its branch. There it was read off the assembled super WFC;
here one can see it happen. We take this to be the mathematical root of the
enhanced symmetry we suspect: the sectors within a branch share one integrand
and one contour, nothing relates a sector to its $\xi_{i,-}$ dressing, and yet the
contours organize themselves by branch all the same. 

\paragraph{The remaining channels.}
The $E_s$ and $E_t$ discontinuities keep the printed form in every sector
checked, $\disc_{1,4,s}\la J^{h}\ra
=\tfrac{(S+T)\,n_H}{ST\,AB}\big|_{(S=0,\beta-\alpha)_b}$, and the quadruple
discontinuity is
$\disc_{1,2,3,4}\la J^+J^+J^+J^+\ra
=2\tfrac{(S+T)\,(\bar1\,\bar2\,\bar3\,\bar4)^2}{ST\,AB}
\big|_{(S=0,\beta-\alpha)_++(T=0,\beta-\alpha)_+}$.

\subsection{Other triple-discontinuity sectors from the hat rule}
The all-plus $\disc_{2,3,4}$ is already given in
eq.~\eqref{eq:D234-allplus}. With the crossed $\disc_{2,4}$ of
eq.~\eqref{eq:hat-D24} now known, the hat relation can also be used in the
forward direction for other helicity sectors. Taking $X=\{2,4\}$ and $i=3$ in
eq.~\eqref{eq:hat-extra-disc},
\begin{equation}
\disc_{2,3,4}\la J^{h}\ra=\disc_{2,4}\la J^{h}\ra
+\hat h_3\disc_{2,4}\la J^{h^{(3)}}\ra\,,
\label{eq:D234-from-D24}
\end{equation}
the same relation reproduces the all-plus line and determines the remaining
sectors using eq.~\eqref{eq:hat-residue}. The column map gives
$\hat h_3:(A,B,C,D)\mapsto(B,A,-D,-C)$ while leaving $S,T$ fixed.
Consequently $\hat h_3$ preserves the $AB$ and $CD$ kernels separately,
with the numerator and residue orientations transported as in
eq.~\eqref{eq:hat-residue}. In eq.~\eqref{eq:D234-allplus}, the $AB$ terms come from
$\disc_{2,4}\la J^+J^+J^+J^+\ra$ and the $AB$ part of the hatted
single-minus term; its $CD$ terms come from the $CD$ part of
$\hat h_3\disc_{2,4}\la J^+J^+J^-J^+\ra$. The same transport gives both
kernel types in the other three sectors. Carrying this out,

\begin{align}
\disc_{2,3,4}\la J^-J^+J^-J^+\ra&=\left.\frac{(S+T)(1\,\bar2\,3\,\bar4)^2}{ST\,AB}\right|_{\begin{smallmatrix}(S=0,\beta-\alpha)_{+}+(T=0,\beta-\alpha)_{+}\\+\tfrac12(A=0,\beta-\alpha)_{+}+\tfrac12(B=0,\beta-\alpha)_{+}\end{smallmatrix}}\nonumber\\
&\quad+\left.\frac{(T-S)(1\,\bar2\,3\,\bar4)^2}{ST\,CD}\right|_{\tfrac12(C=0,\beta-\alpha)_{+}-\tfrac12(D=0,\beta-\alpha)_{+}},\nonumber\\
\disc_{2,3,4}\la J^-J^+J^+J^+\ra&=\left.\frac{(S+T)(1\,\bar2\,\bar3\,\bar4)^2}{ST\,AB}\right|_{\begin{smallmatrix}(S=0,\beta-\alpha)_{-}+(T=0,\beta-\alpha)_{-}\\+\tfrac12(A=0,\beta-\alpha)_{-}+\tfrac12(B=0,\beta-\alpha)_{-}\end{smallmatrix}}\nonumber\\
&\quad+\left.\frac{(T-S)(1\,\bar2\,\bar3\,\bar4)^2}{ST\,CD}\right|_{\tfrac12(C=0,\beta-\alpha)_{-}-\tfrac12(D=0,\beta-\alpha)_{-}},\nonumber\\
\disc_{2,3,4}\la J^+J^+J^-J^+\ra&=\left.\frac{(S+T)(\bar1\,\bar2\,3\,\bar4)^2}{ST\,AB}\right|_{\begin{smallmatrix}(S=0,\beta-\alpha)_{-}+(T=0,\beta-\alpha)_{-}\\+\tfrac12(A=0,\beta-\alpha)_{-}+\tfrac12(B=0,\beta-\alpha)_{-}\end{smallmatrix}}\nonumber\\
&\quad+\left.\frac{(T-S)(\bar1\,\bar2\,3\,\bar4)^2}{ST\,CD}\right|_{\tfrac12(C=0,\beta-\alpha)_{-}-\tfrac12(D=0,\beta-\alpha)_{-}}
\label{eq:hat-D234}
\end{align}
Each line is $\disc_{1,4,s}+\disc_{1,2,t}$ plus the explicitly oriented
flat-pole contributions. Together with eq.~\eqref{eq:D234-allplus}, the
four sectors share this contour pattern. Their component values, together
with those of the two double discontinuities,
were checked numerically as described below eq.~\eqref{eq:super-disc}.

\subsection{Two scalars and two currents}
\label{app:ojoj}

Because any numerator regular at the poles may replace $n_H$
(footnote~\ref{fn:numerator-regular}), the same contours also reach correlators
outside the pure-current family, with no new contour to compute. Putting
$X^2$ with $X=(1\,\bar1\,\bar2\,\bar4)$ in its place in the four sectors
$H\subseteq\{1,3\}$, and adding them with the sign $(-1)^{|H|}$, trades the
currents on legs $1$ and $3$ for the dimension-two scalar and returns the
discontinuities of $\la O_2J^+O_2J^+\ra$. From there the appendix runs as it did for
$\la JJJJ\ra$: the double discontinuities $\disc_{1,3}$ and $\disc_{2,4}$
first, then the triple ones, and the remaining helicity sectors by the same
transport---$\hat h_2$ and $\hat h_4$ conjugate the entries $\bar 2$ and
$\bar 4$ of $X$ through eq.~\eqref{eq:hat-columns}, exactly as they did the
entries of $n_H$---which we do not write out.

One feature does set these apart from
eqs.~\eqref{eq:hat-D13}--\eqref{eq:hat-D234}: both branches contribute here,
and with opposite signs, where above the helicity of the sector selected a
single branch. The five explicitly oriented expressions, in the orientation fixed in
appendix~\ref{app:hat}, are
\begingroup\allowdisplaybreaks
\begin{align}
\disc_{1,3}\la O_2J^+O_2J^+\ra(C)&\nonumber\\
&=\left.\frac{2(S+T)(1\bar1\bar2\bar4)^2}{ST\,AB}\right|_{\begin{smallmatrix}\tfrac12(S=0,\beta-\alpha)_{+}+\tfrac12(T=0,\beta-\alpha)_{+}\\+\tfrac12(A=0,\beta-\alpha)_{+}-\tfrac12(B=0,\beta-\alpha)_{+}\end{smallmatrix}}\nonumber\\
&\quad-\left.\frac{2(S+T)(1\bar1\bar2\bar4)^2}{ST\,AB}\right|_{\begin{smallmatrix}\tfrac12(S=0,\beta-\alpha)_{-}+\tfrac12(T=0,\beta-\alpha)_{-}\\-\tfrac12(A=0,\beta-\alpha)_{-}+\tfrac12(B=0,\beta-\alpha)_{-}\end{smallmatrix}},\nonumber\\
\disc_{2,4}\la O_2J^+O_2J^+\ra(C)&\nonumber\\
&=\left.\frac{2(S+T)(1\bar1\bar2\bar4)^2}{ST\,AB}\right|_{\begin{smallmatrix}\tfrac12(S=0,\beta-\alpha)_{+}+\tfrac12(T=0,\beta-\alpha)_{+}\\+\tfrac12(A=0,\beta-\alpha)_{+}+\tfrac12(B=0,\beta-\alpha)_{+}\end{smallmatrix}}\nonumber\\
&\quad-\left.\frac{2(S+T)(1\bar1\bar2\bar4)^2}{ST\,AB}\right|_{\tfrac12(S=0,\beta-\alpha)_{-}+\tfrac12(T=0,\beta-\alpha)_{-}}\nonumber\\
&\quad-\left.\frac{2(T-S)(1\bar1\bar2\bar4)^2}{ST\,CD}\right|_{\tfrac12(C=0,\beta-\alpha)_{-}-\tfrac12(D=0,\beta-\alpha)_{-}},\nonumber\\
\disc_{2,3,4}\la O_2J^+O_2J^+\ra(C)&\nonumber\\
&=\left.\frac{2(S+T)(1\bar1\bar2\bar4)^2}{ST\,AB}\right|_{\begin{smallmatrix}(S=0,\beta-\alpha)_{+}+(T=0,\beta-\alpha)_{+}\\+\tfrac12(A=0,\beta-\alpha)_{+}+\tfrac12(B=0,\beta-\alpha)_{+}\end{smallmatrix}}\nonumber\\
&\quad+\left.\frac{2(T-S)(1\bar1\bar2\bar4)^2}{ST\,CD}\right|_{\tfrac12(C=0,\beta-\alpha)_{+}-\tfrac12(D=0,\beta-\alpha)_{+}}\nonumber\\
&\quad-\left.\frac{2(S+T)(1\bar1\bar2\bar4)^2}{ST\,AB}\right|_{\begin{smallmatrix}(S=0,\beta-\alpha)_{-}+(T=0,\beta-\alpha)_{-}\\+\tfrac12(A=0,\beta-\alpha)_{-}+\tfrac12(B=0,\beta-\alpha)_{-}\end{smallmatrix}}\nonumber\\
&\quad-\left.\frac{2(T-S)(1\bar1\bar2\bar4)^2}{ST\,CD}\right|_{\tfrac12(C=0,\beta-\alpha)_{-}-\tfrac12(D=0,\beta-\alpha)_{-}},\nonumber\\
\disc_{1,4,s}\la O_2J^+O_2J^+\ra(C)&\nonumber\\
&=\left.\frac{2(S+T)(1\bar1\bar2\bar4)^2}{ST\,AB}\right|_{(S=0,\beta-\alpha)_{+}}\nonumber\\
&\quad-\left.\frac{2(S+T)(1\bar1\bar2\bar4)^2}{ST\,AB}\right|_{(S=0,\beta-\alpha)_{-}},\nonumber\\
\disc_{1,2,t}\la O_2J^+O_2J^+\ra(C)&\nonumber\\
&=\left.\frac{2(S+T)(1\bar1\bar2\bar4)^2}{ST\,AB}\right|_{(T=0,\beta-\alpha)_{+}}\nonumber\\
&\quad-\left.\frac{2(S+T)(1\bar1\bar2\bar4)^2}{ST\,AB}\right|_{(T=0,\beta-\alpha)_{-}}\,.
\label{eq:ojoj-discs}
\end{align}
\endgroup
They coincide with the projections $F_i(C)$ of the super WFC in
section~\ref{sec:4pt-super-wfc}.
Supersymmetry reproduces all of this by a longer route: in
section~\ref{sec:4pt-super-wfc} the component $\la O_2J^+O_2J^+\ra$ is a sum
over the four sectors that differ in legs $1$ and $3$---the same four as in
eqs.~\eqref{eq:hat-D13}--\eqref{eq:hat-D24}---and the conjugate minors
$(1\,\bar1\,\bar2\,\bar4)$ and $(3\,\bar3\,\bar2\,\bar4)$ appearing there
cancel in pairs, so that the sum collapses to the single replacement made
above.

\section{Full three-point super WFCs}
\label{app:full-three-point-super-wfcs}

Section~\ref{sec:3pt-super} builds the transverse three-point super WFC, which
is all the four-point construction needs. The full super WFC also has a
longitudinal part. This appendix constructs it at $\mathcal N=1$ and at
$\mathcal N=2$.

The organizing split is the one of eq.~\eqref{eq:n1-TL-split}: a transverse
part, built only from the superfields' $T$ pieces, and a longitudinal part
carrying at least one $L$ piece. The transverse part carries no independent
contact datum, the contact fields entering it only through
$\chi+\frac12\hat B_c$ and $O_1+16\hat\phi_{3,c}$, and is fixed directly by
the pure-current WFC, so the work is all in the longitudinal part. Appendix~\ref{app:N1} does it for $\mathcal N=1$. The lever
is that the contact field $B_c$ enters the superfield expansion twice, once
alone and once paired with $\chi$: the first occurrence ties
$\la J^-B_c^-B_c^+\ra$ to the longitudinal $\la J^-J_LJ_L\ra$, the second ties
$\la J^-\chi^-B_c^+\ra$ to $\la J^-J^-J_L\ra$, and the Ward--Takahashi identity
fixes both from lower-point data. The one coefficient this does not reach sits
at the total-energy pole, in the transverse part, where $\la JJJ\ra$ has
already fixed it. The outcome is
the complete $\vev{\bm J^-_{1/2}\bm J^-_{1/2}\bm J^+_{1/2}}$ and
$\vev{\bm J^-_{1/2}\bm J^-_{1/2}\bm J^-_{1/2}}$, with the remaining helicities
following by parity; the first reduces at $E_T\to0$ to the $\mathcal N=1$
flat-space superamplitude, and the second carries no total-energy pole, as it
should, since there is no flat amplitude in that helicity configuration.

Appendix~\ref{N2Longs} then turns to $\mathcal N=2$. The same split
applies; the decomposition into $\mathcal N=1$ multiplets places the contact
terms of the $\mathcal N=2$ superfield, and the Ward--Takahashi
identities of the current, scalar and fermion components, with the absence of
an $AA\phi$ vertex, fix the longitudinal sector completely,
eq.~\eqref{eq:n2-L-fixed}.

The discontinuities of section~\ref{sec: discontinuity JJJJ} are designed to
remove exactly the longitudinal content assembled here, which is why
section~\ref{sec:3pt-super} and the four-point construction can work with the
transverse part alone.

\subsection{\texorpdfstring{$\mathcal{N}=1$}{N=1} contact terms and super WFCs}
\label{app:N1}
The full two- and three-point super WFCs can be expressed as linear
combinations of the invariants $\Gamma_2$ and $\Gamma_3^\pm$ of
eqs.~\eqref{eq: N=1Gamma2} and~\eqref{eq:n1-Gamma3}. At two points the longitudinal part is a
one-point function and vanishes, so all contact terms vanish with it and the
Lorentzian-AdS two-point WFCs can be written down directly, one for each
multiplet:

\begin{equation}
\langle \bm J^-_{1/2} \bm J^-_{1/2} \rangle= \frac{\ab{1}{2}}{16E_1}\Gamma_2, \quad
\langle \bm J^+_{1/2} \bm J^+_{1/2} \rangle= \frac{\abBB{1}{2}^2}{256 E_1^2}\xi_{1,-}\xi_{2,-}\Gamma_2, \,
\langle \bm A_0 \bm A_0 \rangle= \left( \frac{1}{16 E_1} -  \frac{\xi_{1,-}\xi_{2,-}\abBB{1}{2}}{256E_1^2} \right) \Gamma_2
\end{equation}

At three points this is no longer automatic, and it pays to split the problem
along eq.~\eqref{eq:n1-TL-split}. Write
\begin{equation}
\vev{\bm J^-_{1/2}\bm J^-_{1/2}\bm J^+_{1/2}}
=\vev{\bm J^-_{1/2}\bm J^-_{1/2}\bm J^+_{1/2}}_{\rm T}
+\vev{\bm J^-_{1/2}\bm J^-_{1/2}\bm J^+_{1/2}}_{\rm L}\,,
\label{eq:n1-3pt-TL}
\end{equation}
the first term built only from the $\bm J^{\pm,T}_{1/2}$ of
eq.~\eqref{eq:n1-TL-split} and the second carrying at least one
$\bm J^{\pm,L}_{1/2}$.

Each $\Gamma$ dressed with a different number of $\xi_-$ factors is a separate
$\xi_-$ sector, and eq.~\eqref{eq:n1-3pt-TL} sorts the sectors for us:
eq.~\eqref{eq:n1-TL-split} puts the transverse piece of a $\bm J^-_{1/2}$ leg at
$\xi_-^0$ and that of a $\bm J^+_{1/2}$ leg at $\xi_-^1$, so with legs $1,2$ of
type $\bm J^-_{1/2}$ and leg $3$ of type $\bm J^+_{1/2}$ the all-transverse term
is the one carrying $\xi_{3,-}$ and nothing else, while every other sector has
at least one longitudinal insertion. We can therefore write down an independent
ansatz for each part. For the transverse part,
\begin{equation}
\vev{\bm J^-_{1/2}\,\bm J^-_{1/2}\,\bm J^+_{1/2}}_{\rm T}=c_3\,\xi_{3-}\Gamma^+_3\,,
\label{eq:n1-ansatz-T}
\end{equation}
and for the longitudinal part,
\begin{align}
\vev{\bm J^-_{1/2}\,\bm J^-_{1/2}\,\bm J^+_{1/2}}_{\rm L}&=
\big(c_1\xi_{1-}+c_2\xi_{2-}\big)\Gamma^+_3
+b\,\xi_{1-}\xi_{2-}\xi_{3-}\Gamma^+_3\nonumber\\
&\quad+d\,\Gamma^-_3
+\big(e_{12}\xi_{1-}\xi_{2-}+e_{23}\xi_{2-}\xi_{3-}+e_{31}\xi_{3-}\xi_{1-}\big)\Gamma^-_3\,.
\label{eq:n1-ansatz-L}
\end{align}
We fix the two in turn.

\paragraph{The transverse part.}
Equation~\eqref{eq:n1-ansatz-T} holds a single coefficient, and it needs no
contact-term analysis at all. The transverse part carries no independent contact datum, $B_c$ entering it
only through $\chi+\frac12\hat B_c$ and $J_L$ not at all, so $c_3$ is read off component by component against the pure-current
Yang--Mills WFC $\la J^-J^-J^+\ra$, which section~\ref{sec: 3ptinversion} has
already reconstructed from its discontinuity. That fixes it outright,
\begin{equation}
\vev{\bm J^-_{1/2}\,\bm J^-_{1/2}\,\bm J^+_{1/2}}_{\rm T}
= \xi_{3-}
\cdot\frac{1}{(\prodE)^{3/2}}
\cdot \prod_{i=1}^3(\ET - 2E_i)
\cdot \frac{\ab{1}{2}^2}{ \ab{1}{3}\abn{23}}\cdot \Gamma^+_3\,,
\label{eq:n1-3pt-T-result}
\end{equation}
and this is the only term of the whole WFC carrying the total-energy pole, as
the flat-space limit below confirms.

\paragraph{The longitudinal part.}
Everything left to determine sits in eq.~\eqref{eq:n1-ansatz-L}, and the
input is the Ward--Takahashi identity, eq.~\eqref{eq:WT-JJJ}: contracting a
leg of the three-current WFC with its momentum gives two-point functions at
shifted momenta, with the polarizations of the other legs retained. Since
$p_{\alpha\beta}J^{\alpha\beta}=E\,J_L$, this fixes every pure-current
component with a longitudinal insertion. With eq.~\eqref{eq:WT-JJJ} taken with
unit coefficient and $J^{\alpha\beta}=J^i\sigma_i^{\alpha\beta}$, the two
needed here are
\ie
\label{eq:WT-inputs}
\la J^-J_LJ_L\ra=-\frac{2\ab{1}{3}\abB{1}{3}}{E_2E_3}\,,\qquad
\la J^-J^-J_L\ra=\frac{2\ab{1}{2}^2(E_1-E_2)}{E_1E_2E_3}\,,
\fe
together with the other components that the same identity gives, namely
$\la J_LJ^-J_L\ra$, $\la J_LJ_LJ^+\ra$, $\la J^-J_LJ^+\ra$ and
$\la J_LJ^-J^+\ra$, and $\la J_LJ_LJ_L\ra=0$, which follows from
transversality of the two-point function. The lever that turns these into
contact terms is the superfield expansion itself: from
eq.~\eqref{superfieldn=1}, $B_c$ appears in both the $\xi_-$ and the $\xi_+$
component of $\bm J^\pm_{1/2}$, once alone and once with $\chi$, so each
structure of eq.~\eqref{eq:n1-ansatz-L} contains one pure-current component
next to its contact-term components, and that one component fixes the
coefficient. Explicitly, $c_1$ and $c_2$ are fixed by $\la J_LJ^-J_L\ra$ and
$\la J^-J_LJ_L\ra$ ($\xi_{j-}\Gamma^+_3$ holds $\la J^-B^-_cB^+_c\ra$ and
$\la J^-J_LJ_L\ra$ side by side), $d$ by $\la J^-J^-J_L\ra$, $b$ by
$\la J_LJ_LJ^+\ra$, $e_{23}$ and $e_{31}$ by $\la J^-J_LJ^+\ra$ and
$\la J_LJ^-J^+\ra$, and $e_{12}$ vanishes with $\la J_LJ_LJ_L\ra$. Every
contact-term component then follows. This does not reach $c_3$, and it should
not: that coefficient belongs to the transverse part, which carries no independent
contact datum and has already been fixed above.

\paragraph{The transverse fermionic component.}
The transverse coefficient can also be checked by adding the contact
contributions to the fermionic WFC $\la J^-\chi^-\chi^+ \ra$ \cite{Chen:2025foq},
\ie
\la J^- (\chi^- + \tfrac{B_c^-}{2E}) (\chi^++\tfrac{B_c^+}{2E}) \ra 
&=
\la J^- \chi^- \chi^+ \ra + \frac{1}{2E_3} \la J^- \chi^- B_c^+ \ra\\
&\quad + \frac{1}{2E_2} \la J^- B_c^- \chi^+ \ra + \frac{1}{4E_2 E_3} \la J^- B_c^- B_c^+ \ra\,,
\fe
This is the component combination multiplying $\xi_{2,+}\xi_{3,-}$.
The displayed expansion fixes the relative weights of its contact terms;
checking $c_3$ through this route additionally requires the independently
normalized fermionic component and the individual contact correlators.

\paragraph{The result.}
With the coefficients of eq.~\eqref{eq:n1-ansatz-L} fixed by the lower-point
data, the longitudinal part is
\begin{align}
    \label{super n1: JJJ--+}
\vev{\bm J^-_{1/2}\,\bm J^-_{1/2}\,\bm J^+_{1/2}}_{\rm L}
&= \frac{E_T}{256\sqrt{\prodE}}\left(\frac{\xi_{2-} (E_T-2E_1)\abn{21}}{E_2 E_3\ab{3}{2}}+\frac{\xi_{1-} (E_T-2E_2)\abn{12}}{E_1 E_3\ab{3}{1}}
 \right)\Gamma^+_3\notag\\[8pt]
&\quad-\frac{\abn{1  2}^2\,\Gamma^-_3}{256\sqrt{\prodE}}\bigg[ \xi_{2-}\xi_{3-}\frac{(E_T-2E_1) }{E_2 \abn{ 2 3} }\Big(\frac{1}{E_3}-\frac{1}{E_1}\Big)\notag\\
&\qquad\qquad\qquad\quad + \xi_{3-}\xi_{1-}\frac{(E_T-2E_2)}{E_1 \abn{ 1 3} }\Big(\frac{1}{E_3}-\frac{1}{E_2}\Big)\bigg]
\notag\\[8pt]
&\quad-\frac{\ET (E_T-2E_1)(E_T-2E_2)}{2048\sqrt{\prodE}}\cdot\frac{\xi_{1-}\xi_{2-} \xi_{3-}\abn{2 1}}{E_1 E_2 \abn{{3}{2}}\abn{1 {3}}} \Gamma^+_3
+\frac{(E_2{-}E_1)\abn{ 1 2}^2 \Gamma^-_3}{32(\prodE)^{3/2}}\,.
\end{align}
Equations~\eqref{eq:n1-3pt-T-result} and~\eqref{super n1: JJJ--+} together are
the full $\vev{\bm J^-_{1/2}\bm J^-_{1/2}\bm J^+_{1/2}}$. None of the
longitudinal terms carries a total-energy pole, as anticipated. The two
$\Gamma^-_3$ terms of eq.~\eqref{super n1: JJJ--+} appear to carry one, through
the $1/E_T$ part of $\Gamma^-_3$; but $\ab{i}{j}\abBB{i}{j}=E_T(E_T-2E_k)$
turns each $\abBB{i}{j}/E_T$ into $(E_T-2E_k)/\ab{i}{j}$, and with the overall
$\ab{1}{2}^2$ these terms stay finite at $E_T\to0$. The absence of
the $\xi_{1,-}\xi_{2,-}$ term reflects $\la J_L J_L J_L \ra=0$. One can
verify that the pure transverse vector component $\la J^-J^-J^+\ra$ is correctly
reproduced within the expansion of the first sector, set by
$\la J (\chi+\frac{B_c}{2E}) (\chi+\frac{B_c}{2E}) \ra$. Since
eq.~\eqref{eq:n1-3pt-T-result} is the only part carrying the total-energy pole,
the flat-space limit sees the transverse part alone, and with
eq.~(\ref{eq: SUSYFlat}),
\begin{align}
\vev{\bm J^-_{1/2}\,\bm J^-_{1/2}\,\bm J^+_{1/2}}\big|_{E_T\rightarrow 0}
&=\vev{\bm J^-_{1/2}\,\bm J^-_{1/2}\,\bm J^+_{1/2}}_{\rm T}\big|_{E_T\rightarrow 0}\nonumber\\
&=64\xi_{3-}\frac{1}{E_T\prod_{i=1}^3 E_i^{\frac{1}{2}}}\frac{\langle12\rangle^2}{\langle 13\rangle \langle23\rangle}(\xi_1\langle23\rangle{+}{\rm cyclic})\,,
\end{align}
where the flat-space superamplitude of $\mathcal{N}=1$ SYM emerges. We perform the same analysis in the all-minus helicity sector
$\la \bm J^-_{1/2} \bm J^-_{1/2} \bm J^-_{1/2} \ra$, where all three legs are of
type $\bm J^-_{1/2}$ and the all-transverse term is therefore the one with no
$\xi_-$ at all. It is again fixed by the pure-current WFC,
\begin{equation}
\vev{\bm J^-_{1/2} \,\bm J^-_{1/2}\,\bm J^-_{1/2}}_{\rm T}
= \frac{\ab{2}{3}\ab{2}{1}\ab{1}{3}}{\prodE^{\frac{3}{2}}}\,\Gamma^-_3\,,
\label{eq:n1-3pt-mmm-T}
\end{equation}
and the contact terms of the longitudinal part, found to be consistent sector by
sector, give
\begin{align}
\vev{\bm J^-_{1/2} \,\bm J^-_{1/2}\,\bm J^-_{1/2}}_{\rm L}
&= \frac{E_T}{256\,\prodE^{\frac{3}{2}}}\Big( \xi_{1-}
\left(E_3-E_2\right)\ab{2}{3}
 + \xi_{2-}
\left(E_3-E_1\right)\ab{1}{3}\notag\\
&\qquad\qquad\qquad\quad + \xi_{3-}\left(E_2-E_1\right)\ab{1}{2}\Big)\Gamma^+_3 \notag\\[6pt]
&\quad+\frac{\Gamma^-_3}{256\,\prodE^{\frac{3}{2}}} \Big(\xi_{1-}\xi_{2-}E_3\ab{2}{3}\ab{3}{1}
 + \xi_{2-}\xi_{3-}E_1\ab{2}{1}\ab{1}{3}\notag\\
&\qquad\qquad\qquad\quad + \xi_{3-}\xi_{1-}E_2\ab{1}{2}\ab{2}{3}\Big)
\label{eq:n1-3pt-mmm-L}
\end{align}
Once again, the absence of $\xi_{1,-}\xi_{2,-}\xi_{3,-}$---the all-longitudinal
sector---reflects $\la J_L J_L J_L \ra=0$. Correspondingly, there are no total energy poles, as there is no flat-space amplitude for this helicity configuration. Other helicity configurations are obtained via parity reflection: 
\ie
\label{eq:n1-parity}
 \bm J^-_{1/2} \leftrightarrow \bm J^+_{1/2}, \quad 8\xi_{1,\pm} \leftrightarrow \xi_{2,\pm} \xi_{3,\pm}, \text{cyclic},\quad  8\xi_\pm^0 \leftrightarrow \xi_{1,\pm} \xi_{2,\pm} \xi_{3,\pm},\quad  \lambda \leftrightarrow \bar\lambda
\fe
The map is exact up to one constant for each number $n_L$ of longitudinal
legs: applied to eq.~\eqref{super n1: JJJ--+} it gives the $n_L=1$ and
$n_L=2$ parts of $\vev{\bm J^+_{1/2}\bm J^+_{1/2}\bm J^-_{1/2}}_{\rm L}$
times $-8$ and $\tfrac18$, and applied to eq.~\eqref{eq:n1-3pt-mmm-L} those of
$\vev{\bm J^+_{1/2}\bm J^+_{1/2}\bm J^+_{1/2}}_{\rm L}$ times $-\tfrac18$ and
$8$. The signs come from the opposite signs of $J_L$ in $\bm J^\pm_{1/2}$,
eq.~\eqref{superfieldn=1}.

In summary: the transverse part of eq.~\eqref{eq:n1-3pt-TL} is the pure-current
WFC and needs no contact-term analysis, while the longitudinal part is
determined by the longitudinal $J_L$ WFCs, which are lower-point data. The only
coefficient the longitudinal data does not reach is the one at the
total-energy pole, and it is supplied by the transverse part. No further input
is required. At $\mathcal N=2$ the scalar multiplet $\bm A_0$ sits in the same
superfield as the currents, and its longitudinal components are fixed together
with theirs, to which we now turn.

\subsection{$\mathcal N=2$ longitudinal WFCs}
\label{N2Longs}

At three points the split of section~\ref{sec:N2superfield} reads
\begin{equation}
\langle \bm{J}_0 \bm{J}_0 \bm{J}_0 \rangle
=\langle \bm{J}_0 \bm{J}_0 \bm{J}_0 \rangle_{\text{T}}
+\langle \bm{J}_0 \bm{J}_0 \bm{J}_0 \rangle_{\text{L}}\,,
\label{eq:n2-3pt-TL}
\end{equation}
the transverse part built from $\bm J_0^{T}=(\bm J_0^{-},\bm J_0^{+})$ alone and
the longitudinal part carrying at least one $\bm J_0^{I,L}$. The transverse
part is eq.~\eqref{eq: N=23pt}. This appendix is about the longitudinal part.
The reduction to $\mathcal N=1$ below places the contact terms of the
$\mathcal N=2$ superfield inside $\mathcal N=1$ multiplets; the longitudinal
part itself is then fixed directly at $\mathcal N=2$ by the Ward--Takahashi
identities. (Two points need nothing: the
$\mathcal N=2$ two-point super WFC is eq.~\eqref{eq:2pt WFC}, and its
longitudinal part is a one-point function and vanishes.)

\paragraph{Reduction to $\mathcal N=1$.}
In the second set of Grassmann variables, $\xi^{(2)}_\pm$, the component
identity is\footnote{The superscript $(2)$ labels the second set of Grassmann
variables and the corresponding fermionic components of
$\bm J^{\pm,(2)}_{1/2}$.}
\begingroup\allowdisplaybreaks
\begin{align}
\bm J_0
&=\frac{1}{2\sqrt E}\bm J_{1/2}^{-,(2)}
+\frac{\xi^{(1)}_++\xi^{(1)}_-}{8}\bm A_0^{(2)}
+\frac{\xi^{(1)}_+-\xi^{(1)}_-}{2E}\bar{\bm A}_0^{(2)}\notag\\
&\quad+\frac{\xi^{(1)}_+\xi^{(1)}_-}{16\sqrt E}\bm J_{1/2}^{+,(2)},
\label{eq:n2-to-n1}\\[4pt]
\bm A_0^{(2)}
&=\frac{\chi^{(1),-}}{2\sqrt E}
-\frac{\xi_-^{(2)}\xi_+^{(2)}}{16\sqrt E}\big(\chi^{(1),+}+\hat B_c^{(1),+}\big)
+\frac{\xi_+^{(2)}-\xi_-^{(2)}}8 \big(O_1+16\hat\phi_{3,c}+\hat C^{12}_c/E\big)\notag\\
&\quad+\frac{\xi_+^{(2)}+\xi_-^{(2)}}{2E}\big(O_2-4E\hat\phi_{3,c}\big),\notag\\
\bar{\bm A}_0^{(2)}
&=\frac{\sqrt E}{8}\big(\chi^{(1),-}+\hat B_c^{(1),-}\big)
+\frac{\sqrt E\,\xi_-^{(2)}\xi_+^{(2)}}{64}\chi^{(1),+}\notag\\
&\quad+\frac{E\big(\xi_+^{(2)}+\xi_-^{(2)}\big)}{32} \big(O_1+16\hat\phi_{3,c}-\hat C^{12}_c/E\big)
+\frac{\xi_+^{(2)}-\xi_-^{(2)}}{8}\big(O_2+4E\hat\phi_{3,c}\big).
\label{eq:n2-to-n1-components}
\end{align}
\endgroup
Every contact term of \eqref{N=2-Superfield} now sits inside one of the four
$\mathcal N=1$ multiplets: the two current multiplets are those of
eq.~\eqref{superfieldn=1} in the second set of variables, with $J_L$ replaced by
$J_L-\hat C^{11}_c$ in $\bm J^{-,(2)}_{1/2}$ and by $J_L+\hat C^{11}_c$ in
$\bm J^{+,(2)}_{1/2}$, and the scalar contacts $\hat\phi_{3,c}$ and $\hat C^{12}_c$
enter only through the scalar slots of $\bm A_0^{(2)}$ and $\bar{\bm A}_0^{(2)}$. The scalar multiplets carry the
same $O_1$ and $O_2$ as \eqref{N=2-Superfield}, with no further operator:
the descendant relation holds up to an explicit contact remainder,
\begin{equation}
\bar{\bm A}_0^{(2)}-\tfrac14D^{\alpha,(2)}D_\alpha^{(2)}\bm A_0^{(2)}
=\frac{\sqrt E}{8}\hat B_c^{(1),-}
+E\big(\xi_+^{(2)}-\xi_-^{(2)}\big)\hat\phi_{3,c}
-\frac{\xi_+^{(2)}+\xi_-^{(2)}}{16}\hat C^{12}_c
-\frac{\sqrt E\,\xi_-^{(2)}\xi_+^{(2)}}{64}\hat B_c^{(1),+}\,.
\label{eq:descendant-remainder}
\end{equation}
Eq.~\eqref{eq:n2-to-n1} agrees component by component
with \eqref{N=2-Superfield}; collecting its $\xi_-$ powers also reproduces
\eqref{eq: N=2SuperField}.

Appendix~\ref{app:N1} fixes correlators of the components shared by the two
current multiplets: $J^\pm$, $J_L\mp\hat C^{11}_c$,
$\chi^{(2)\pm}+\tfrac12\hat B_c^{(2)\pm}$ and $\hat B_c^{(2)\pm}$.
Both blocks enter with a common factor, $1/(2\sqrt E)$ for
$\bm J^{-,(2)}_{1/2}$ and $\xi^{(1)}_+\xi^{(1)}_-/(16\sqrt E)$ for
$\bm J^{+,(2)}_{1/2}$, so their components are read off directly.
The longitudinal solution below is obtained directly at $\mathcal N=2$, from
the component coefficients of \eqref{eq: N=2SuperField} and the
Ward--Takahashi identities; it does not fix the normalization of the
transverse Grassmannian integral.

\paragraph{The invariant basis.}
Since the two supercharges act on separate sets of $\xi_+$ variables, the
three-point basis is
\begin{equation}
p(\xi_-)\,\Gamma^a_3(\xi^{(1)}_+)\Gamma^b_3(\xi^{(2)}_+),
\qquad a,b\in\{+,-\}.
\end{equation}
Here $p$ is a polynomial in the six $\xi^I_{i,-}$, odd for the mixed
products and even otherwise. Longitudinal structures have degree one in
$\xi_{i,-}$ on at least one leg and are restricted to SO$(2)$ singlets.
Imposing the Ward--Takahashi identity on every component with a longitudinal
insertion determines this part completely.

\paragraph{The result.}
The mixed products form the doublet
\begin{equation}
V^I=\big(\Gamma^+_3(\xi^{(1)}_+)\Gamma^-_3(\xi^{(2)}_+),\;
\Gamma^-_3(\xi^{(1)}_+)\Gamma^+_3(\xi^{(2)}_+)\big).
\end{equation}
The SO$(2)$-invariant longitudinal basis has sixty structures and
fifty-four independent directions. All of them are fixed, and the result is
\begingroup\allowdisplaybreaks
\begin{align}
\label{eq:n2-L-fixed}
\langle \bm{J}_0 \bm{J}_0 \bm{J}_0 \rangle_{\text{L}}
&=\frac{1}{2^{14}\prodE^{2}}\sum_{\rm cyc}\Big[E_T(E_k-E_j)\ab{j}{k}\,\xi^I_{i,-}V^I\notag\\
&\qquad+E_k\ab{j}{k}\ab{k}{i}\,\Gamma^{--}_3\,(\xi^I_{i,-}\xi^I_{j,-})\Big]\notag\\
&\quad-\frac{1}{2^{20}\prodE^{2}}\sum_{\rm cyc}\Big[E_T(E_k-E_j)\abBB{j}{k}\,\xi^2_{j,-}\xi^2_{k,-}\xi^I_{i,-}V^I\notag\\
&\qquad+E_k\abBB{j}{k}\abBB{k}{i}\,\Gamma^{++}_3\,\xi^2_{k,-}(\xi^I_{i,-}\xi^I_{j,-})\Big]\notag\\
&\quad+\frac{E_T}{2^{17}\prodE}\sum_{\rm cyc}\xi^2_{k,-}\ab{i}{j}\Big[\frac{E_T-2E_j}{E_i\ab{i}{k}}\Big(\frac1{E_k}-\frac1{E_j}\Big)\xi^I_{i,-}\epsilon_{IJ}V^J\notag\\
&\qquad-\frac{E_T-2E_i}{E_j\ab{j}{k}}\Big(\frac1{E_k}-\frac1{E_i}\Big)\xi^I_{j,-}\epsilon_{IJ}V^J\Big]\notag\\
&\quad+\frac{E_T^2}{2^{17}\prodE}\sum_{\rm cyc}\frac{E_T-2E_i}{E_jE_k}\Big[\frac{\Gamma^{++}_3}{\ab{k}{j}}-\frac{\Gamma^{--}_3\,\xi^2_{i,-}}{\abBB{k}{j}}\Big](\xi^I_{j,-}\epsilon_{IJ}\xi^J_{k,-})\,,
\end{align}
\endgroup
where the sums run over the cyclic permutations $(i,j,k)$ of $(1,2,3)$. The
first cyclic sum is the lift of eq.~\eqref{eq:n1-3pt-mmm-L}; the second is its
parity image; the last two sums lift eq.~\eqref{super n1: JJJ--+} and its
relabelings, with the extra $E_T/\ab{\cdot}{\cdot}$ coming from the $\xi_+$ that
the pure-current component takes from the linear term of $\Gamma^+_3$.

The conditions that fix it are all physical. Every component with a
longitudinal insertion obeys the Ward--Takahashi identity, which needs only the
two-point functions of eq.~\eqref{eq:2pt WFC}: the pure-current
$\la J_L JJ\ra$, with one, two or three longitudinal legs, $\la J_L O_aO_a\ra$ with $\la O_1O_1\ra=2/E$ and
$\la O_2O_2\ra=E/8$, and $\la J_L\chi^I\chi^J\ra$, diagonal in $I,J$, with the
fermion two-point function $q^{\alpha\beta}/E$; and the tree-level action has no
$AA\phi$ vertex, so $\la J\,J\,O_a\ra=0$ with $J^\pm$ or $J_L$ on the other two
legs. Pure-current components alone fix nineteen of the fifty-four directions;
the scalar and fermion identities and $\la JJO_a\ra=0$ fix the rest. The $133$
conditions are consistent and leave no freedom, and their unique solution is
eq.~\eqref{eq:n2-L-fixed}: the complementary contractions, namely the
$\epsilon$ partners of the first two sums, the $\delta$ partners of the last two
and all twelve triple insertions, have vanishing coefficients. (The twelve
triple insertions obey six linear SO$(2)$ identities, which is how sixty
structures give fifty-four directions.) The Yukawa-type components
$\la O_a\chi^I\chi^J\ra$ and every contact component are then predictions. The
consistency of the conditions requires the full contact content of
eq.~\eqref{N=2-Superfield}: with $\hat C^{IJ}_c$ set to zero, or with other
relative weights of $\hat\phi_{3,c}$ and $\hat B^{I+}_c$, the scalar and fermion
identities cannot be met together. Eq.~\eqref{eq:n2-L-fixed} is annihilated by
$Q^{\alpha,I}$ and invariant under SO$(2)$, and its all-longitudinal sector, one
$\xi^I_{i,-}$ on each leg, is absent. What the discontinuities of
section~\ref{sec: discontinuity JJJJ} use is the transverse part, which is
complete.

\section{Explicit super-Grassmannian invariants}
\label{app:invariants}

This appendix collects the explicit superinvariants produced by the fermionic
delta function of eq.~\eqref{eq: SUSYDelta} acting on the test functions of
section~\ref{sec:gen-superinv}.

\subsection{Two and three points}
\label{app:inv-23}
Let us illustrate the construction in the simplest non-trivial case, namely $n=2$. For test functions containing an even number of $\xi_{i,+}^2$ pairs, we obtain
\begin{align}
    &\phantom{cc}\hat \delta \left(C_{2} \cdot  \Omega\cdot  \Xi_2^I \right)  \cdot 1
    =\Biggl[ \frac{1}{4} (1 \bar 1)+ \frac{1}{4} (2 \bar 2)+\frac{1}{16}( \bar 1 \bar 2)  \xi_{1,+}^I \xi_{2,+}^I \Biggr]^2\,,\label{eq:2pt-SUSY-invariant-1}\\
    &   \hat \delta \left(C_{2} \cdot  \Omega\cdot  \Xi_2^I \right)  \cdot  \xi_{1,+}^2  \xi_{2,+}^2
    =-\Biggl[ 4\,  (1 2 )+ \frac{1}{4} \big(  (1 \bar 1)+  (2 \bar 2)\big)  \xi_{1,+}^I \xi_{2,+}^I \Biggr]^2\,.\label{eq:2pt-SUSY-invariant-2}
\end{align}
%Although these two expressions appear rather different, they in fact represent the same supersymmetric invariant up to a helicity-dependent kinematic prefactor in momentum sapce. More precisely, both are proportional to the two-point invariant \eqref{} introduced earlier,
%\begin{align}
%    \hat \delta \left(C_{2} \cdot  \Omega\cdot  \Xi_2^I \right)  \cdot 1 \sim   \hat \delta \left(C_{2} \cdot  \Omega\cdot  \Xi_2^I \right)  \cdot  \xi_{1,+}^2  \xi_{2,+}^2 \sim \Gamma_{2}^{++}
%\end{align}
For a test function containing an odd number of $\xi_{i,+}^2$ pairs, we instead find
\begin{align}
    \hat \delta \left(C_{2} \cdot  \Omega\cdot  \Xi_2^I \right)  \cdot \xi_{1,+}^2 &=\Biggl[-\frac{1}{4}\big( (1 \bar 1)- (2 \bar 2) \big) \xi_{1,+}^I  -\frac{1}{2}  (1 \bar 2)  \xi_{2,+}^I  \Biggr]^2\,.\label{eq:2pt-SUSY-invariant-3}
\end{align}

At three points, test functions containing an even number of $\xi_{i,+}^2$ pairs give
\begin{align}\label{eq: OG3SUSY}
     \hat \delta \left(C_{3} \cdot  \Omega\cdot  \Xi_3^I \right)  \cdot 1
    &=\Biggl[\frac{1}{16} \Big((2 \bar 2\bar  1)+ (3 \bar 3 \bar 1) \Big) \xi_{1,+}^I+\frac{1}{16}\Big((1 \bar 1 \bar 2)+ (3 \bar 3 \bar 2)\Big) \xi_{2,+}^I \nonumber  \\
    &\phantom{ccc}+\frac{1}{16}\Big((1 \bar 1 \bar 3)+ (2 \bar 2 \bar 3)\Big) \xi_{3,+}^I+ \frac{1}{64}(\bar 1\bar 2\bar 3) \xi_{1,+}^I\xi_{2,+}^I\xi_{3,+}^I\Biggr]^2\,,\nonumber\\
    \hat \delta \left(C_{3} \cdot  \Omega\cdot  \Xi_3^I \right)  \cdot \xi_{1,+}^2 \xi_{2,+}^2
    &=-\Biggl[\frac{1}{2} \Big((2  1\bar  1)- (2 3 \bar 3 ) \Big) \xi_{1,+}^I-\frac{1}{2}\Big((1 2 \bar 2)- (1 3 \bar 3 )\Big) \xi_{2,+}^I \nonumber  \\
    &\phantom{ccc}-(1 2 \bar 3) \xi_{3,+}^I- \frac{1}{16} \Big( (1 \bar 1 \bar 3)+(2 \bar 2 \bar 3) \Big)\xi_{1,+}^I\xi_{2,+}^I\xi_{3,+}^I \Biggr]^2\,.
\end{align}
For test functions containing an odd number of $\xi_{i,+}^2$ pairs, we instead obtain
\begin{align}\label{eq: OG3SUSY1}
     \hat \delta \left(C_{3} \cdot  \Omega\cdot  \Xi_3^I \right)  \cdot \xi_{1,+}^2 
    &=-\Bigg[\frac{1}{2} \Big( ( 1 2 \bar 2)+ ( 1 3 \bar 3) \Big)+\frac{1}{16}\Big(( \bar 2 1 \bar 1 )- (\bar 2 3 \bar 3 ) \Big) \xi_{1,+}^I  \xi_{2,+}^I \nonumber \\
    &\phantom{ccccc}+\frac{1}{16} \Big((\bar 3 1 \bar 1)- (\bar 3 2 \bar 2)\Big) \xi_{1,+}^I \xi_{3,+}^I+\frac{1}{8} (1 \bar 2 \bar 3) \xi_{2,+}^I\xi_{3,+}^I\Bigg]^2 \,,\nonumber\\
    \hat \delta \left(C_{3} \cdot  \Omega\cdot  \Xi_3^I \right)  \cdot \prod_{i=1,2,3}  \xi_{i,+}^2 
   & =\Bigg[8 ( 1 2 3) +\frac{1}{2}\Big((1 \bar 1 3)+ (2 \bar 2 3) \Big) \xi_{1,+}^I  \xi_{2,+}^I + \frac{1}{2}\Big((2 \bar 2 1)+ (3 \bar 3 1) \Big)   \nonumber  \\
    &\phantom{cccccccc} \cdot \xi_{2,+}^I\xi_{3,+}^I +\frac{1}{2} \Big((1 \bar 1 2)+ (3 \bar 3 2)\Big) \xi_{3,+}^I \xi_{1,+}^I\Bigg]^2 \,.
\end{align}
Recall that the positive branch kills test functions with an odd number of $\xi_+^2$ pairs, while the negative branch kills those with an even number. In the explicit formulae above, these vanishings follow from the minor identities on the two branches.

\subsection{Four points}
\label{app:inv-4}
Following the $n=2,3$ construction, we build the four-point superinvariants. Since $\operatorname{OG}(4,8)$ is not fully localized by external kinematics, an additional residue condition must be specified before the Grassmannian superinvariant can be translated into momentum space. We first discuss the top cell of $\operatorname{OG}(4,8)$ and then turn to codimension-one factorization cells, where the corresponding supersymmetric invariants can be written directly in momentum space.

Let us begin with test functions containing an even number of $\xi_{i,+}^2$ pairs, which give
%\begingroup
%\allowdisplaybreaks
\begin{align}\label{SUSY inv 4pt: top----}
    & \hat \delta( C_4 \cdot \Omega\cdot \Xi_4^I )\cdot 1 \nonumber  \\
    =\, &\Biggl[
\frac{1}{16} \Big( (1 \bar 1 2 \bar 2)+(3 \bar 3 4 \bar 4)+(1 \bar 1 4\bar 4) + (2\bar 2 3 \bar 3) + (1 \bar 1 3 \bar 3 ) +(2 \bar 2 4 \bar 4 ) \Bigr) +\frac{1}{64} \Big((\bar1 \bar 2 3 \bar 3)+(\bar 1 \bar 2 4 \bar 4)\Big)  \nonumber  \\
    &\phantom{cccc} \cdot \xi_{1,+}^I \xi_{2,+}^I +\frac{1}{64} \Big((\bar 1 \bar 3 2 \bar 2) + (\bar 1 \bar 3 4 \bar 4)\Big) \xi_{1,+}^I \xi_{3,+}^I + \frac{1}{64} \Big( (\bar 1\bar 4 2 \bar 2) + (\bar 1\bar 4 3 \bar 3) \Big) \xi_{1,+}^I \xi_{4,+}^I  \nonumber  \\
    & \phantom{ccc}+ \frac{1}{64} \Big( (\bar 2 \bar 3 1 \bar 1) + (\bar 2 \bar 3 4 \bar 4) \Big) \xi_{2,+}^I \xi_{3,+}^I+\frac{1}{64} \Big( (\bar 2 \bar 4  1 \bar 1) + (\bar 2 \bar 4  3 \bar 3) \Big) \xi_{2,+}^I \xi_{4,+}^I \nonumber  \\
    & \phantom{ccc}+  \frac{1}{64}\Big( (\bar 3 \bar 4 1 \bar 1)+ (\bar 3 \bar 4 2 \bar 2)\Big) \xi_{3,+}^I \xi_{4,+}^I+\frac{1}{256} (\bar 1\bar 2\bar 3\bar 4 ) \xi_{1,+}^I \xi_{2,+}^I \xi_{3,+}^I \xi_{4,+}^I\Biggr]^2\,;
\end{align}
Note that the first term is simply $A$ in the positive branch. Similar configurations appear in 
%\endgroup
\begin{align}
    \label{SUSY inv 4pt: top--++}
    &\hat\delta (C_4 \cdot \Omega \cdot \Xi_4^I)\cdot  \xi_{1,+}^2 \xi_{2,+}^2  \nonumber \\
    =-\, & \Biggl[- (12  3 \bar 3)- (12  4 \bar 4) +\frac{1}{16}\Big( (1\bar 1 2 \bar 2)+(3 \bar 3 4 \bar 4 )-(1 \bar 1 3 \bar 3) -(2\bar 2 4 \bar 4) -(1 \bar 1 4 \bar 4) -(2\bar 2 3 \bar 3) \Big)  \nonumber  \\
    &\phantom{cccc}\cdot \xi_{1,+}^I \xi_{2,+}^I +\frac{1}{8} \Big( (1\bar 1 2 \bar 3)- (4\bar 4 2 \bar 3) \Big) \xi_{1,+}^I \xi_{3,+}^I+\frac{1}{8} \Big( (1\bar 1 2 \bar 4)-(3 \bar 3 2 \bar 4 )\Big) \xi_{1,+}^I \xi_{4,+}^I\nonumber \\
    &\phantom{cc}-\frac{1}{8} \Big( (2 \bar 2 1 \bar 3)-(4\bar 4 1 \bar 3 ) \Big) \xi_{2,+}^I \xi_{3,+}^I-\frac{1}{8} \Big( (2 \bar 2 1 \bar 4 )- (3 \bar 3  1 \bar 4 )\Big) \xi_{2,+}^I \xi_{4,+}^I \nonumber  \\
    & \phantom{cc}-\frac{1}{4}  (1  2 \bar 3 \bar 4 ) \xi_{3,+}^I \xi_{4,+}^I-\frac{1}{64} \Big( (1 \bar 1 \bar 3 \bar 4 )+(2 \bar 2 \bar 3 \bar 4 )\Big) \xi_{1,+}^I \xi_{2,+}^I \xi_{3,+}^I \xi_{4,+}^I\Biggr]^2\,;
\end{align}
Here the helicity-neutral term becomes $D$. Since different terms in the $\xi$ expansion project onto different component WFCs, the Grassmannian representation naturally produces denominators involving different sign choices of $S,T,U$. We also have
\begin{align}\label{SUSY inv 4pt: top++++}
    & \hat\delta (C_4 \cdot \Omega \cdot \Xi_4^I)\cdot \prod_{i=1,2,3,4} \xi_{i,+}^2  \nonumber \\
  =\,& \Biggl[ 16\, (1234) + \Big( (34 1 \bar 1) + (34 2 \bar 2 )\Big) \xi_{1,+}^I  \xi_{2,+}^I -\Big( (2 4 1 \bar 1) +  (24 3\bar 3)\Big) \xi_{1,+}^I  \xi_{3,+}^I 
   \nonumber  \\
  &\phantom{cc}+ \Big( (23 1\bar 1)+(234 \bar 4 )\Big) \xi_{1,+}^I  \xi_{4,+}^I + \Big( (1 4 2 \bar 2) +  (14 3\bar 3)\Big) \xi_{2,+}^I  \xi_{3,+}^I\nonumber  \\
  &\phantom{cc}-  \Big( (13 2\bar 2) + (134 \bar 4 )\Big) \xi_{2,+}^I  \xi_{4,+}^I+ \Big( (12 3\bar 3)+(12 4 \bar 4 )\Big) \xi_{3,+}^I  \xi_{4,+}^I+\frac{1}{16} \Big( (1 \bar 1 2 \bar 2)\nonumber \\
  &\phantom{cc}+(3 \bar 3 4 \bar 4) +(1 \bar 1 4\bar 4)+(2\bar 2 3 \bar 3)+(1 \bar 1 3 \bar 3 )+(2 \bar 2 4 \bar 4 ) \Big) \xi_{1,+}^I  \xi_{2,+}^I \xi_{3,+}^I  \xi_{4,+}^I\Biggr]^2 \,. 
\end{align}

For test functions containing an odd number of $\xi_{i,+}^2$ pairs, we instead obtain
\begin{align}\label{SUSY inv 4pt: top-+++}
    &  \hat\delta (C_4 \cdot \Omega \cdot \Xi_4^I)\cdot  \xi_{1,+}^2  \nonumber \\
   =\,&\Biggl[\frac{1}{16} \Big((1 \bar1 2 \bar 2 )-(3 \bar 3 4 \bar 4 )+(1 \bar 1 3 \bar 3)-(2 \bar 2 4 \bar 4 )+ (1 \bar 1 4 \bar 4)-(2\bar 2 3 \bar 3 )\Big) \xi_{1,+}^I\nonumber\\
   &\phantom{ccc}+\frac{1}{8} \Big((1 \bar2 3 \bar 3 )+(1 \bar 2 4 \bar 4)\Big) \xi_{2,+}^I+\frac{1}{8} \Big((1 \bar3 2 \bar 2 )+(1 \bar 3 4 \bar 4)\Big) \xi_{3,+}^I+\frac{1}{8}\Big((1 \bar4 2 \bar 2 )+(1 \bar 4 3 \bar 3)\Big) \nonumber \\
   &\phantom{ccc}\cdot  \xi_{4,+}^I+\frac{1}{64} \Big( (1 \bar 1\bar 2\bar 3)-(4 \bar 4 \bar 2 \bar3  )\Big) \xi_{1,+}^I  \xi_{2,+}^I \xi_{3,+}^I+\frac{1}{64} \Big( (1 \bar 1 \bar 2 \bar 4)-(3 \bar 3 \bar2 \bar 4)\Big) \xi_{1,+}^I  \xi_{2,+}^I \xi_{4,+}^I\nonumber \\
   &\phantom{ccc}+\frac{1}{64} \Big( (1 \bar 1 \bar 3 \bar 4)-(2 \bar 2 \bar3 \bar 4)\Big) \xi_{1,+}^I  \xi_{3,+}^I \xi_{4,+}^I +\frac{1}{32} (1\bar 2 \bar 3 \bar 4) \xi_{2,+}^I  \xi_{3,+}^I \xi_{4,+}^I\Biggr]^2 \,;
\end{align}
\begin{equation}\label{SUSY inv 4pt: top+---}
    {\begin{split}
        &\,  \hat\delta (C_4 \cdot \Omega \cdot \Xi_4^I)\cdot \prod_{i=2,3,4} \xi_{i,+}^2  \\
=-\,& \Biggl[ -2 (\bar 1  2 3 4) \xi_{1,+}^I  - \Big( (\bar 2 234)-(\bar 1 134)\Big) \xi_{2,+}^I - \Big( (\bar 3 234)-(\bar 1 2 1 4)\Big) \xi_{3,+}^I \\
&\phantom{cccc}- \Big( (\bar 4 234 )-(\bar 1 231 )\Big)  \xi_{4,+}^I-\frac{1}{8} \Big( (\bar 1  4 2 \bar 2)+(\bar 1  4 3 \bar 3)\Big) \xi_{1,+}^I  \xi_{2,+}^I  \xi_{3,+}^I\\
&\phantom{cccc}+\frac{1}{8} \Big( (\bar 1 3 2 \bar 2)+(\bar 1 3 4 \bar 4)\Big)\xi_{1,+}^I  \xi_{2,+}^I  \xi_{4,+}^I
-\frac{1}{8} \Big( (\bar 1 2  3 \bar 3)+(\bar 1 2  4  \bar 4)\Big)\xi_{1,+}^I  \xi_{3,+}^I  \xi_{4,+}^I\\
&\phantom{ccc}-\frac{1}{16} \Big( (1\bar 1  2\bar 2) - (3\bar 3 4 \bar 4) + (1 \bar 1  3 \bar 3) -(2\bar 2 4 \bar 4) + (1 \bar 1 4 \bar 4)-(2 \bar 2 3 \bar 3)\Big)  \xi_{2,+}^I  \xi_{3,+}^I  \xi_{4,+}^I\Biggr]^2\,. 
    \end{split}}
\end{equation}

We next obtain momentum-space expressions for the four-point invariants by imposing the vanishing of an appropriate minor. 

\subsubsection*{Flat-space-limit cells}
We begin with the cell relevant to the flat-space limit, which satisfies
\begin{align}
    A\equiv S+T+U=0\,.
\end{align}
In this locus, each of the positive and negative branches further splits into two solutions labeled as $\{\alpha,\beta\}$. For the positive branch, these are

\begin{align}\label{eq: amplitude cell 4pt 1}
 \{ {+,\alpha} \}:  \left(
\begin{array}{cccccccc}
 0 & 0 & 0 & 0 & 1 & 0 & \frac{\langle 41\rangle }{\langle34\rangle} & \frac{\langle 13\rangle}{\langle 34\rangle } \\
 0 & 0 & 0 & 0 & 0 & 1 & \frac{\langle 42\rangle }{\langle 34\rangle} & \frac{\langle 23\rangle }{\langle 34\rangle } \\
 -\frac{\langle 41\rangle }{\langle 34\rangle} & -\frac{\langle 42\rangle}{\langle 34\rangle} & 1 & 0 & 0 & 0 & 0 & \frac{E_T}{\langle 34\rangle } \\
 -\frac{\langle 13\rangle }{\langle 34 \rangle} & -\frac{\langle 2 3\rangle}{\langle 34\rangle } & 0 & 1 & 0 & 0 & -\frac{E_T}{\langle 34\rangle } & 0 \\
\end{array}
\right)\,,\quad  \{ {+,\beta} \}:   \left(
\begin{array}{cccccccc}
 0 & -\frac{E_T}{\langle \bar 1\bar 2\rangle } & 0 & 0 & 1 & 0 & -\frac{\langle \bar 2\bar 3\rangle }{\langle \bar 1\bar2\rangle} & -\frac{\langle \bar 2\bar 4\rangle}{\langle \bar 1 \bar 2\rangle } \\
 \frac{E_T}{\langle \bar 1\bar 2\rangle } & 0 & 0 & 0 & 0 & 1 & -\frac{\langle \bar 3 \bar 1\rangle }{\langle \bar 1 \bar 2\rangle} & -\frac{\langle \bar 4\bar 1\rangle }{\langle \bar 1\bar 2\rangle } \\
 \frac{\langle \bar 2\bar3\rangle }{\langle \bar 1\bar 2\rangle} & \frac{\langle \bar 3\bar 1\rangle}{\langle \bar 1\bar 2\rangle} & 1 & 0 & 0 & 0 & 0 & 0 \\
 \frac{\langle \bar 2\bar 4\rangle }{\langle \bar 1\bar 2 \rangle} & \frac{\langle \bar 4 \bar 1\rangle}{\langle \bar 1\bar 2\rangle } & 0 & 1 & 0 & 0 & 0 & 0 \\
\end{array}
\right)
\end{align}

For the negative branch, the two solutions are
\begingroup
\allowdisplaybreaks
\begin{align}\label{eq: amplitude cell 4pt 3}
  \{-,\alpha\}: \left(
\begin{array}{cccccccc}
 1 & -\frac{\langle  \bar 1 2\rangle}{E_T-2E_1} & -\frac{\langle  \bar 1 3\rangle}{E_T-2E_1} & -\frac{\langle  \bar 1 4\rangle}{E_T-2E_1} & 0 & 0 & 0 & 0 \\
 0 & 0 & -\frac{\langle23\rangle}{E_T-2E_1} & -\frac{\langle24\rangle}{E_T-2E_1}  & \frac{\langle  \bar 1 2\rangle}{E_T-2E_1} & 1 & 0 & 0 \\
 0 & \frac{\langle23\rangle}{E_T-2E_1} & 0 & -\frac{\langle34\rangle}{E_T-2E_1} & \frac{\langle  \bar 1 3\rangle}{E_T-2E_1} & 0 & 1 & 0 \\
 0 & \frac{\langle24\rangle}{E_T-2E_1} & \frac{\langle34\rangle}{E_T-2E_1} & 0 & \frac{\langle  \bar 1 4\rangle}{E_T-2E_1} & 0 & 0 & 1 \\
\end{array}
\right)\,
\end{align}
\endgroup
\begin{align}
  \{-,\beta\}:\left(
\begin{array}{cccccccc}
 0 & 0 & 0 & 0 & 1 & \frac{\langle 1 \bar 2 \rangle}{E_T-2 E_1} & \frac{\langle 1 \bar 3 \rangle}{E_T-2 E_1} & \frac{\langle 1 \bar 4 \rangle}{E_T-2 E_1} \\
 -\frac{\langle 1 \bar 2 \rangle}{E_T-2 E_1} & 1 & 0 & 0 &0 & 0 & \frac{\langle  \bar 2 \bar 3\rangle}{E_T-2E_1}  & \frac{\langle  \bar 2 \bar 4\rangle}{E_T-2E_1}   \\
 -\frac{\langle 1 \bar 3 \rangle}{E_T-2 E_1} & 0 & 1 & 0 & 0 & -\frac{\langle \bar 2\bar 3 \rangle}{E_T-2E_1} & 0 & \frac{\langle \bar 3\bar 4 \rangle}{E_T-2E_1} \\
 -\frac{\langle 1 \bar 4 \rangle}{E_T-2 E_1} & 0 & 0 & 1 & 0 & -\frac{\langle \bar2 \bar 4 \rangle}{E_T-2 E_1} & -\frac{\langle \bar 3 \bar 4 \rangle}{E_T-2 E_1} & 0 \\
\end{array}
\right)\,.
\end{align}

Substituting these solutions into the top-cell expressions~\eqref{SUSY inv 4pt: top----}--\eqref{SUSY inv 4pt: top+---}, one finds that only four inequivalent supersymmetric invariants are produced:
 \begingroup
\allowdisplaybreaks
\begin{align}
&\Gamma_{4,\alpha}^{++}\Big|_{A=0}
:=
\Biggl[
\frac{E_T}{8}\xi_{1,+}^I \xi_{2,+}^I\xi_{3,+}^I\xi_{4,+}^I
-\langle 34\rangle 
\xi_{1,+}^I\xi_{2,+}^I
+ \langle 24\rangle \xi_{1,+}^I\xi_{3,+}^I- \langle 23\rangle \xi_{1,+}^I\xi_{4,+}^I
\nonumber 
\\
&\phantom{ccccccccccccccccc}-\langle 14\rangle \xi_{2,+}^I\xi_{3,+}^I +\langle 13\rangle\xi_{2,+}^I\xi_{4,+}^I
-\langle 12\rangle \xi_{3,+}^I\xi_{4,+}^I
\Biggr]^2\,;\\
&\Gamma_{4,\beta}^{++}\Big|_{A=0}:=\Biggl[
8 E_T -
\langle \bar1\bar2\rangle \xi_{1,+}^I\xi_{2,+}^I
-\langle \bar1\bar3\rangle \xi_{1,+}^I\xi_{3,+}^I
-\langle \bar1\bar4\rangle \xi_{1,+}^I\xi_{4,+}^I
-\langle \bar2\bar3\rangle \xi_{2,+}^I\xi_{3,+}^I\nonumber\\
& \phantom{cccccccccccccccccccccccc}
-\langle \bar2\bar4\rangle \xi_{2,+}^I\xi_{4,+}^I
-\langle \bar3\bar4\rangle \xi_{3,+}^I\xi_{4,+}^I
\Biggr]^2\,;\\
&\Gamma_{4,\alpha}^{--}\Big|_{A=0}
:=
\Biggl[
(E_T-2E_1)\xi_{2,+}^I\xi_{3,+}^I\xi_{4,+}^I
-\langle 4\bar1\rangle \xi_{1,+}^I\xi_{2,+}^I\xi_{3,+}^I+\langle 3\bar1\rangle
\xi_{1,+}^I\xi_{2,+}^I\xi_{4,+}^I\nonumber\\
&\phantom{cccccccccccccccccc}-\langle 2\bar1\rangle\xi_{1,+}^I\xi_{3,+}^I\xi_{4,+}^I
-8\langle 34\rangle \xi_{2,+}^I
-8\langle 42\rangle \xi_{3,+}^I
-8\langle 23\rangle\xi_{4,+}^I
\Biggr]^2\, ;\\
&\Gamma_{4,\beta}^{--}\Big|_{A=0}
:=
\Biggl[
8(E_T-2E_1)\xi_{1,+}^I
-
\langle \bar2\bar3\rangle \xi_{1,+}^I\xi_{2,+}^I\xi_{3,+}^I
-\langle \bar2\bar4\rangle \xi_{1,+}^I\xi_{2,+}
^I\xi_{4,+}^I
 \nonumber \\
&\phantom{ccccccccccccccccccc}
-\langle \bar3\bar4\rangle\xi_{1,+}^I\xi_{3,+}^I\xi_{4,+}^I
+8\langle 1 \bar2 \rangle \xi_{2,+}^I
+8\langle 1 \bar3 \rangle \xi_{3,+}^I
+8\langle 1 \bar4 \rangle \xi_{4,+}^I
\Biggr]^2.
\end{align}
\endgroup

The first two invariants, $\Gamma_{4,i}^{++}\big|_{A=0}$ for $i=\alpha,\, \beta$, can be identified with the squares of the four-point even invariant found in the $\mathcal{N}=1$ analysis of~\cite{Jain:2023idr}. In contrast, the remaining two invariants, $\Gamma_{4,i}^{--}\big|_{A=0}$ for $i=\alpha,\,\beta$, do not coincide directly with the corresponding invariant in the $\mathcal{N}=1$ analysis. Rather, the invariant discussed in~\cite{Jain:2023idr} arises from the condition
\begin{align}
   C\equiv-S+T-U=0\,.
\end{align}

\subsubsection*{Factorization cells}

The second class of cells captures factorization channels. Let us focus on the $s$-channel, 
\begin{align}
    S=0\,,
\end{align}
from which the $t$-channel  follows by the exchange $2\leftrightarrow 4$.

On this locus, the positive branch splits into two solutions:
\begin{align}
       C^+_{4,\alpha}= \begin{pmatrix}
\lambda_1^\alpha & \lambda_2^\alpha  &\lambda_3^\alpha & \lambda_4^\alpha &\bar{\lambda}_1^\alpha & \bar{\lambda}_2^\alpha &\bar \lambda_3^\alpha & \bar \lambda_4^\alpha \\
0 &0   &\frac{\langle 1 2 \rangle\langle 3 s \rangle}{\langle  s \bar s \rangle} & \frac{\langle 1 2 \rangle\langle 4 s \rangle}{\langle  s \bar s \rangle}  &\langle 2 s\rangle &\langle  s 1\rangle   &\frac{\langle 1 2 \rangle\langle \bar 3 s \rangle}{\langle  s \bar s \rangle}  & \frac{\langle 1 2 \rangle\langle \bar 4 s \rangle}{\langle s \bar s \rangle}\\
0 &0  &\frac{-\langle \bar 4 \bar s \rangle\langle s \bar s\rangle+ \langle\bar3\bar4\rangle\langle3 \bar s\rangle}{\langle s \bar s\rangle}  &\frac{\langle \bar 3 \bar s \rangle\langle s \bar s\rangle+\langle\bar3\bar4\rangle\langle 4 \bar s \rangle}{\langle s \bar s\rangle} &0 &0 &\frac{\langle\bar3\bar4\rangle\langle\bar 3\bar s\rangle}{\langle s \bar s\rangle} &\frac{ \langle\bar3\bar4\rangle\langle \bar4 \bar s \rangle}{\langle s \bar s\rangle} 
\end{pmatrix}
\end{align}

\begin{align}
       C^+_{4,\beta}= \begin{pmatrix}
\lambda_1^\alpha & \lambda_2^\alpha  &\lambda_3^\alpha & \lambda_4^\alpha &\bar{\lambda}_1^\alpha & \bar{\lambda}_2^\alpha &\bar \lambda_3^\alpha & \bar \lambda_4^\alpha \\
\langle \bar 2 \bar s \rangle  &\langle \bar s \bar 1\rangle   &-\frac{\langle \bar 1 \bar  2 \rangle\langle 3 \bar s \rangle}{\langle  s \bar s \rangle} & -\frac{\langle \bar 1 \bar 2 \rangle\langle 4 \bar s \rangle}{\langle  s \bar s \rangle}  &0 &0   &-\frac{\langle \bar 1 \bar 2 \rangle\langle \bar 3 \bar s \rangle}{\langle  s \bar s \rangle}  & -\frac{\langle \bar 1 \bar 2 \rangle\langle \bar 4 \bar s \rangle}{\langle s \bar s \rangle}\\
0 &0  &-\frac{ \langle  3 4 \rangle\langle 3 s \rangle}{\langle s \bar s\rangle}  &-\frac{\langle3 4\rangle\langle 4  s \rangle}{\langle s \bar s\rangle} &0 &0 &\frac{-\langle 4 s \rangle \langle s \bar s \rangle -\langle 3 4\rangle\langle\bar 3 s\rangle}{\langle s \bar s\rangle} &\frac{ \langle 3 s \rangle \langle s \bar s \rangle-\langle 3 4 \rangle\langle \bar4  s \rangle}{\langle s \bar s\rangle} 
\end{pmatrix}
\end{align}
Similarly, the negative branch also admits two solutions:
\begin{align}
       C^-_{4,\alpha}= \begin{pmatrix}
\lambda_1^\alpha & \lambda_2^\alpha  &\lambda_3^\alpha & \lambda_4^\alpha &\bar{\lambda}_1^\alpha & \bar{\lambda}_2^\alpha &\bar \lambda_3^\alpha & \bar \lambda_4^\alpha \\
0 &0   &\frac{\langle 1 2 \rangle\langle 3 s \rangle}{\langle  s \bar s \rangle} & \frac{\langle 1 2 \rangle\langle 4 s \rangle}{\langle  s \bar s \rangle}  &\langle 2 s\rangle &\langle  s 1\rangle   &\frac{\langle 1 2 \rangle\langle \bar 3 s \rangle}{\langle  s \bar s \rangle}  & \frac{\langle 1 2 \rangle\langle \bar 4 s \rangle}{\langle s \bar s \rangle}\\
0 &0  &-\frac{ \langle  3 4 \rangle\langle 3 s \rangle}{\langle s \bar s\rangle}  &-\frac{\langle3 4\rangle\langle 4  s \rangle}{\langle s \bar s\rangle} &0 &0 &\frac{-\langle 4 s \rangle \langle s \bar s \rangle -\langle 3 4\rangle\langle\bar 3 s\rangle}{\langle s \bar s\rangle} &\frac{ \langle 3 s \rangle \langle s \bar s \rangle-\langle 3 4 \rangle\langle \bar4  s \rangle}{\langle s \bar s\rangle} 
\end{pmatrix}
\end{align}

\begin{align}
       C^-_{4,\beta}= \begin{pmatrix}
\lambda_1^\alpha & \lambda_2^\alpha  &\lambda_3^\alpha & \lambda_4^\alpha &\bar{\lambda}_1^\alpha & \bar{\lambda}_2^\alpha &\bar \lambda_3^\alpha & \bar \lambda_4^\alpha \\
\langle \bar 2 \bar s \rangle  &\langle \bar s \bar 1\rangle   &-\frac{\langle \bar 1 \bar  2 \rangle\langle 3 \bar s \rangle}{\langle  s \bar s \rangle} & -\frac{\langle \bar 1 \bar 2 \rangle\langle 4 \bar s \rangle}{\langle  s \bar s \rangle}  &0 &0   &-\frac{\langle \bar 1 \bar 2 \rangle\langle \bar 3 \bar s \rangle}{\langle  s \bar s \rangle}  & -\frac{\langle \bar 1 \bar 2 \rangle\langle \bar 4 \bar s \rangle}{\langle s \bar s \rangle}\\
0 &0  &\frac{-\langle \bar 4 \bar s \rangle\langle s \bar s\rangle+ \langle\bar3\bar4\rangle\langle3 \bar s\rangle}{\langle s \bar s\rangle}  &\frac{\langle \bar 3 \bar s \rangle\langle s \bar s\rangle+\langle\bar3\bar4\rangle\langle 4 \bar s \rangle}{\langle s \bar s\rangle} &0 &0 &\frac{\langle\bar3\bar4\rangle\langle\bar 3\bar s\rangle}{\langle s \bar s\rangle} &\frac{ \langle\bar3\bar4\rangle\langle \bar4 \bar s \rangle}{\langle s \bar s\rangle} 
\end{pmatrix}
\end{align}
where the exchanged momentum is $
p_s^{\alpha\beta}=p_1^{\alpha\beta}+p_2^{\alpha\beta}=\lambda_s^{(\alpha}\bar{\lambda}_s^{\beta)}.$

Substituting  them  into \eqref{SUSY inv 4pt: top----}--\eqref{SUSY inv 4pt: top+---}, one obtains the following compact expressions for the inequivalent
$s$-channel invariants:
\begin{equation}
\begin{aligned}
    \Gamma_{4,\alpha}^{++}\Big|_{S=0}
    :=\Biggl[&-2\vev{12}E_{3,4,s}+\frac14E_{1,2,\bar s}E_{3,4,s}\,\xi_{1,+}^{I}\xi_{2,+}^{I}
    -\frac14\vev{2s}\vev{\bar s\bar3}\,\xi_{1,+}^{I}\xi_{3,+}^{I}\\
    &-\frac14\vev{2s}\vev{\bar s\bar4}\,\xi_{1,+}^{I}\xi_{4,+}^{I}
    +\frac14\vev{1s}\vev{\bar s\bar3}\,\xi_{2,+}^{I}\xi_{3,+}^{I}
    +\frac14\vev{1s}\vev{\bar s\bar4}\,\xi_{2,+}^{I}\xi_{4,+}^{I}\\
    &+\frac14\vev{12}\vev{\bar3\bar4}\,\xi_{3,+}^{I}\xi_{4,+}^{I}
    -\frac1{32}\vev{\bar3\bar4}E_{1,2,\bar s}\,\xi_{1,+}^{I}\xi_{2,+}^{I}\xi_{3,+}^{I}\xi_{4,+}^{I}\Biggr]^2,
\end{aligned}
\end{equation}
\begin{equation}
\begin{aligned}
    \Gamma_{4,\beta}^{++}\Big|_{S=0}
    :=\Biggl[&-2\vev{34}E_{1,2,\bar s}+\frac14\vev{34}\vev{\bar1\bar2}\,\xi_{1,+}^{I}\xi_{2,+}^{I}
    -\frac14\vev{4s}\vev{\bar s\bar1}\,\xi_{1,+}^{I}\xi_{3,+}^{I}\\
    &+\frac14\vev{3s}\vev{\bar s\bar1}\,\xi_{1,+}^{I}\xi_{4,+}^{I}
    -\frac14\vev{4s}\vev{\bar s\bar2}\,\xi_{2,+}^{I}\xi_{3,+}^{I}
    +\frac14\vev{3s}\vev{\bar s\bar2}\,\xi_{2,+}^{I}\xi_{4,+}^{I}\\
    &+\frac14E_{1,2,\bar s}E_{3,4,s}\,\xi_{3,+}^{I}\xi_{4,+}^{I}
    -\frac1{32}\vev{\bar1\bar2}E_{3,4,s}\,\xi_{1,+}^{I}\xi_{2,+}^{I}\xi_{3,+}^{I}\xi_{4,+}^{I}\Biggr]^2,
\end{aligned}
\end{equation}
\begin{equation}
\begin{aligned}
    \Gamma_{4,\alpha}^{--}\Big|_{S=0}
    :=\Biggl[&-\vev{2s}\vev{34}\,\xi_{1,+}^{I}+\vev{1s}\vev{34}\,\xi_{2,+}^{I}
    -\vev{12}\vev{s4}\,\xi_{3,+}^{I}+\vev{12}\vev{s3}\,\xi_{4,+}^{I}\\
    &+\frac18E_{1,2,\bar s}\vev{s4}\,\xi_{1,+}^{I}\xi_{2,+}^{I}\xi_{3,+}^{I}
    -\frac18E_{1,2,\bar s}\vev{s3}\,\xi_{1,+}^{I}\xi_{2,+}^{I}\xi_{4,+}^{I}\\
    &+\frac18E_{3,4,s}\vev{2s}\,\xi_{1,+}^{I}\xi_{3,+}^{I}\xi_{4,+}^{I}
    -\frac18E_{3,4,s}\vev{1s}\,\xi_{2,+}^{I}\xi_{3,+}^{I}\xi_{4,+}^{I}\Biggr]^2,
 \end{aligned}
\end{equation}
\begin{equation}
\begin{aligned}
    \Gamma_{4,\beta}^{--}\Big|_{S=0}
    :=\Biggl[&-E_{3,4,s}\vev{\bar1\bar s}\,\xi_{1,+}^{I}-E_{3,4,s}\vev{\bar2\bar s}\,\xi_{2,+}^{I}
    -E_{1,2,\bar s}\vev{\bar s\bar3}\,\xi_{3,+}^{I}-E_{1,2,\bar s}\vev{\bar s\bar4}\,\xi_{4,+}^{I}\\
    &+\frac18\vev{\bar1\bar2}\vev{\bar s\bar3}\,\xi_{1,+}^{I}\xi_{2,+}^{I}\xi_{3,+}^{I}
    +\frac18\vev{\bar1\bar2}\vev{\bar s\bar4}\,\xi_{1,+}^{I}\xi_{2,+}^{I}\xi_{4,+}^{I}\\
    &+\frac18\vev{\bar3\bar4}\vev{\bar1\bar s}\,\xi_{1,+}^{I}\xi_{3,+}^{I}\xi_{4,+}^{I}
    +\frac18\vev{\bar3\bar4}\vev{\bar2\bar s}\,\xi_{2,+}^{I}\xi_{3,+}^{I}\xi_{4,+}^{I}\Biggr]^2.
\end{aligned}
\end{equation}
Here we introduced the shorthand
\begin{equation}
    E_{1,2,\bar s}=E_1+E_2-E_s,
    \qquad
    E_{3,4,s}=E_3+E_4+E_{s}.
\end{equation}
The corresponding $t$-channel expressions are obtained by the exchange $2\leftrightarrow 4$.

\bibliographystyle{utphys}
\bibliography{bib}
    
\end{document}